\documentclass[useAMS,usenatbib]{mn2e}
\usepackage{graphicx}
\usepackage{txfonts}
\usepackage{fixabs}

\def\Pr{\mathop{Pr}}
\let\tilde=\widetilde
\title[Parameterisation effects in SZ analysis]{Parameterisation effects in the
analysis of AMI Sunyaev--Zel'dovich observations \thanks{We request that any reference to this paper cites"AMI Consortium: Olamaie et~al. 2011"}}
\label{firstpage}
\author[AMI Consortium: Olamaie et~al.]{AMI Consortium:
 Malak Olamaie,$^{1}$\thanks{Email:mo323@mrao.cam.ac.uk}
 Carmen Rodr\'{i}guez-Gonz\'{a}lvez,$^{1}$\newauthor
 Matthew L. Davies,$^{1}$
 Farhan Feroz,$^{1}$
 Thomas M. O. Franzen,$^{1}$
 Keith J. B. Grainge,$^{1,2}$\newauthor
 Michael P. Hobson,$^{1}$
 Natasha Hurley-Walker,$^{1}$
 Anthony N. Lasenby,$^{1,2}$\newauthor
 Guy G. Pooley,$^{1}$
 Richard D. E. Saunders,$^{1,2}$
 Anna M. M. Scaife,$^{1,3}$\newauthor
 Michel Schammel,$^{1}$
 Paul F. Scott,$^{1}$
 Timothy W. Shimwell,$^{1}$
 David J. Titterington,$^{1}$\newauthor
 Elizabeth M. Waldram,$^{1}$
 and Jonathan T. L. Zwart$^{1,4}$\\
$^{1}$ Astrophysics Group, Cavendish Laboratory,
       19 J. J. Thomson Avenue, Cambridge, CB3 0HE\\
$^{2}$ Kavli Institute for Cosmology Cambridge, Madingley Road,
       Cambridge, CB3 0HA\\
$^{3}$ Dublin Institute for Advanced Studies,
       31 Fitzwilliam Place, Dublin 2, Ireland\\
$^{4}$ Columbia Astrophysics Laboratory, Columbia University,
       550 West 120th Street, New York, NY 10027, USA}
\begin{document}

\date{Accepted 12/12/11; Received 22/11/10}
\pagerange{\pageref{firstpage}--\pageref{lastpage}}
\pubyear{2011}

\maketitle

\vspace*{-100pt}

\begin{abstract}
Most Sunyaev--Zel'dovich (SZ) and X-ray analyses of galaxy clusters try to constrain the cluster total mass ($M_{\rm {T}}(r)$) and/or gas mass ($M_{\rm {g}}(r)$) using parameterised models derived from 
both simulations and imaging observations, and assumptions of spherical symmetry and 
hydrostatic equilibrium. By numerically exploring the probability distributions of 
the cluster parameters given the simulated interferometric SZ data in the context of 
Bayesian methods, and assuming a $\beta$-model for the electron number density $n_
{\rm e}(r)$ described by two shape parameters $\beta$ and $r_{\rm c}$, we 
investigate the capability of this model and analysis to return the simulated 
cluster input quantities via three parameterisations. In parameterisation I  we
assume that the gas temperature is an independent free parameter and assume 
hydrostatic equilibrium, spherical geometry and an ideal gas equation of state. We 
find that  parameterisation I can hardly constrain the cluster parameters and fails 
to recover the true values of the simulated cluster. In particular it 
overestimates $M_{\rm T}(r_{\rm 200})$ and $T_{\rm g}(r_{\rm 200})$ ($M_{\rm T}(r_{\rm 200})=(6.43\pm 5.43)\times 10^{15}\,\rm{M_\odot}$ and 
$T_{\rm g}(r_{\rm 200})=(10.61 \pm 5.28)\,\rm{keV}$ ) compared to the corresponding  
values of the simulated cluster ($M_{\rm T}(r_{\rm 200})=5.83\times 10^{14}\,\rm{M_
\odot}$ and $T_{\rm g}(r_{\rm 200})=5 \,\rm{keV} $). We then investigate 
parameterisations II and III in which $f_{\rm g}(r_{\rm 200})$ replaces temperature   
as a main variable; we do this because $f_{\rm g}$ may vary significantly less from 
cluster to cluster than temperature. In parameterisation II we relate $M_{\rm T}(r_
{\rm 200})$ and $T_{\rm g}$ assuming hydrostatic equilibrium. We find that 
parameterisation II can constrain the cluster physical parameters but the 
temperature estimate is biased low ( $M_{\rm T}(r_{\rm 200})= (6.8 \pm 2.1)\times 10^
{14}\,\rm{M_\odot}$ and $T_{\rm g}(r_{\rm 200})=(3.0 \pm 1.2)\,\rm{keV}
$). In parameterisation III, the virial theorem (plus the assumption that all the 
kinetic energy of the cluster is the internal energy of the gas) replaces 
the hydrostatic equilibrium assumption because we consider it more robust both in 
theory and in practice. We find that parameterisation III results in 
unbiased estimates of the cluster properties ($M_{\rm T}(r_{\rm 200})= (4.68 \pm 
1.56)\times 10^{14}\,\rm{M_\odot}$ and $T_{\rm g}(r_{\rm 200})= (4.3 \pm 
0.9)\,\rm{keV}$). We generate a second simulated cluster using a generalised
NFW (GNFW) pressure profile and analyse it with an entropy based model to take into
account the temperature gradient in our analysis and  improve the cluster gas
density distribution. This model also constrains the cluster physical
parameters and the results show a radial decline in the gas temperature as
expected. The mean cluster total mass estimates are also within $1 \sigma$
from the simulated cluster true values: $M_{\rm T}(r_{\rm 200})= (5.9 \pm
3.4)\times 10^{14}\,\rm{M_\odot}$ and $T_{\rm g}(r_{\rm 200})= (7.4 \pm
2.6)\,\rm{keV}$ using parameterisation II and $M_{\rm T}(r_{\rm 200})= (8.0 \pm
5.6)\times 10^{14}\,\rm{M_\odot}$ and $T_{\rm g}(r_{\rm 200})= (5.98 \pm
2.43)\,\rm{keV}$ using parameterisation III. However, we find that for at least 
interferometric SZ analysis in practice at the present time, there is no differences 
in the AMI visibilities between the two models. This may of course change as the 
instruments improve. 
\end{abstract}
\begin{keywords}
  galaxies: clusters: general -- cosmic microwave background --
  cosmology: observations -- methods: data analysis
\end{keywords}
\clearpage
\section{Introduction}
Clusters of galaxies contain large reservoirs of hot, ionized gas. This plasma,
although invisible in the optical waveband, can be observed in both X-ray and
microwave bands of the electromagnetic spectrum through thermal Bremsstrahlung
radiation and its scattering of the cosmic microwave background (CMB)
respectively. This inverse Compton scattering results in a decrement in the
intensity of CMB photons in the direction of the cluster at frequencies $<
218$~GHz, and is known as the Sunyaev--Zel'dovich (SZ) effect
\citep{1970CoASP...2...66S, 1999PhR...310...97B, 2002ARA&A..40..643C}.
%
%
To describe the full spectral behaviour of the SZ effect, one needs to consider three 
main components. These include the thermal SZ effect caused by
thermal (random) motion of scattering electrons, including thermal weakly 
relativistic electrons, the kinematic SZ effect caused by peculiar velocity of the 
cluster with respect to Hubble flow, and relativistic effects caused by presence of 
the energetic nonthermal electrons in the hot plasma of the cluster that are 
responsible for synchrotron emission of radio halos or relics. However, since the  
last two processes have significantly smaller effects on the overall spectral 
distortion at cm wavelengths, we only consider the thermal SZ effect in this paper. 
Moreover, we ignore the effects of weakly relativistic thermal electrons, which are 
negligible at cm wavelengths.

A main science driver for studying clusters through their thermal SZ signal
arises from the fact that SZ surface brightness is independent of redshift.
This provides us with a powerful opportunity to study galaxy clusters out to
high redshift. However, estimating the physical properties of the clusters
depends strongly on the model assumptions. In this paper we aim to show how
employing different parameterisations for a cluster model affects the
constraints on cluster properties. These tasks are conveniently carried out
through Bayesian inference using a highly efficient parameter space sampling
method: nested sampling (Skilling 2004). This sampling method is employed using
the package \textsc{Multinest} (Feroz \& Hobson 2008; Feroz, Hobson \& Bridges
2009).\textsc{ Multinest} explores the high dimensional parameter space and
calculates both the probability distribution of cluster parameters and the
Bayesian evidence. This algorithm is employed to analyse real multi-frequency
SZ observations made by the Arcminute Microkelvin Imager (AMI), (AMI
Consortium: Zwart et~al.\ 2008).

The rest of the paper is organised as follows. In Section~2, we describe the
AMI telescope. In Section~3, we discuss Bayesian inference. Section~4 gives
details of how we model interferometric SZ data. In Section 5, we describe the
modelling of the SZ signal using both the isothermal $\beta$-model and an "entropy"-
GNFW pressure model. Section 6 outlines the assumptions needed to estimate cluster 
physical parameters and describes how different parameterisations introduce 
different constraints and biases in the resulting marginalised posterior probability 
distributions. In Section 7, we describe how to generate a simulated SZ cluster in a 
consistent manner for both models, and in Section 8, we present our results.
Finally, Section 9 summarises our conclusions.
\section{The Arcminute Microkelvin Imager (AMI)}
AMI comprises two arrays: the Small Array (SA) and the Large Array (LA) located
at the Mullard Radio Astronomy Observatory near Cambridge.
The SA consists of ten 3.7-m diameter equatorially--mounted antennas surrounded
by an aluminium groundshield to suppress ground-based interference and to ensure 
that the sidelobes from the antennas do not terminate on warm emitting material. The 
LA consists of eight 13-m diameter antennas. A summary of the technical
details of AMI is given in Table~\ref{tab:antenna}. Further details of the 
instrument are in AMI Consortium: Zwart et~al.\ (2008).
\begin{table}
\centering
%
\caption{AMI technical summary.\label{tab:antenna}}
\begin{tabular}{@{}lcc@{}}\hline
                           & SA                  & LA                \\\hline
Antenna Diameter           & 3.7~m               & 12.8~m            \\
Number of Antennas         & 10                  & 8                 \\
Baseline Lengths (current) & 5--20~m             & 18--110~m         \\
Primary Beam at 15.7~GHz   & $20\farcm1$         & $5\farcm5$          \\
Synthesized Beam           & $\approx 3\arcmin$  & $\approx 30\arcsec$    \\
Flux Sensitivity           & 30~mJy~s$^{-1/2}$   & 3~mJy~s$^{-1/2}$  \\
Observing Frequency        & 13.5--18~GHz        & 13.5--18~GHz        \\
Bandwidth                  & 3.7~GHz             & 3.7~GHz           \\
Number of Channels         & 6                   & 6                 \\
Channel Bandwidth          & 0.75~GHz            & 0.75~GHz          \\\hline
\end{tabular}
%
\end{table}
\section{Bayesian Inference}
Bayesian inference has been shown to provide an efficient and robust approach
to parameter estimation in astrophysics and cosmology by offering consistent
procedures for the estimation of a set of parameters $\mbox{\boldmath
$\Theta$}$ within a model (or hypothesis) $H$ using the data $\mbox{\boldmath
$D$}$ without loss of information. Bayes' theorem states that:
\begin{equation}
  \Pr(\mbox{\boldmath $\Theta$}|\mbox{\boldmath $D$}, H) =
    \frac{\Pr(\mbox{\boldmath $D$}|\mbox{\boldmath $\Theta$}, H)
    \Pr(\mbox{\boldmath $\Theta$}|H)}{\Pr(\mbox{\boldmath $D$}|H)},
\end{equation}
where $\Pr(\mbox{\boldmath $\Theta$}|\mbox{\boldmath $D$}, H)\equiv
P(\mbox{\boldmath $\Theta$})$ is the posterior probability distribution of the
parameters, $\Pr(\mbox{\boldmath $D$}|\mbox{\boldmath $\Theta$}, H)\equiv
\mathcal{L}(\mbox{\boldmath $\Theta$})$ is the likelihood, $\Pr(\mbox{\boldmath
$\Theta$}|H)\equiv \pi(\mbox{\boldmath $\Theta$})$ is the prior probability
distribution and $\Pr(\mbox{\boldmath $D$}|H)\equiv \mathcal{Z}$ is the
Bayesian evidence.

Bayesian inference in practice often divides into two parts: parameter estimation 
and model selection. In parameter estimation, the normalising evidence factor is 
usually ignored, since it is independent of the parameters
$\mbox{\boldmath$\Theta$}$, and inferences are obtained by taking samples
from the unnormalised posterior distributions using sampling techniques. The
posterior distribution can be subsequently marginalised over each parameter to
give individual parameter constraints.

In contrast to parameter estimation, for model selection the evidence takes the
central role and is simply the factor required to normalise the posterior over
$\mbox{\boldmath$\Theta$}$:
\begin{equation}
  \mathcal{Z}=\int\mathcal{L}(\mbox{\boldmath $\Theta$})\pi(\mbox{\boldmath $\Theta
$})\rm {d}^D\mbox{\boldmath$\Theta$},
\end{equation}
where $D$ is the dimensionality of the parameter space. The question of model
selection between two models $H_0$ and $H_1$ is then decided by comparing their
respective posterior probabilities, given the observed data set
$\mbox{\boldmath $D$}$, via the model selection ratio
\begin{equation}
  R=\frac{\Pr(H_1|\mbox{\boldmath $D$})}{\Pr(H_0|\mbox{\boldmath $D$})}=\frac{\Pr
(\mbox{\boldmath $D$}|H_1)\Pr(H_1)}{\Pr(\mbox{\boldmath $D$}|
  H_0)\Pr(H_0)}=\frac{\mathcal{Z}_1}{\mathcal{Z}_0}\frac{\Pr(H_1)}{\Pr(H_0)},
\end{equation}
where $\Pr(H_1)/\Pr(H_0)$ is the \textit{a priori} probability ratio for the
two models. It should be noted that the evaluation of the multidimesional
integral in the Bayesian evidence is a challenging numerical task which can be
tackled by using \textsc{Multinest}. This Monte-Carlo method is targeted at the
efficient calculation of the evidence, but also produces posterior inferences
as a by-product. This method is also very efficient in sampling from posteriors
that may contain multiple modes or large (curving) degeneracies.
\section{Modelling interferometric SZ data}
In the cluster plasma the central optical depth $\tau$ is typically between
$0.001{-}0.01$ and the temperature $T$ varies from $10^7{-}10^8$~K. Thus the
observed SZ surface brightness in the direction of electron reservoir may be
described as
\begin{equation}
  \delta I_\nu=T_{\rm CMB}yf(\nu)\frac{\partial B_\nu}{\partial T}\Big\vert_{T=T_
{\rm CMB}}.
\end{equation}
Here $B_\nu$ is the blackbody spectrum, $T_{\rm CMB}=2.73 $~K (Fixsen et~al. 1996) 
is the temperature of the CMB radiation,
$f(\nu)=\left(x\frac{e^x+1}{e^x-1}-4\right)(1 + \delta (x , T_{\rm e})$ is the 
frequency dependence of thermal SZ signal, $x=\frac{h_{\rm p}\nu}{k_{\rm B}T_{\rm 
CMB}}$, $h_{\rm p}$ is Planck's constant, $\nu$ is the frequency and $\rm{k_{\rm B}}
$ is Boltzmann's constant. $\delta (x , T_{\rm e})$ takes into
account the relativistic corrections in the study of the thermal SZ effect which is 
due to the presence of thermal weakly relativistic electrons in the ICM and is 
derived by solving the Kompaneets equation up to the higher orders (Rephaeli~1995, 
Itoh et~al. 1998, Nozawa et~al. 1998, Pointecouteau et~al. 1998 and Challinor and 
Lasenby 1998). It should be noted that at 15 GHz (AMI observing frequency) $x= 0.3$ 
and therefore the relativistic correction, as shown by Rephaeli
(1995), is negligible for $k_{\rm B}T_{\rm e} \leq 15\, \rm{keV}$. The dimensionless
parameter $y$, known as Comptonisation parameter, is the integral of the number
of collisions multiplied by the mean fractional energy change of photons per
collision, along the line of sight
\begin{eqnarray}
 y &=& \frac{\sigma_{T}}{m_{\rm e}c^2} \int_{-\infty}^{+\infty}{n_{\rm e}(r)k_{\rm B}
T_{\rm e}(r){\rm d}l}\\
&=& \frac{\sigma_{T}}{m_{\rm e}c^2} \int_{-\infty}^{+\infty}{P_{\rm e}(r){\rm d}l},
\end{eqnarray}
where $n_{\rm e}(r)$, $P_{\rm e}(r)$ and $T_{\rm e}$ are the electron number 
density, pressure and temperature at radius $r$ respectively. $\sigma_{\rm T}$ is 
Thomson scattering cross-section, $m_{\rm e}$ is the electron mass, $c$ is the speed 
of light and $\rm{d}l$ is the line element along the line of sight. 
It should be noted that in equation (6) we have used the ideal gas equation of state.

An interferometer like AMI operating at a frequency $\nu$ measures samples
from the complex visibility plane $\tilde{I}_\nu({\bf u})$. These are given by
a weighted Fourier transform of the surface brightness $I_\nu({\bf x})$, namely
\begin{equation}
  \tilde{I}_\nu({\bf u})=\int{A_\nu({\bf x})I_\nu({\bf x})\exp(2\pi i{\bf u\cdot x})
{\rm d}{\bf x}},
\end{equation}
where ${\bf x}$ is the position relative to the phase centre, $A_\nu({\bf x})$
is the (power) primary beam of the antennas at observing frequency $\nu$
(normalised to unity at its peak) and ${\bf u}$ is the baseline vector in units
of wavelength. In our model, the measured visibilities are defined as
\begin{equation}
  V_\nu({\bf u})=\tilde{S}_\nu({\bf u}) + N_\nu({\bf u}),
\end{equation}
where the signal component, $\tilde{S}_\nu({\bf u})$, contains the contributions
from the SZ cluster and identified radio point sources whereas the generalised
noise part, $N_\nu({\bf u})$, contains contributions from  background of 
unsubtracted radio point sources, primary CMB anisotropies and instrumental noise.

We assume a Gaussian distribution for the generalised noise. This component then 
defines the likelihood function for the data
\begin{equation}
  \mathcal{L}(\mbox{\boldmath $\Theta$})=\frac{1}{Z_N}\exp\left(-\frac{1}{2}\chi^2
\right),
\end{equation}
where $\chi^2$ is the standard statistic quantifying the misfit between the
observed data $\mbox{\boldmath $D$}$ and the predicted data
$\mbox{\boldmath $D$}^p(\mbox{\boldmath$\Theta$})$:
\begin{equation}
  \chi^2=\sum_{\nu , \nu' }(\mbox{\boldmath $D$}_\nu -\mbox{\boldmath $D$}_\nu^p)^T
(\mbox{\boldmath $C$}_{\nu , \nu'})^{-1}(\mbox{\boldmath $D$}_{\nu'}-\mbox{\boldmath 
$D$}_{\nu'}^p),
\end{equation}
where $\nu$ and $\nu'$ are channel frequencies. $\mbox{\boldmath $C$}$ is the 
generalised noise covariance matrix
\begin{equation}
 \mbox{\boldmath $C$}=\mbox{\boldmath $C$}^{\rm {rec}}_{\nu , \nu'} +\mbox
{ \boldmath $C$}^{\rm {CMB}}_{\nu , \nu'} +\mbox{ \boldmath $C$}^{\rm {conf}}_{\nu , 
\nu'},
\end{equation}
 and the normalisation factor $Z_N$ is given by
\begin{equation}
  Z_N=(2\pi)^{(2N_{\rm {vis}})/2}|\mbox{\boldmath $C$}|^{1/2},
\end{equation}
where $N_{vis}$ is the total number of visibilities. It should be noted that
since the main goal of this paper is to demonstrate the effect of different
parameterisations in modelling the SZ cluster signal, we ignore the contributions
due to subtracted and unsubtracted radio point sources so that the non--Gaussian 
nature of these sources is irrelevant. Moreover, the simulations, used in our 
analysis do not include extragalactic radio sources or diffuse foreground emission 
from the galaxy. The effects of the former have already been addressed in Feroz 
et~al. (2009), and here we wish to concentrate on the different parameterisation of
the cluster. We also note that foreground galactic emission is unlikely to be a 
major contaminant since our interferometric observations resolve out large-scale 
emission.
\section{Analysing the SZ Signal: $\beta$-model versus GNFW model}
As may be seen from  equations (5) and (6), in order to calculate the $y$ parameter 
and therefore to model the SZ signal, we need to assume either density and
temperature profiles (Feroz et~al.\ 2009; AMI Consortium: Zwart et~al.\ 2010; AMI
Consortium: Rodr\'{i}guez-Gonz\'{a}lvez et~al.\ 2011) or a pressure profile
(Nagai et~al.\ 2007; Mroczkowski et~al.\ 2009; Arnaud et~al.\ 2010;
Plagge et~al.\ 2010 and Planck Collaboration 2011d) for the plasma content of the 
galaxy cluster. It is also possible to assume a profile for the gas ``entropy'' and 
then derive the distribution of gas pressure assuming hydrostatic equilibrium 
(Allison et~al.\ 2011). Indeed, in general, one may choose to model the SZ signal by 
assuming parameterised functional forms for any two linearly independent functions of
the ICM thermodynamic quantities.

Following our previous analysis methodology (Feroz et~al.\ 2009; AMI Consortium: 
Zwart et~al.\ 2010; AMI Consortium: Rodr\'{i}guez-Gonz\'{a}lvez et~al.\ 2011), we
first review the application of the isothermal $\beta$-model in modelling the SZ 
effect and extracting the cluster physical parameters demonstrating the impact of 
different parameterisations on the inferred cluster properties within a 
model.  We then repeat our analysis for the Generalised NFW 
(GNFW) pressure profile, first presented in Nagai et~al.\ (2007), together with the 
entropy profile presented in Allison et~al.\ (2011) to model the SZ effect and 
derive the cluster physical parameters. This approach has potential advantages.
It not only removes the assumption of isothermality but also leads to a density 
profile that is more consistent with the results of the both numerical analysis of 
hydrodynamical simulations (Voit et~al. 2003; Nagai et~al. 2006; Kravtsov 2006; 
Hallman et~al. 2007) and deep X-ray observations of galaxy clusters (Pratt \& Arnaud 
2002 ; Vikhlinin et~al. 2006). 
\subsection{Isothermal $\beta$-model}
This model assumes a $\beta$-profile for
electron number density (Cavaliere and Fusco- Femiano 1976, 1978) and a
constant temperature throughout the cluster
\begin{eqnarray}
n_{\rm e}(r) &=& \frac{n_{\rm e}(0)}{\left(1+\frac{r^2}{r^2_{\rm c}}\right)^{3
\beta/2}},\nonumber \\
T_{\rm e}(r) &=& T_{\rm g}(r)=\mbox{constant}.
\end{eqnarray}
Here $n_{\rm e}(0)$ is the central electron number density, $T_{\rm e}$ is the 
electron temperature, which is assumed to be the same as the gas temperature,
$T_{\rm g}$, and $r_{\rm c}$ is the core radius. It should be noted that in our 
model, $\beta$ is considered as a free fitting parameter (Plagge et~al.\ 2010;
AMI Consortium: Zwart et~al.\ 2010 ) and is not fixed to for example: $\langle\beta_
{\rm fit}\rangle=2/3$ (Sarazin 1988).

Using this isothermal $\beta$-model, we can then calculate a map of the $y$
parameter on the sky along the line of sight by solving the integral in equation (5) 
analytically (Birkinshaw et~al.\ 1999)
\begin{equation}
y(s)=y_0\left(1 +\frac{s^2}{r^2_{\rm c}}\right)^{(1-3\beta)/2} ,
\end{equation}
where $\beta>1/3$, $s$ is the projected distance from the centre of the cluster on 
the sky such that $r^2=s^2+l^2$ and $y_0$ is the central Comptonisation parameter
\begin{equation}
  y_0=\frac{\sqrt{\pi}\sigma_{\rm T}k_{\rm B}T_{\rm g}n_{\rm e}(0)r_{\rm c}}{m_{\rm 
e}c^2}\frac
{\Gamma (\frac{3\beta}{2}-\frac{1}{2})}{\Gamma (\frac{3\beta}{2})}.
\end{equation}
The integral of the $y$ parameter over the solid angle $\Omega$ subtended by
the cluster is denoted by $Y_{\rm SZ}$, and is proportional to the volume
integral of the gas pressure. It is thus a good estimate for the total thermal
energy content of the cluster and its mass (see e.g. Bartlett \& Silk 1994). Thus 
determining the normalisation and the slope of $Y_{\rm SZ}-M$ relation have been the 
subject of studies of the SZ effect (da Silva et~al.\ 2004; Nagai 2006; Kravtsov 
2006; Plagge et~al.\ 2010; Andersson et~al. 2011; Arnaud et~al 2010; Planck 
Collaboration 2011d,e,f,g,h). In particular, Andersson et~al. (2011) investigated the 
$Y_{\rm SZ}-Y_{\rm X}$ scaling relation within a sample of 15
clusters observed by South Pole Telescope (SPT), Chandra and XMM Newton and found a 
slope of close to unity ($0.96 \pm 0.18$). Similar studies were carried out by 
Planck Collaboration (Planck Collaboration 2011g) using a sample of 62 nearby ($z < 
0.5$) clusters observed by both Planck and XMM--Newton satellites. The results are 
consistent with predictions from X-ray studies (Arnaud et~al. 2010) and the
ones presented in Andersson et~al. (2011). These studies at low redshifts where the 
data are available from both X-ray and SZ observations of galaxy clusters are 
crucial to calibrate the $Y_{\rm SZ}-M$ relation and such a relation can then be scaled 
and used to determine masses of SZ selected clusters at high redshifts in order to 
constrain cosmology.

We calculate the $Y_{\rm SZ}$ parameter for the isothermal $\beta$-model in
both cylindrical and spherical geometries. Assuming azimuthal symmetry, $Y_{\rm cyl}$ reads
\begin{eqnarray}
Y_{\rm cyl}(R)&=& \frac{\sigma_{T}}{m_{\rm e}c^2}\int_{-\infty}^{+\infty}{\rm {d}l}\,
\int_{0}^{R}{P_{\rm e}(r)2\pi s \, \rm {d}s} \\
              &=& \int_{0}^{R}{y(s)2\pi s \, \rm {d}s}  \\
              &=& \left\{
\begin{array}{l l}
  \frac{\pi y_0 {r^2_{\rm c}}}{\frac{3}{2}-\frac{3}{2}\beta}\left\{(1 +(R/r_{\rm c})
^2)^{(3-3\beta)/2} - 1 \right \} & \mbox{ $\beta \neq$  1}\\
\\
  \pi y_0 {r^2_{\rm c}}\ln[1 +(R/r_{\rm c})^2] & \mbox{ $\beta$ =1},\\ \end{array} 
\right.  \nonumber
\end{eqnarray}
where $R$ is the projected radius of the cluster on the sky.

The integrated $y$ parameter in the case of assuming spherical geometry $Y_{\rm
sph}$, is given by integrating the plasma pressure within a spherical volume of
\mbox{radius} $r$
\begin{eqnarray}
Y_{\rm sph}(r)&=& \frac{\sigma_{\rm T}}{m_{\rm e}c^2}\int_{0}^{r}{P_{\rm e}(r')4\pi 
r^{'2}\rm {d}r'}  \\
              &=&\frac{\sigma_{\rm T}k_{\rm B}T_{\rm g}n_{\rm e}(0)}{m_{\rm e}c^2}
\int_{0}^{r}{\frac{4\pi r^{'2}\rm{d}r'}{\left(1+\frac{r^{'2}}{r^2_{\rm c}}\right)^{3
\beta/2}}}.
\end{eqnarray}
It should be noted that there is an analytical solution for the above integral 
provided that the upper limit is infinity and $\beta > 1$. However, since we
study the cluster to a finite extent and $\beta$ varies over a wide range including $
\beta < 1$, we calculate $Y_{\rm sph}$ numerically.
\subsection{GNFW Pressure Profile}
As the SZ surface brightness is proportional to the line of sight integral of the 
electron pressure, assuming a pressure profile for the hot plasma within the cluster 
to model the SZ effect seems a reasonable choice. In this context, Nagai et~al. 
(2007) analysed the pressure profiles of a series of simulated clusters (Kravtsov 
et~al. 2005) as well as a sample of relaxed real clusters presented in
Vikhlinin et~al. (2005 , 2006). They found that the pressure profiles of all of 
these clusters can be described by a generalisation of the Navarro, Frenk, and White 
(Navarro et~al. 1997) (NFW) model used to describe the dark matter halos of simulated 
clusters. The GNFW pressure profile (Nagai et~al. 2007) is described as
\begin{equation}
 P_{\rm e}(r) = \frac{P_{\rm {ei}}}{\left(\frac{r}{r_{\rm p}}\right)^c\left(1+\left
(\frac{r}{r_{\rm
p}}\right)^{a}\right)^{(b-c)/a}},
\end{equation}
where $P_{\rm {ei}}$ is the normalisation coefficient of the pressure profile, $r_
{\rm p}$ is the scale radius and the parameters $( a,  b,  c)$ describe the 
slopes of the pressure profile at $r\approx r_{\rm p}$, $r> r_{\rm p}$ and $r \ll r_
{\rm p}$ respectively. We fix the values for the slopes to the ones given in Arnaud 
et~al. (2010): $( a,  b,  c)=(1.0620,5.4807, 0.3292)$. Arnaud et~al. (2010) 
derived the pressure profiles for the  REXCESS cluster sample from XMM-Newton 
observations (B\"ohringer et~al. 2007; Pratt et~al. 2010 and Arnaud et~al. 2010) 
within $r_{500}$. These pressure profiles also match (within $r_{500}$) three sets 
of different numerical simulations (Borgani et~al. 2004; Piffaretti \&
Valdarini 2008; Nagai et~al. 2007). They thus derived an analytical function, the so-
called universal pressure profile with above mentioned parameters. This 
profile has been successfully tested against SZ data from SPT (Plagge et~al. 2010) 
and the Planck survey data (Planck Collaboration 2011d).

We calculate the map of the $y$ parameter on the sky along the line of sight by 
solving the integral in equation (6) numerically. However, we note that the central 
Comptonisation parameter $y_0$ has an analytical solution
\begin{equation}
 y_0=\frac{2\sigma_{\rm T}P_{\rm {ei}}r_{\rm p}}{m_{\rm e}c^2}\frac{1}{ a}\frac
{\Gamma(\frac{1- c}{ a})\Gamma(\frac{ b -1}{ a})}{\Gamma(\frac{ b - 
 c}{ a})}.
\end{equation}
Similarly, to calculate the thermal energy content of the cluster within a sphere 
with finite radius we use equation (18). In this context, Anaud 
et~al. (2010) have shown that the pressure profile flattens at the radius of $5r_
{500}$ and used this to define the boundary of the cluster. One can thus use this 
radius to define the total volume integrated SZ signal. 
\begin{equation}
Y_{\rm tot} = Y_{5r_{500}}=\frac{\sigma_{\rm T}}{m_{\rm e}c^2}\int_{0}^{5r_{500}}{P_
{\rm e}(r')4\pi r^{'2}\rm {d}r'} 
\end{equation}
%
\section{Estimating cluster physical parameters}
To study the physical parameters of the cluster, such as its total mass and gas
mass, we have to make some assumptions about the dynamical state of the
cluster. The most widely-used assumptions are: that the gas distribution is in
hydrostatic equilibrium with the cluster total gravitational potential
dominated by dark matter, and that both dark matter and the plasma are
spherically symmetric and have the same centroid.

The cluster mass $M_T(r_X)$ is also defined as the total amount of matter
internal to radius $r_X$ within which the mean density of the cluster is $X$
times the critical density at the cluster redshift. Mathematically, the
assumption of hydrostatic equilibrium applies everywhere inside the
cluster and relates total cluster mass internal to radius $r$ to the gas
pressure gradient at that radius and hence to the density and temperature
gradients respectively
\begin{equation}
M_{\rm T}(r)=-\frac{k_{\rm B}T_{\rm g}(r)r}{\mu G}\left[\frac{{\rm d}\ln \varrho_
{\rm g}(r)}{{\rm d}\ln r} + \frac{{\rm d}\ln T_{\rm g}(r)}{{\rm d}\ln r}\right],
\end{equation}
where $\mu=0.6m_{\rm p}$ (Sarazin 1988) is the mean mass per gas particle, $m_{\rm p}$ 
is the proton mass and $G$ is the universal gravitational constant. Assuming spherical 
geometry, it is also possible to calculate the gas mass and total mass internal to 
radius $r_X$
\begin{eqnarray}
 M_{\rm g}(r_X) &=&  4\pi \int_{0}^{r_X} r^2 \varrho_{\rm g}(r){\rm d}r,         \\
 M_{\rm T}(r_X) &=&  \frac{4\pi}{3}r^3_X(X \varrho_{\rm crit}(z)).
\end{eqnarray}
Here $\varrho_{\rm {crit}}(z)=\frac{3{H(z)}^2}{8\pi G}$ is the critical density
of the universe at the cluster redshift $z$ and $H(z)=H_0\sqrt{(\Omega_{\rm M}
+ \Omega_{\rm \Lambda} Q)(1+z)^3+\Omega_{\rm R}(1+z)^4 +\Omega_{\rm K}(1+z)^2}$
is the Hubble parameter at redshift $z$, where $H_{\rm 0}=100h \,
\rm{km~s^{-1}Mpc^{-1}}$ is the Hubble constant now. $\Omega_{\rm M}$ measures the
present mean mass density including baryonic and nonbaryonic dark matter in
non-relativistic regime, $\Omega_{\rm \Lambda}$ takes into account the present
value of the dark energy, $\Omega_{\rm R}$ measures the current energy density
in the CMB and the low mass neutrinos, $\Omega_{\rm K}$ describes the curvature of the 
universe and $Q=(1+z)^{3[w_{\rm 0}+w_{\rm a}]}\exp\left [\frac{-3w_{\rm a}z}{1+z}\right]
$ is the dark energy equation of state.

Moreover, it has been long known that the total mass of the cluster is strongly
correlated with its mean temperature. This arises
from both X-ray observations of galaxy clusters (Voit \& Ponman 2003) and the
fact that the gravitational heating is the dominant process in the clusters within 
the hierarchical structure formation scenario (Kaiser 1986 ; Sarazin 2008).

Assuming virialisation and that all cluster kinetic energy is in gas 
internal energy suggests that $T \propto M^{2/3}$, where $T$ is the mean
gas temperature within the virial radius and $M$ is the cluster total mass
internal to that radius. However, an extensive range of studies based both on 
observations of galaxy clusters and on numerical simulations have been carried out 
aiming  to determine the proportionality coefficient of such
relation (Evrard et~al.\ 1996; Eke et~al. 1998; Voit 2000; Yoshikawa et~al. 2000; 
Finoguenov et~al. 2001; Afshordi \& Renyue 2002; Evrard et~al.\ 2002; Sanderson 
et~al. 2003; Borgani et~al. 2004; Voit 2005; Arnaud et~al. 2005; Vikhlinin et~al. 
2006; Afshordi et~al.\ 2007; Maughan et~al. 2007 and Nagai et~al. 2007). Finoguenov 
et~al (2001) studied the observational mass-temperature relation of two sets of 
cluster samples. In their first sample they used the assumption of isothermality 
whereas in the second set they knew the temperature gradient of the clusters within 
the sample. In both samples, they found that the discrepancy from the self-
similarity in the M-T relation is more pronounced in the low mass clusters ($k_{\rm 
B}T_{\rm g} < 3.5~\rm{keV})$ as non-gravitational processes become more dominant in 
these clusters. Similar results were obtained by Arnaud et~al. (2005) when they 
analysed a sample of 10 nearby ($z\le 0.15$) relaxed clusters in the temperature
range $2-9\,\rm{keV}$. They showed that the slope of the M-T relation for hot 
clusters is consistent with self-similar expectation while for low temperature (low 
mass) clusters the slope is significantly higher. Studies of the observational mass-
X-ray luminosity relation (Maughan et~al. 2007) also show that the scatter in the $L_
{\rm X}-M_{\rm 500}$ relation is dominated by cluster cores and is almost 
insensitive to the merger status of the cluster. Theoretical studies based on 
the adiabatic simulations and the hydrodynamical simulations of cluster formation 
with gravitational heating only also verify the slope of $3/2$ in M-T relation 
(Evrard et~al.\ 1996; Eke et~al. 1998; Voit 2000; Yoshikawa et~al. 2000) while 
numerical simulations which take into account the non-gravitational heating
processes and the effect of the radiative cooling of the gas (Borgani et~al. 2004; 
Nagai et~al. 2007) do predict a slightly higher slope. Moreover, almost all of the 
above mentioned studies do agree that the discrepancy in the slope of the M-T 
relation could also be due to the different procedures used for estimating masses in 
simulations and observational analyses.

In this paper we therefore decided to follow the approach given in Voit (2005). This 
is based on using the virial theorem to relate a collapsing top-hat density 
perturbation model to a singular truncated isothermal sphere. It also takes into 
account the finite boundary pressure and assumes all kinetic energy is internal 
energy of the hot plasma. This gives
\begin{equation}
 k_{\rm B}T_{\rm g}(r_{X}) =\frac{\mu}{2}\left(\frac{X}{2}\right)^{1/3}[GM_{\rm T}(r_
{X})H(z)]^{2/3}.
\end{equation}
It should be noted that above relation assumes that the virialisation occurs at
$r_{X}$.

Based on the above assumptions, one can adopt different parameterisations to study 
physical properties of the cluster within a particular model (e.g. using either 
the assumption of hydrostatic equilibrium or the M-T relation). These
different approaches shed light on the realism of the assumptions made throughout 
the analysis and reveal different biases and constraints associated with them. In a 
single- frequency observation that at least partially resolves the cluster, the best 
one can hope to achieve in constraining an empirical model of the SZ decrement is to 
estimate the central position of the cluster (the position of the decrement) and two further parameters--i.e. shape and scale parameters. The interpretation of such constraints does however depend on the 
particular parrameterisation. 

Hence in the following sections we discuss 
different possible parameterisations within two models: isothermal $\beta$-model and 
``entropy''-GNFW pressure model. In doing so we try to disentangle the thermal pressure built-in correlation between pairs of physical parameters that lead to the SZ effect intensity--i.e. ($T_g\, , \, M_g$),($f_g\, , \, M_T$), etc.  
\subsection{Isothermal $\beta$-Model}
Generally, there are two different parameterisations that one could use in the analysis 
of the cluster SZ effect and deriving its physical parameters. However, the assumption of 
isothermality provides another form of parameterisation where the gas temperature is 
assumed as an input free parameter along with the assumption of hydrostatic equilibrium 
(e.g. isothermal $\beta$-model or isothermal GNFW model).

In the following sections we discuss our three different parameterisations for the 
isothermal $\beta$-model within our Bayesian framework. It should be noted that in 
all of these parameterisations, we employ physically-based sampling parameters. Such 
parameters reveal the structure of degeneracies in the cluster parameter space more 
clearly than parameters that just describe the $y$-map such as angular core radius $
\theta_{\rm c}$, shape $\beta$ and central temperature decrement $\Delta T_{\rm 0}$. 
We also note that throughout our analysis we impose the additional constraint that 
the cluster has a non-zero $r_{\rm 500}$.
\subsubsection{Parameterisation I}
Our sampling parameters for this case are $\mbox{\boldmath$\Theta$}_{\rm c}\equiv (x_
{\rm c}, y_{\rm c}, r_{\rm c}, \beta,T_{\rm g}, M_{\rm g}(r_{\rm 200}), z)$, where 
$x_{\rm c}$ and $y_{\rm c}$ are cluster projected position on the sky, $r_{\rm c}$ 
and $\beta$ are the parameters defining the density profile, $T_{\rm g}$ is the gas 
temperature, $M_{\rm g}(r_{\rm 200})$ is the gas mass internal to radius $r_{\rm 200}
$ and $z$ is the cluster redshift. It should be noted that AMI can typically measure the 
overdensity radii  $r_{500}$ and $r_{200}$ for $z>0.15$. However, we choose to work in 
terms of an overdensity radius of $r_{\rm 200}$ since the constrains from AMI data on the 
cluster physical parameters are stronger at this radius and this radius is approximately 
the virial radius. We further assume that the priors on sampling parameters are separable 
(Feroz et~al.\ 2009) such that
\begin{equation}
\pi(\mbox{\boldmath$\Theta$}_{\rm c})=\pi(x_{\rm c})\,\pi(y_{\rm c})\,\pi(r_{\rm c})\,
\pi(\beta)\,\pi(T_{\rm g})\,\pi(M_{\rm g}(r_{\rm 200}))\,\pi(z).
\end{equation}
This implies that parameterisation I ignores the known apriori correlation between the 
cluster total mass and gas temperature. We use Gaussian priors on cluster position 
parameters, centred on the pointing centre and with standard deviation of 1 arcmin. 
We adopt uniform priors on the cluster core radius, $\beta$ and the gas temperature. As 
mentioned in Feroz et al. (2009), for SZ pointed observations, where we know the cluster 
redshift from optical studies and possibly the gas mass fraction from X-ray studies, we 
can assume a separable prior on the gas mass and redshift, namely, $\pi(M_{\rm g}, z)
\propto \pi(M=M_{\rm g}/f_{\rm g}, z)=\pi(M=M_{\rm g}/f_{\rm g})\pi(z)$, (Feroz et~al.\ 
2009), where each factor has some simple functional form such that their product gives a 
reasonable approximation to a known mass function e.g. the Press--Schechter (Press 
\& Schechter 1974) mass function. We will assume such a form in our analysis where $
\pi(M_{\rm g})$ will be taken to be uniform in $\log$ in the range $(M_{\rm g})_{\rm 
min} = 10^{12}\, {h^{-2}M_\odot}$ to $(M_{\rm g})_ {\rm max} =5\times10^{14}\, 
 {h^{-2}M_\odot}$ and the redshift is fixed to the cluster redshift. A summary of the 
priors and their ranges for this parameterisation is presented in Table 2.

Having established our physical sampling parameters, modelling the SZ signal is 
performed through the calculation of the $y$ parameter which requires the knowledge 
of parameters describing the 3-D plasma density and its temperature, namely $r_{\rm 
c}$, $\beta$, $n_{\rm e}(0)$ and $T_{\rm g}$. We sample from $r_{\rm c}$ , $
\beta$ and $T_{\rm g}$ but as shown below, deriving $n_{\rm e}(0)$ requires 
employing the assumptions of hydrostatic equilibrium, isothermality and spherical 
geometry right from the beginning of the analysis.

Substituting the isothermal $\beta$-model into the equation of hydrostatic equilibrium 
equation (23), we can then relate the  $M_{\rm {T}}(r_{200})$ to our
model parameters as well as to the temperature
\begin{equation}
 M_{\rm {T}}(r_{200})=\frac{3\beta r^3_{200}}{{r^2_{\rm c}}+r^2_{200}} \frac{k_{\rm B}T_
{g}}{\mu G},
\end{equation}
where we have used $\varrho_{\rm g}(r)=\mu_{\rm e} n_{\rm e}(r)$ with $\mu_{\rm e}
=1.14m_{\rm p}$ (Jones et~al. 1993; Mason \& Myers 2000) defined as the mean gas 
mass per electron. By combining equations (25) and (28) at $r_{\rm 200}$, we first 
calculate the overdensity radius of $r_{\rm 200}$ and since $M_{\rm g}(r_{\rm 200})$ 
is also one of our sampling parameters we can recover the central electron number 
density by rearranging equation (24):
\begin{equation}
 r_{\rm 200}=\sqrt{\frac{9\beta k_{\rm B}T_{g}}{4\pi\mu G (200 \varrho_{\rm {crit}}
(z))} - r_{\rm c}^2},
\end{equation}
\begin{equation}
 n_{\rm e}(0)=\frac{M_{\rm g}(r_{\rm 200})}{4\pi \mu_{\rm e} \displaystyle \int_{0}^
{r_{\rm 200}} \frac{r'^2 {\rm d}r'}{\left(1+\frac{r^{'2}}{r^2_{\rm c}}\right)^{3
\beta/2}}}.
\end{equation}
For cluster physical parameters we use the value of the overdensity radius of
$r_{\rm 200}$ (equation 29) to calculate the cluster total mass internal to
$r_{\rm 200}$ assuming spherical geometry, (equation 25). The gas mass fraction
at $r_{\rm 200}$ is then simply $f_{\rm g}(r_{\rm 200})=M_{\rm g}(r_{\rm
200})/M_T(r_{\rm 200})$. As the central electron number density and plasma
temperature are assumed to be constants, we can in principle calculate cluster
physical parameters in any overdensity radius other than $r_{\rm 200}$ by
assuming that the hydrostatic equilibrium holds everywhere in the
cluster. In this paper we study the cluster properties at two overdensity
radii $r_{\rm 200}$ and $r_{\rm 500}$. Extracting cluster physical parameters
at $r_{\rm 500}$ in particular enables us to compare our results with the
results obtained from X-ray analysis of the clusters of galaxies. $r_{\rm 500}$
is calculated by equating equations (23) and (25) and setting $X=500$.  $M_{\rm g}(r_
{\rm500})$ and $M_T(r_{\rm 500})$ are then derived using equations (24) and (25) 
respectively.
\begin{table}
 \caption{Summary of the priors on the sampling parameter set in parameterisation I. 
Note that $N(\mu ,\sigma)$ represents a Gaussian probability distribution with mean $
\mu$ and standard deviation of $\sigma$ and $U(a,b)$ represents a uniform 
distribution between $a$ and $b$.}
  \begin{tabular}{@{}ll@{} }
  \hline
Parameter                           &\qquad \qquad Prior \\\hline
$x_{\rm c}$ , $y_{\rm c}\qquad$     &\qquad \qquad $N(0 \,\, , \, \, 60)\arcsec $ \\
$r_{\rm c}\qquad$      &\qquad \qquad $U(10 \,\, , \, \,  1000)\,h^{-1}\rm{kpc}$  \\
$\beta\qquad$          & \qquad \qquad $U(0.34 \,\, , \, \, 2.5)$             \\
$\log M_{\rm g}(r_{\rm 200})\qquad$  &\qquad \qquad $U(12 \,\, , \, \, 14.5)\, h^{-2}\rm
{M_\odot}$             \\
$T_{\rm g}\qquad$ &\qquad \qquad  $U(0 \,\,  , \, \, 20)\,\rm{keV}$     \\\hline
\end{tabular}
\end{table}
\subsubsection{Parameterisations II and III}
Parameterisation I does not take into account the correlation between the cluster 
total mass and its mean gas temperature. However, as mentioned earlier, observations 
of galaxy clusters and theoretical studies have both shown that there is a strong 
correlation between these two cluster parameters. We have already used this
parameterisation in the analyses of 7 clusters out to the virial radius (AMI 
Consortium: Zwart et~al. 2010). We found that using this parameterisation along side 
the assumption of isothermality led to strong biases in the estimation of cluster 
parameters. This implies that, in the absence of a measured temperature profile, we 
should eliminate gas temperature from the list of our sampling parameters and 
instead sample from either $M_T(r_{\rm 200})$ or $M_{\rm g}(r_{\rm 200})$,and $f_
{\rm g}(r_{\rm 200})$. We choose total mass as a sampling parameter since this is 
consistent with our cluster detection algorithm and analysis (AMI Consortium: 
Shimwell et~al.\ 2010). This form of parameterisation then allows us to
calculate gas temperature either by using isothermal hydrostatic equilibrium 
(parameterisation II), or virial relation, (parameterisation III).

Our sampling parameters for these two parameterisations are $\mbox{\boldmath$\Theta$}
_{\rm c}\equiv (x_{\rm c}, y_{\rm c}, r_{\rm c}, \beta,
M_{\rm T}(r_{\rm 200}), f_{\rm g}(r_{\rm 200}), z)$ which are assumed to be 
independent for the same reasons described in previous section such that
\begin{equation}
 \pi(\mbox{\boldmath$\Theta$}_{\rm c})=\pi(x_{\rm c})\,\pi(y_{\rm c})\,\pi(r_{\rm c})\,\pi(\beta)\,\pi(M_T(r_{\rm 200}))\,\pi(f_{\rm g}(r_{\rm 200}))\,\pi(z).
\end{equation}
The priors on $x_{\rm c}$, $y_{\rm c}$, $r_{\rm c}$, $\beta$ and $z$ are the same as 
for parameterisation I. The prior on $M_{\rm T}(r_{\rm 200})$ is taken to be uniform 
in log$M$ in the range $M_{\rm {min}} = 10^{14}\, h^{-1}\rm{M_ \odot}$ to $M_{\rm 
{max}} = 5\times10^{15}\, h^{-1}\rm{M_\odot}$ and the prior of $f_{\rm g}(r_{\rm 
200})$ is set to be a Gaussian centred at the WMAP7 best fit value: $f_{\rm g}=0.12$ 
with a width of $0.016$ (Komatsu et~al.\ 2011; Larson et~al.\ 2011 and AMI 
Consortium: Rodr\'{i}guez-Gonz\'{a}lvez et~al.\ 2011). A summary of the priors
and their ranges for these two parameterisations are presented in Table 3. We 
calculate $M_{\rm g}(r_{\rm 200})$ from the definition of gas mass fraction and $r_
{\rm 200}$ is determined assuming spherical geometry, equation (25). Central 
electron number density is then calculated using equation (30).

For parameterisation II, the gas temperature at $r_{\rm 200}$ is estimated assuming 
hydrostatic equilibrium using equation (28) and is assumed to be constant
throughout the cluster
\begin{eqnarray}
  k_{\rm B} T_{\rm g}(r_{\rm 200}) &=&\frac{\mu GM_{\rm T}(r_{\rm 200})(r_{\rm 200}
^2 +r_{\rm c}^2)}{3\beta r_{\rm 200}^3}  \nonumber \\
                                   &=& \frac{(4 \pi \mu G)(200 \varrho_{\rm crit}(z))
(r^2_{\rm 200}+r^2_{\rm c})}{9 \beta},
\end{eqnarray}
where the last form is derived by substituting for $M_{\rm T}(r_{\rm 200})$ using 
equation (25). We refer to relation (32) as the HSE mass-temperature relation. 
Similar to parameterisation I, once temperature and central electron number density 
are determined, we can calculate cluster physical properties at the overdensity 
radius of $r_{\rm 500}$ using equations (23), (24) and (25) and setting $X=500$.

For parameterisation III, we calculate the mean gas temperature within a virial 
radius of $r_{\rm 200}$ using the mass-temperature relation described in
equation (26), which is then assumed to be constant throughout the cluster
\begin{equation}
 k_{\rm B} T_{\rm g}(r_{\rm 200}) = 8.2
\left(\frac{M_T(r_{\rm 200})}{10^{15}h^{-1}M_{\odot}}\right)^{2/3}
\left(\frac{H(z)}{H_0}\right)^{2/3}   \rm {keV}.
\end{equation}
We refer to this relation as the virial mass-temperature relation. This also
implies that virialisation occurs at $r_{\rm 200}$. In our analysis, we use this 
relation to determine the mean gas temperature and once this is determined we repeat 
the same procedure carried out for parameterisations I and II to obtain cluster 
physical properties at the overdensity radius $r_{\rm 500}$.
\begin{table}
\caption{Summary of the priors on the sampling parameter set in parameterisations II 
and III.}
\begin{tabular}{@{}ll@{} }
\hline
Parameter       &\qquad \qquad Prior                                      \\
\hline
$x_{\rm c}$ , $y_{\rm c}\qquad$ &\qquad \qquad $N(0 \,\, , \, \,60)\arcsec$  \\
$r_{\rm c}\qquad$ &\qquad \qquad $U(10 \,\, , \, \, 1000)\,h^{-1}\rm{kpc}$ \\
$\beta\qquad$                &\qquad \qquad$U(0.34 \,\, , \, \, 2.5)$ \\
$\log M_{\rm T}(r_{\rm 200})\qquad$ &\qquad \qquad $U(14 \,\, , \, \, 15.5)\,h^
{-1}\rm{M_\odot}$  \\
$f_{\rm g}(r_{\rm 200})\qquad$ &\qquad \qquad $N( 0.084 \,\, , \, \, 0.016)\,{h^{-1}}$ \\ \hline
\end{tabular}
\end{table}
\subsection{Entropy-GNFW Pressure Model}
As it was mentioned in Section~5.2 the choice of the GNFW pressure profile to model 
the SZ signal is reasonable as the SZ surface brightness is proportional the line of 
sight integral of the electron pressure. However, in order to link the gravitational 
potential shape to the baryonic physical properties of the ICM, one has to make 
assumptions on the radial profile of another thermodynamical quantity. Among the 
thermodynamical quantities of the ICM, entropy has proved to be an important gas 
property within the cluster. Entropy is conserved during the adiabatic collapse of 
the gas into the cluster gravitational potential well, however, it will be affected 
by any non-gravitational processes such as radiative cooling, star formation, energy 
feed back from supernovae explosions and Active Galactic Nuclei (AGN) activities. It 
therefore keeps a record of the thermodynamic history of the ICM (Voit 2000; 2003, 
2005; Ponmann et~al. 1999; Pratt \& Arnaud 2002; Allison et~al. 2011).

Moreover, for a gravitationally collapsed gas in hydrostatic equilibrium , entropy 
profile is expected to have an approximate power law distribution ($ \approx r^{1.1}
$) (Lloyd-Davies et~al. 2000; Voit 2005; Nagai et~al. 2007; Pratt 2010). However, 
there is a large deviation from self-similarity in the entropy radial profile in the 
inner region of the cluster ($r< 0.1r_{200}$) due to the impact of all of the non-
gravitational mechanisms described above on the thermodynamics of the ICM 
(Finoguenov et~al. 2002; Ponman et~al. 1999, 2003; Lloyd-Davies et~al. 2000; Pratt 
2010). In the inner region, the results of the non-radiative simulations and 
simulations that take into account AGN activities plus preheating models predict a 
flat core in the entropy distribution due to entropy mixing (Wadsley et~al. 2008; 
Mitchell et~al. 2009). The observed entropy profiles using X-ray telescopes also 
flatten in the inner regions in general while having similar external slopes (Pratt 
et~al. 2006). In the outskirt of the cluster (out to virial radius and beyond), on 
the other hand, the results of the latest numerical simulations (Nagai 2011; Nagai et~al. 2011) 
and observational studies of the clusters using \textit{Suzaku} and XMM- Newton 
satellites at large radii including A1795 (Bautz et~al. 2009), PKS 0745-191 (George 
et~al. 2009), A2204 (Reiprich et~al. 2009), A1413 (Hoshino et~al. 2010), A1689 
(Kawaharada et~al. 2010), Virgo cluster (Urban et~al. 2011) and Perseus cluster 
(Simionescu et~al. 2011) also  show that behaviour of ICM entropy  deviates from the 
prediction of a spherically systematic shock heated gas model (Tozzi \& Norman 
2001). According to these studies major sources of this deviation may be due to 
incomplete virialisation, departure from hydrostatic equilibrium, gas motion and gas clumping.

In this context and to derive the cluster physical parameters we decided
to adopt the entropy profile presented in Allison et~al. (2011) which is a $\beta$-
model like profile:

\begin{equation}
K_{\rm e}(r) = K_{\rm {ei}}\left(1+\frac{r^2}{r^2_{\rm c}}\right)^{\alpha},
\end{equation}
where $K_{\rm e}(r)$ is the plasma entropy at radius $r$, $K_{\rm {ei}}$ is the 
normalisation coefficient of the entropy profile, $r_{\rm c}$ and $\alpha$ are the 
parameters defining the shape of the profile at different radii. Assuming an entropy 
profile with above form guarantees the flat shape in the inner region and a power 
law distribution at the larger radii (up to $r_{500}$) where $ K_{\rm e}(r)\propto r^{2\alpha}$ with $\alpha \sim 0.55$. We note that in order to take into account the  behaviour of entropy at the clusters outskirts we need to modify the assumed entropy profile with additional parameter and/or component. However, as the studies of this kind in understanding the physics of the cluster outskirts and accurate measurements of the ICM profiles in the cluster outer regions are still ongoing, we do not study a modified form of our assumed entropy profile here. We of course aim to consider a more general form in our future analyses.

As for using the GNFW profile, one has indeed to make an assumption on either the 
density, temperature  or the entropy profile shape in order to link the 
gravitational potential shape to the baryonic physical properties of the ICM. In 
this paper, we decided to work with an assumption of entropy profile for all the 
reasons given above. The combination of the GNFW pressure and the " $\alpha$--model" 
entropy profiles can then fully describe the large-scale properties of clusters as 
they determine the form of the dark matter potential well in addition to the 
structure of the ICM.

To relate the entropy to the other thermodynamical quantities inside the ICM we use 
the definition of the entropy given in the astronomy literature. For an adiabatic 
monatomic gas,
\begin{eqnarray}
 K_{\rm e} &=& k_{\rm B}Tn_{\rm e}^{-2/3}    \\
 P_{\rm e} &=& K_{\rm e}n_{\rm e}^{5/3},
\end{eqnarray}
which is related to the true thermodynamic entropy per gas particle via $S=\frac{3}
{2}k_{\rm B}\ln(K_{\rm e})+S_0$ where $S_0$ is a constant (Voit 2005).

Using equations (20),(34),(35) and (36) we can derive the 3-D radial profiles of the 
electron number density and the temperature,
\begin{eqnarray}
n_{\rm e}(r) &=& n_{\rm {ei}}\left(\frac{r}{r_{\rm {p}}}\right)^{(\frac{-3}{5}){
c}}\left[1+\left(\frac{r}{r_{\rm p}}\right)^{ a}\right]^{-\frac{3}{5}\left
( \frac{{ b} - { c}}{{ a}}\right)}\left[1 + \left(\frac{r}{r_{ c}}\right)
^2 \right]^{-\frac{3}{5}\alpha}\\
k_{\rm B}T_{\rm e}(r) &=& k_{\rm B}T _{\rm {ei}}\left(\frac{r}{r_{\rm {p}}}\right)^
{(\frac{-2}{5}){ c}}\left[1+\left(\frac{r}{r_{\rm p}}\right)^{ a}\right]^{-
\frac{2}{5}\left( \frac{{ b} - { c}}{{ a}}\right)}\left[1 + \left(\frac{r}
{r_{ c}}\right)^2 \right]^{\frac{3}{5}\alpha}
\end{eqnarray}
where
\begin{equation}
 n_{\rm {ei}}=\left(\frac{P_{\rm {ei}}}{K_{\rm {ei}}}\right)^{\frac{3}{5}}
\end{equation}
and
\begin{equation}
 k_{\rm B}T _{\rm {ei}}=P_{\rm {ei}}^{2/5}K_{\rm {ei}}^{3/5}
\end{equation}
are the normalisation coefficients for the electron number density and the 
temperature profiles respectively. We note that the above derived electron number 
density has components that take into account both the fit for the inner slope 
of the cuspy cluster density profiles and the steepening at larger radii ($r\ge r_
{500})$ (Pratt \& Arnaud 2002, Vikhlinin et~al. 2006).

As using this model to analyse the cluster SZ signal removes the assumption of 
isothermality, parameterisation I which assumes a single core temperature as a 
free input parameter can not be used in the analysis using entropy-GNFW model or any non-
isothermal model. We therefore study the cluster SZ signal and its physical properties 
using parameterisations II and III.

Our sampling parameters for this model are $\mbox{\boldmath$\Theta$}_{\rm c}\equiv 
(x_{\rm c}, y_{\rm c}, r_{\rm c}, \alpha,M_{\rm T}(r_{\rm 200}), f_{\rm g}(r_{\rm 
200}), z)$. A summary of the priors and their ranges for the "entropy"-GNFW pressure 
model is presented in \mbox{Table 4}.
\begin{equation}
\pi(\mbox{\boldmath$\Theta$}_{\rm c})=\pi(x_{\rm c})\,\pi(y_{\rm c})\,\pi(r_{\rm c})
\,\pi(\alpha)\,
\pi(M_T(r_{\rm 200}))\,\pi(f_{\rm g}(r_{\rm 200}))\,\pi(z).
\end{equation}
\begin{table}
\caption{Summary of the priors on the sampling parameter set in the entropy-GNFW 
pressure model.}
\begin{tabular}{@{}ll@{} }
\hline
Parameter     &\qquad \qquad Prior                                \\
\hline
$x_{\rm c}$ , $y_{\rm c}\qquad$ &\qquad \qquad $N(0 \,\, , \, \,60)\arcsec$   \\
$r_{\rm c}\qquad$  &\qquad \qquad $U(10 \,\, , \, \, 1000)\,h^{-1}\rm{kpc}$ \\
$\alpha\qquad$     &\qquad \qquad$U(0.0 \,\, , \, \, 1.0)$             \\
$\log M_{\rm T}(r_{\rm 200})\qquad$ &\qquad \qquad $U(14 \,\, , \, \,  15.5)\,h^{-1}\rm{M_
\odot}$\\
$f_{\rm g}(r_{\rm 200})\qquad$ &\qquad \qquad $N( 0.084 \,\, , \, \, 0.016)\,{h^
{-1}}$\\
$r_{\rm p}\qquad$ &\qquad \qquad $U(0.001 \,\, , \, \, 3)\,h^{-1}\rm{Mpc}$ \\
\hline
\end{tabular}
\end{table}
Sampling from $M_{\rm T}(r_{\rm 200})$ in both parameterisations leads to the 
estimation of $r_{\rm 200}$ assuming spherical geometry for the cluster, equation 
(25). Sampling from $M_{\rm T}(r_{\rm 200})$ and $f_{\rm g}(r_{\rm 200})$ also 
allows us to calculate $M_g(r_{\rm200})$. $n_{\rm {ei}}$ is then
\begin{equation}
 n_{\rm ei}=\frac{M_{\rm g}(r_{\rm 200})}{4\pi \mu_{\rm e} \displaystyle \int_{0}^{r_
{\rm 200}} r'^2 \left(\frac{r}{r_{\rm {p}}}\right)^{(\frac{-3}{5}){ c}}\left[1+
\left(\frac{r}{r_{\rm p}}\right)^{ a}\right]^{-\frac{3}{5}\left
( \frac{{ b} - { c}}{{ a}}\right)}\left[1 + \left(\frac{r}{r_{ c}}\right)
^2 \right]^{-\frac{3}{5}\alpha} dr'}.
\end{equation}
In parameterisation II we substitute the electron number density and GNFW pressure 
profiles in the assumption of hydrostatic equilibrium ($\frac{1}{\rho_{\rm gas}}\frac
{dP_{\rm gas}}{dr}=-\frac{{\rm G}M_{\rm tot}}{r^2})$ at $r_{200}$ and 
derive $P_{\rm ei}$. The normalisation coefficients for the temperature and entropy 
profiles, $k_{\rm B}T _{\rm {ei}}$ and $ K_{\rm {ei}}$, are then calculated by 
solving the equations (39) and (40) simultaneously.

In parameterisation III we calculate $k_{\rm B}T_{\rm e}(r_{\rm 200})$ using virial 
M-T relation, equation(33). $k_{\rm B}T _{\rm {ei}}$ is then calculated by 
substituting the values derived for $r_{\rm 200}$ and $k_{\rm B}T_{\rm e}(r_{\rm 
200})$ in temperature profile given in equation (38). Similarly, the normalisation 
coefficients for the pressure and entropy profiles, $P_{\rm {ei}}$ and $ K_{\rm {ei}}
$, are then calculated by solving the equations (39) and (40) simultaneously.

In order to estimate the cluster physical parameters at $r_{\rm 500}$ we use the 
definition of gas concentration parameter $c_{\rm 500}$ to estimate $r_{\rm 500}$,
\begin{equation}
c_{\rm 500}=\frac{r_{\rm 500}}{r_{\rm p}}.
\end{equation}
We fix $c_{\rm 500}$ to the value given in Arnaud et~al. (2010) ($c_{\rm 500}=1.156
$). With knowledge of $r_{\rm 500}$ and all the four normalisation coefficients $(P_
{\rm {ei}},K_{\rm {ei}},k_{\rm B}T _{\rm {ei}},n_{\rm {ei}} )$ we can calculate $M_
{\rm T}(r_{\rm 500}), M_{\rm g}(r_{\rm 500}), f_{\rm g}(r_{\rm 500})$ and $k_{\rm B}
T _{\rm {500}}$.
\section{Simulated AMI SA data}
In generating simulated SZ skies and observing them with a model AMI SA, we have 
used the methods outlined in Hobson \& Maisinger (2002) and Grainge et~al.\ (2002). 

Generating a simulated cluster SZ signal using the isothermal $\beta$-model requires 
the input parameters of $z$, $T_{\rm g}$, $n_{\rm e}(0)$, $r_{\rm c}$, and $\beta$; 
this set of parameters fully describes the Comptonisation $y$ parameter. However, in 
order to verify the results of our analysis and to see if our methodology is capable 
of recovering the true values associated with the simulated cluster, it is 
instructive to estimate the cluster physical parameters using the three 
parameterisation discussed. We note that any parameterised model within the 
hierarchical structure formation of the universe for the ICM, including the 
isothermal $\beta$-model, introduces constraints and biases
in the inferred cluster parameters. Moreover, we now show that it is possible to get 
different cluster physical parameters with the same set of input model parameters 
derived using the two different mass-temperature relations described in Section 
6.1.2.

For example, if we consider parameterisation II we can use the HSE mass-temperature 
relation to calculate $r_{\rm 200}$ given in equation (29). $M_{\rm g}(r_{\rm 200})$ 
and $M_{\rm T}(r_{\rm 200})$ are then calculated applying spherical geometry
assumptions described in equations (24) and (25) respectively. We can also determine 
$f_{\rm g}(r_{\rm 200})$.

However, if we consider parameterisation III we first calculate $M_{\rm T}(r_{\rm 
200})$ using the virial mass-temperature relation given in equation (33). $r_{\rm 
200}$ and $M_{\rm g}(r_{\rm 200})$ are then estimated assuming spherical geometry 
for the cluster, (equations 25 and 24). A numerical example that leads to different 
results in cluster parameters is given in Table 5.
\begin{table}
\caption{An example of input cluster parameters for the isothermal $\beta$-model that 
lead to inconsistent results using different parameterisations assuming $h=0.7$.}
\begin{tabular}{@{}c  c c c c@{} }
\hline
Input & Assumed  & Derived   & Parameterisation& Parameterisation \\
parameter &value & parameter & I,II& III\\\hline
$x_{\rm c}$      & $0$ &$r_{\rm 200}$  & $1.56\,\rm{Mpc}$ &$1.71\,\rm{Mpc}$  \\
$y_{\rm c}$      & $0$ &  & &  \\
$r_{\rm c}$      & $200\,\rm{kpc}$ & $M_{\rm T}(r_{\rm 200})$ &$5.83\times 10^{14}\, \rm
{M_\odot}$ & $7.67\times 10^{14}\,\rm{M_\odot}$ \\
$\beta$          &$0.95$ &  & &            \\
$T_{\rm g}$      &$5\,\rm{keV}$&$M_{\rm g}(r_{\rm 200})$  & $5.91\times 10^{13}\,\rm{M_
\odot}$&$6.28\times 10^{13}\,\rm{M_\odot}$             \\
$n_{\rm e}(0)$   & $10^4\,\rm{m^{-3}}$ &  & &                \\
$z$              &$0.3$ &$f_{\rm g}(r_{\rm 200})$  &$0.102$ & $0.082$\\ \hline
\end{tabular}
\end{table}
To address this issue, we studied how $f_{\rm g}(r_{\rm 200})$ varies as a function 
of $r_{\rm c}$ and $\beta$ while $n_{\rm e}(0)$, $T_{\rm g}$ and $z$ are fixed for 
both the HSE and virial M-T relations. Clearly, to obtain consistent results from 
both parameterisations one should select the values of $r_{\rm c}$ and $\beta$ for 
which the corresponding gas mass fraction ratio is one.

It should be noted that the same study may be carried out by investigating the 
variation of the ratio of gas mass fractions with either $n_{\rm e}(0)$ or $T_{\rm g}
$ while keeping $r_{\rm c}$ and $\beta$ constant. However we find that the ratio is 
not sensitive to variation of these two parameters.

Given the above, we decided to generate a simulated cluster with input parameters 
given in Table 6, which leads to consistent physical parameters for the cluster, at 
both $r_{\rm 200}$ and $r_{\rm 500}$ in both parameterisations. 
We assume that our cluster target is at declination $\delta = +$40$\degr$ observed 
for hour angles between $-$ 4 and $+$ 4 with 2-s sampling for four days and 8 hours 
per day. We calculate the Comptonisation $y$ parameter on a grid of 512 $\times$ 512 
pixels with pixel size of 30\arcsec. A realisation of the primary CMB is calculated 
using a power spectrum of primary anisotropies which was generated for $l<8000$ 
using CAMB (Lewis, Challinor \& Lasenby 2000), with a $\rm{\Lambda CDM}$ cosmology: $
\Omega_{\rm M}=0.3$, $\Omega_{\rm \Lambda}=0.7$, $\sigma_{\rm 8}=0.8$, $H_{\rm 0}=70
\,\rm{km\,s^{-1}Mpc^{-1}}$, $w_{\rm 0}=-1$ and $w_{\rm a}=0$. The CMB realisation is 
then co-added to the cluster in brightness temperature. It should be noted that in 
our simulation we did not include extragalactic radio sources, or diffuse foreground 
emission from the galaxy as we have already addressed the effects of the former in 
Feroz et~al. 2009 and the foreground galactic emission is unlikely to be a major 
contaminant since our interferometric observations resolve out such large scale 
emission. The map is scaled by the primary beam appropriate to the
measured value in the frequency channel and transformed into the Fourier plane. The 
resulting distribution is sampled at the required visibility points and thermal
noise of $0.54$ Jy per channel per baseline in one second which is appropriate to 
the measured sensitivity of the SA is added. Fig. 1 shows a map of the SZ
temperature decrement of the first simulated cluster generated using the isothermal $
\beta$-model.
\begin{table}
\caption{Cluster parameters for the isothermal $\beta$-model assuming $h=0.7$.}
\begin{tabular}{@{}cccc@{} }
\hline
Input     & Assumed  &Derived & Parameterisation\\
Parameter&  value   &Parameter& I,II,III\\\hline
$x_{\rm c}$    & $0$ &$r_{\rm 200}$ & $1.56\,\rm{Mpc}$\\
$y_{\rm c}$    & $0$ &$M_{\rm T}(r_{\rm 200})$ &$5.83\times 10^{14}\,\rm{M_\odot}$ \\
$r_{\rm c}$    & $155\,\rm{kpc}$ &$M_{\rm g}(r_{\rm 200})$ &$6.36\times 10^{13}\,\rm
{M_\odot}$  \\
$\beta$        &$0.79$ & $f_{\rm g}(r_{\rm 200})$&$0.109$             \\
$T_{\rm g}$    &$5\,\rm{keV}$ & $r_{\rm 500}$& $0.98\,\rm{Mpc}$    \\
$n_{\rm e}(0)$ & $10^4\,\rm{m^{-3}}$ & $M_{\rm T}(r_{\rm 500})$& $3.64\times 10^{14}
\,\rm{M_\odot}$\\
$z$         &$0.3$ &$M_{\rm g}(r_{\rm 500})$ & $4.13\times 10^{13}\,\rm{M_\odot}$ \\ 
            &      & $f_{\rm g}(r_{\rm 500})$&$0.11$\\\hline
\end{tabular}
\end{table}
\begin{figure}
\includegraphics[width=80mm,clip=]{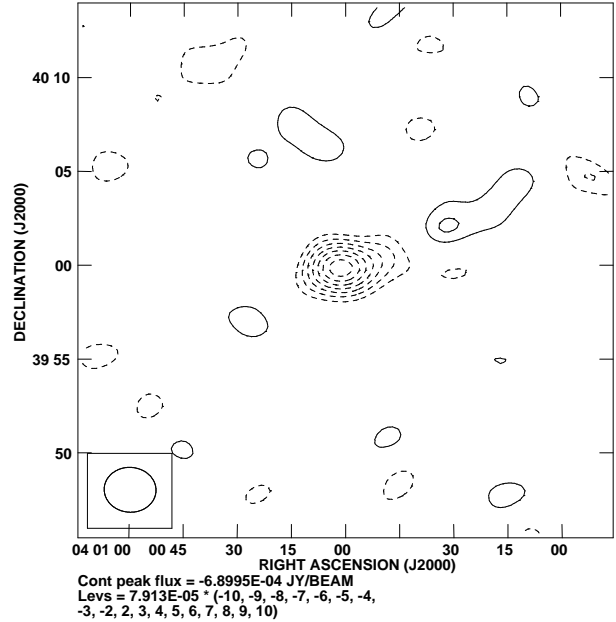}
\caption{Map of the simulated AMI SZ temperature decrement generated
with the isothermal-$\beta$ model and parameters given in Table 6. Contours at $...-3, -2, 2, 3 ...$ times noise ($
\sigma =79~\muup$Jy~beam$^{-1}$), negative contours are dashed. The
coordinates are J2000.0.}
\end{figure}

To generate the second simulated cluster we use the GNFW pressure profile to 
calculate the Comptonisation y parameter. The input parameters for this model, ($P_
{\rm ei}$, $r_{\rm p}$), were selected to represent a cluster with the same physical
parameters at $r_{\rm 200}$ and the same noise level as the first cluster.
\begin{table}
\caption{Cluster parameters for the entropy-GNFW model assuming $h=0.7$.}
\begin{tabular}{@{}cccc@{} }
\hline
Input     & Assumed  & Derived   & "Entropy"-GNFW \\
parameter & value    & parameter & model \\
\hline
$x_{\rm c}$ & $0$ &$r_{\rm 200}$ &$1.56\,\rm{Mpc}$\\
$y_{\rm c}$ & $0$ &$M_{\rm T}(r_{\rm 200})$ &$5.83\times 10^{14}\,\rm{M_\odot}$\\
$r_{\rm p}$ & $0.85\,\rm{Mpc}$ &$M_{\rm g}(r_{\rm 200})$&$6.36\times 10^{13}\,\rm{M_
\odot}$ \\
$P_{\rm ei}$ & $37647.51\,\rm{keV}\rm{m^{-3}}$ & $f_{\rm g}(r_{\rm 200})$ &$0.109$  \\
$z$ &$0.3$& $r_{\rm 500}$& $0.98\,\rm{Mpc}$\\
$r_{\rm c}$ &$155\,\rm{kpc}$&$M_{\rm T}(r_{\rm 500})$ & $3.64\times 10^{14}\,\rm{M_
\odot}$ \\
$\alpha$ &$0.55$ & $M_{\rm g}(r_{\rm 500})$   & $4.08\times 10^{13}\,\rm{M_\odot}
$       \\
$T_{\rm 200}$ &$5\,\rm{keV}$ & $f_{\rm g}(r_{\rm 500})$   & $0.11$  \\
              &              &$T_{\rm 500}$ & $6.95\,\rm{keV}$ \\
\hline
\end{tabular}
\end{table}
Although the parameters $r_{\rm c}$, $\alpha$, $z$, and $T_{\rm e}(r_{\rm 200})$ do 
not contribute to the calculation of Comptonisation y parameter directly, their 
values were used to derive the parameters describing the GNFW pressure
profile by following the steps described in section 6.2 to ensure that they 
represent the cluster with required physical parameters at $r_{\rm 200}$. The 
cluster physical parameters at different radii will be different from the first 
cluster due to the different models describing the ICM and relaxing the assumption 
of isothermality in the second cluster. A summary of the cluster
parameters is presented in Table 7. 
Fig. 2 shows a map of the SZ temperature decrement of the second simulated cluster, 
generated using the GNFW model.
\begin{figure}
\includegraphics[bb=49 116 565 649,width=80mm,clip=]{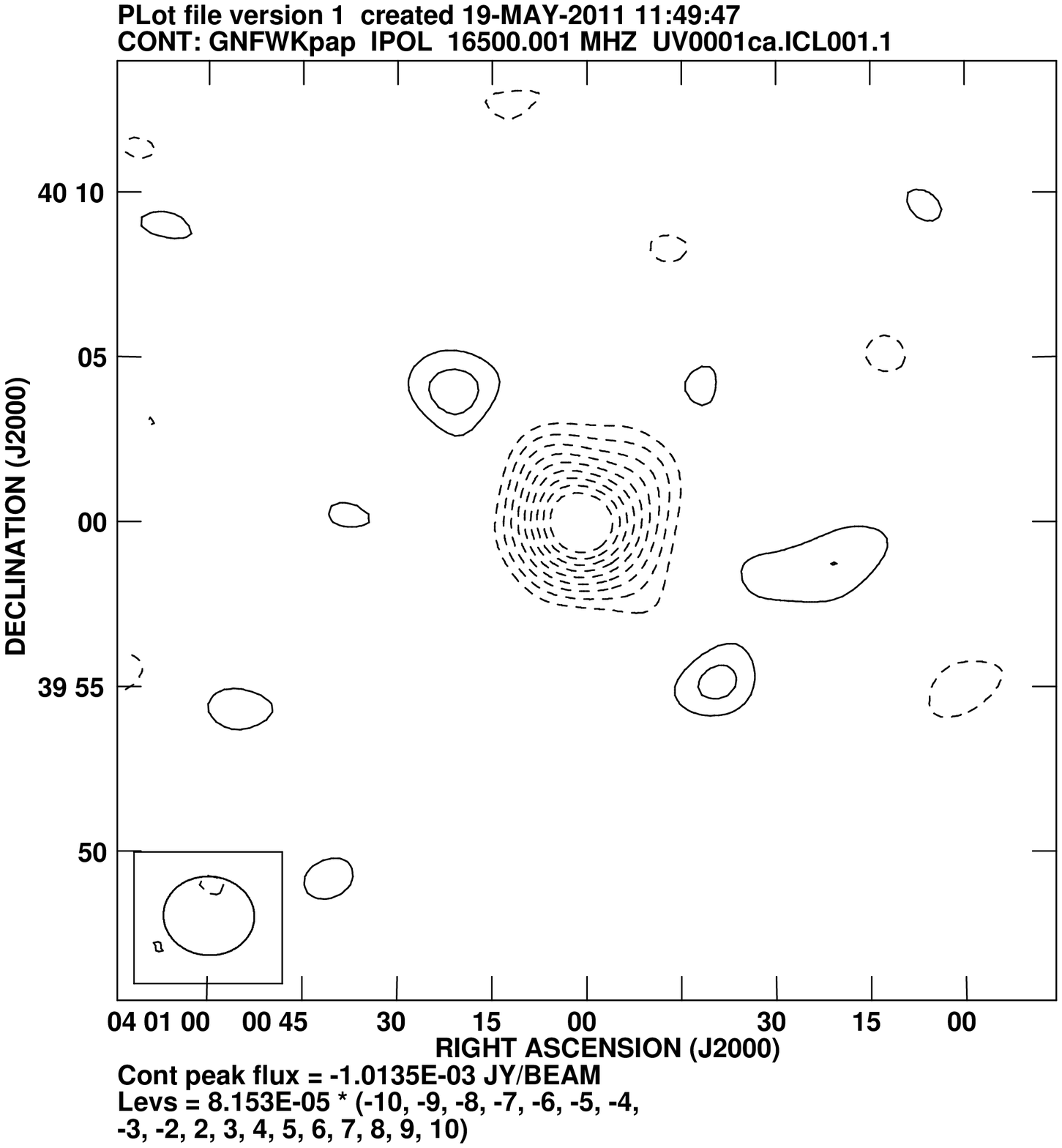}
\caption{Map of the simulated AMI SZ temperature decrement generated
with GNFW model and parameters given in Table 7. Contours at $...-3, -2, 2, 3 ...$ 
times noise ($\sigma =81.5~\muup$Jy~beam$^{-1}$), negative contours are dashed. The 
coordinates are J2000.0.}
\end{figure}
\section{Analysis and results}
In this section we present the results of our analysis for all three
parameterisations within the context of the isothermal $\beta$-model and the 
entropy-GNFW pressure model. In each case we first study our methodology in the 
absence of data. This can be carried out by setting the likelihood to
a constant value and hence the algorithm explores the prior space. This
analysis is crucial for understanding the underlying biases and constraints
imposed by the priors and the model assumptions. Along with the analysis
done using the simulated AMI data, this approach reveals the constraints that
measurements of the SZ signal place on the cluster physical parameters and the
robustness of the assumptions made. It should be noted that in all the plots of
probability distributions, we explicitly include the dimensionless Hubble
parameter $h=H_0/(\rm{100\, km\,s^{-1}Mpc^{-1})}$  with $h$ set to 1.0.
\subsection{Analysis using isothermal $\beta$-model-Parameterisation I}
\begin{figure}
  \includegraphics[width=80mm]{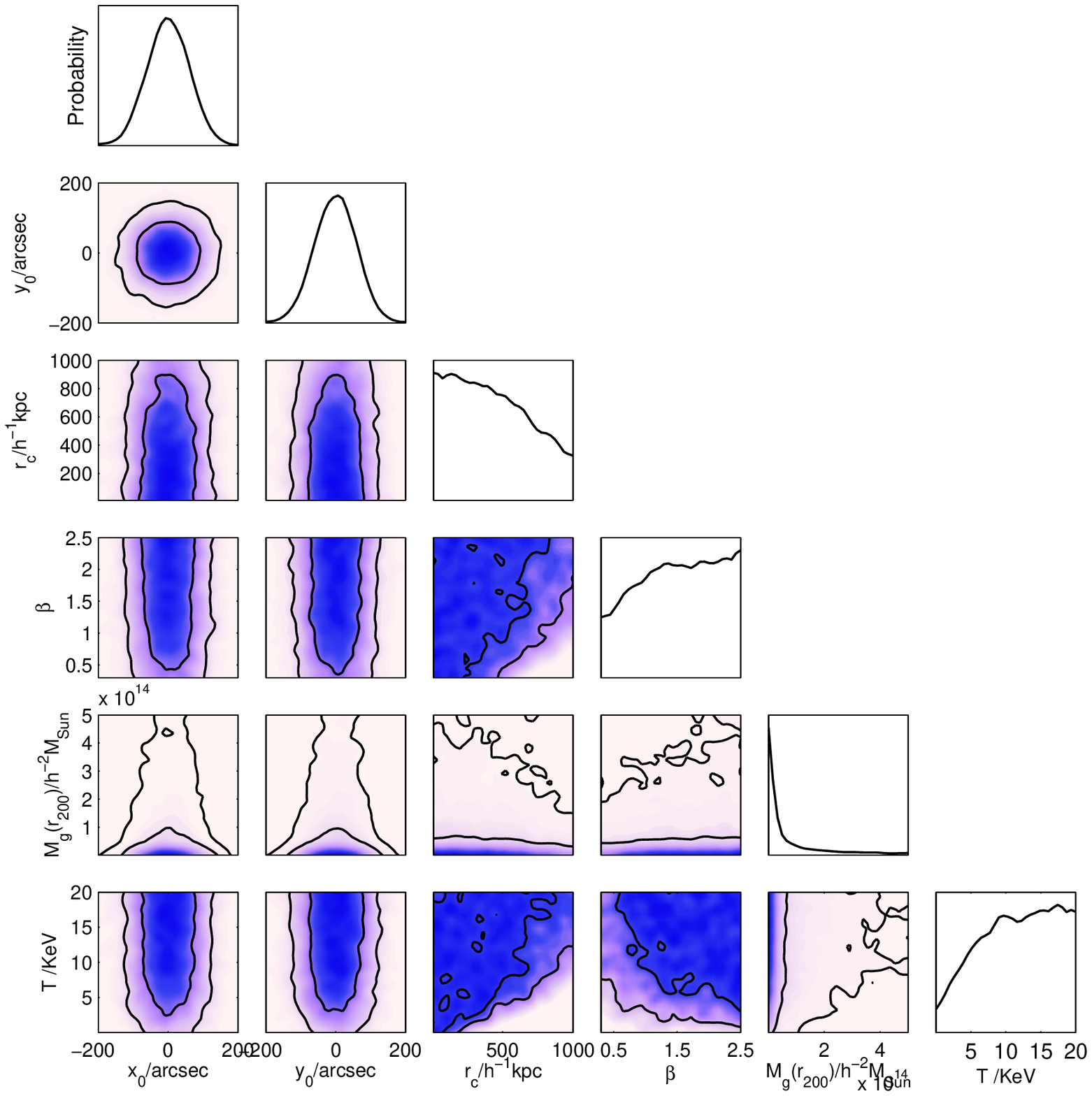}
  \caption{Marginal distributions for the sampling parameters with no data for 
isothermal $\beta$-model --parameterisation I.}
  \medskip
  \includegraphics[width=80mm]{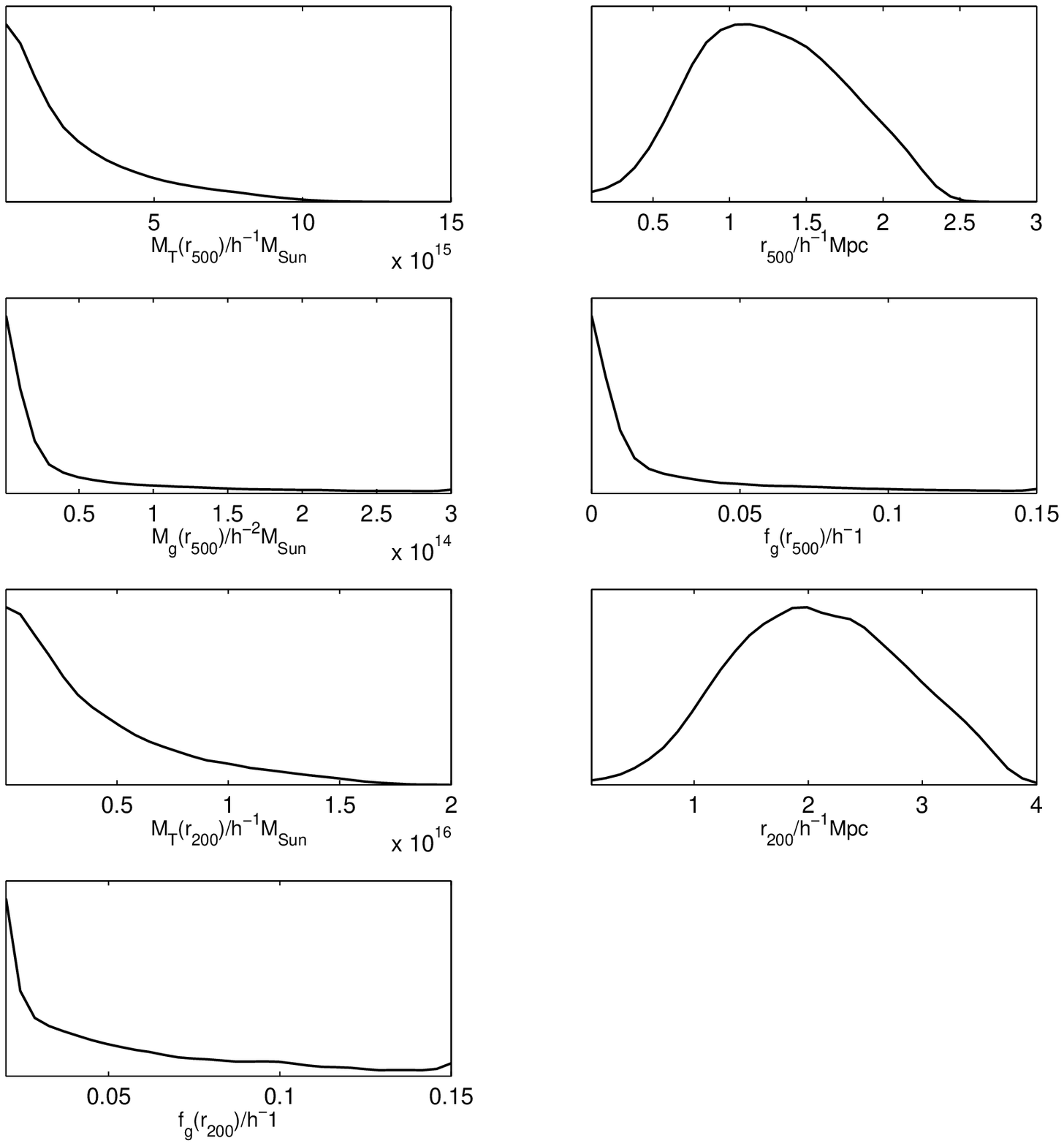}
  \caption{Marginal distributions for the derived cluster physical parameters with no 
data for isothermal $\beta$-model --parameterisation I.}
\end{figure}
\begin{figure}
  \includegraphics[width=80mm]{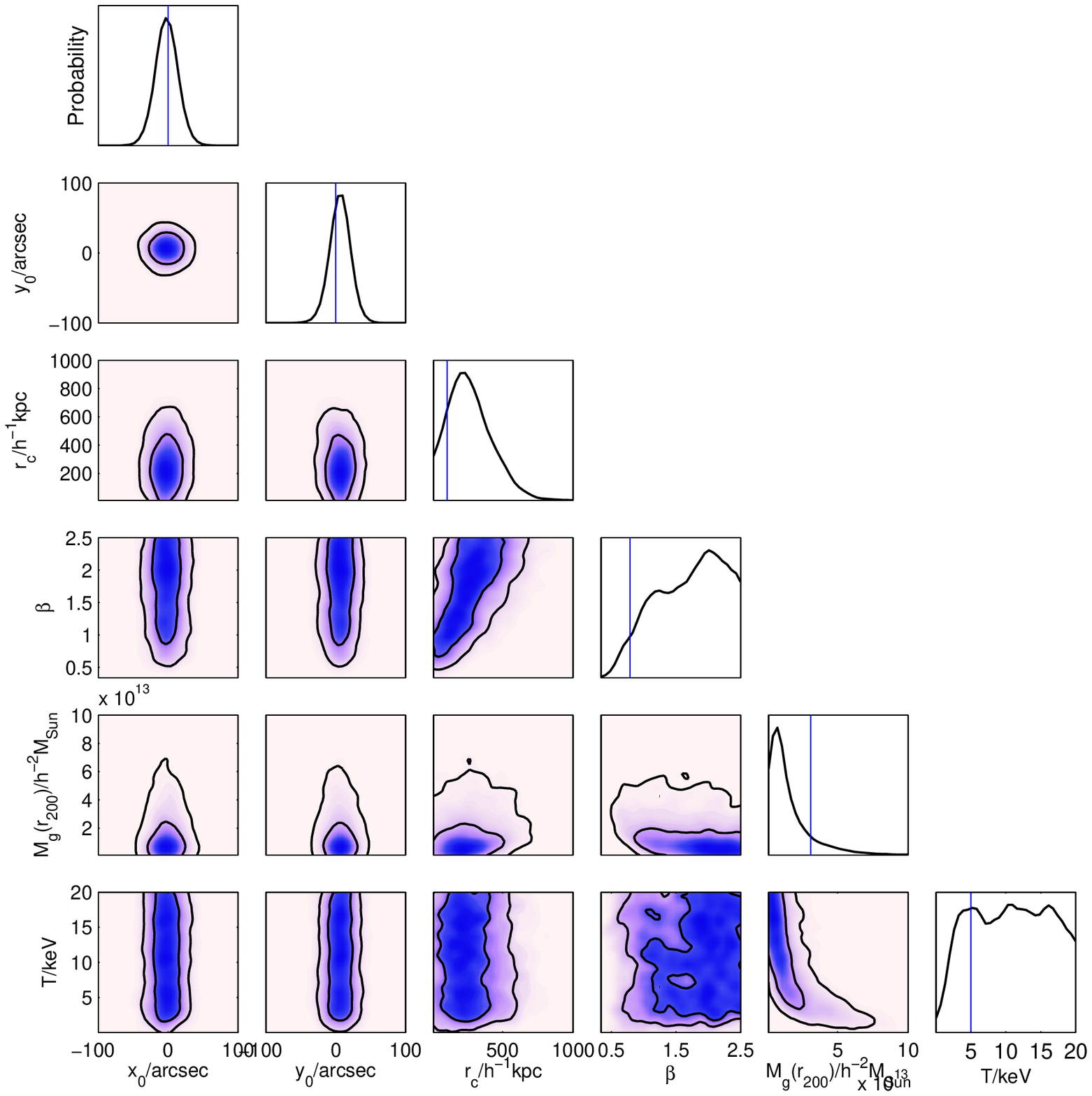}
  \caption{Marginal distributions of the sampling parameters for simulated data for 
isothermal $\beta$-model --parameterisation I.}
  \medskip
  \includegraphics[width=80mm]{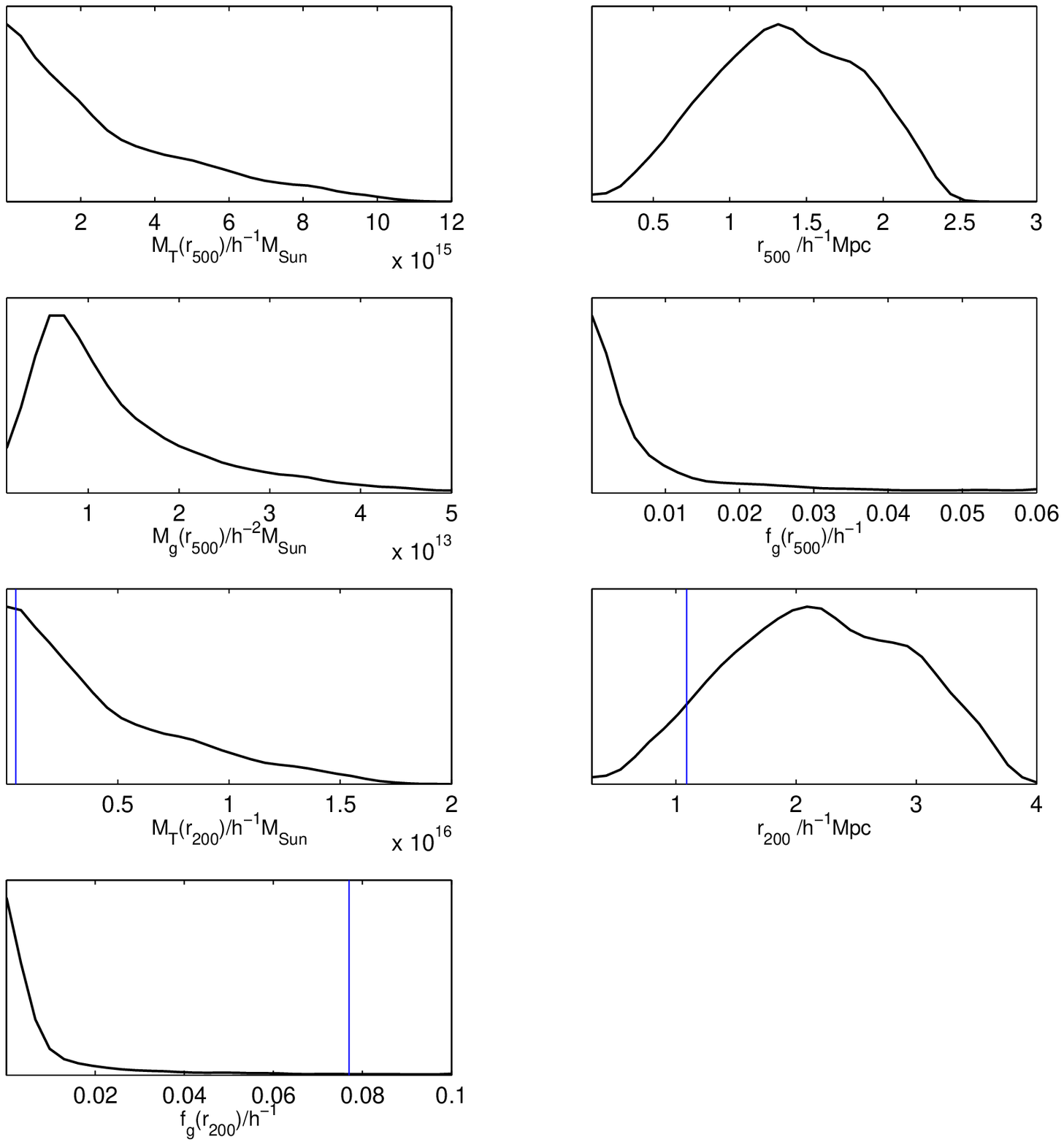}
  \caption{Marginal distributions of the derived cluster physical parameters with 
simulated data for isothermal $\beta$-model -- parameterisation I.}
\end{figure}
Figs. 3 and 4 represent the results of a priors-only analysis showing the sampling and 
derived parameters respectively. \mbox{1-D}
marginalised posterior distributions of sampling parameters in Fig. 3 show that we were 
able to recover the assumed prior probability distributions for cluster
position and the gas mass. However, this parameterisation clearly prefers higher 
temperature and $\beta$ and the probability distribution for $r_{\rm c}$ falls as we
go towards higher $r_{\rm c}$. This feature in particular creates a void region in the 
2-D marginalised probability distributions of $\beta - r_{\rm c}$ and $T_{\rm g}-r_{\rm
c}$ at higher $r_{\rm c}$ which implies that low mass clusters are unlikely to have 
high $r_{\rm c}$ and low $\beta$. This effect is a direct result of imposing the 
constraint that $r_{\rm {500}}>0$. Moreover, as may be seen from Fig. 4, this
choice of priors drives the posterior probability distributions of both the gas mass 
and the gas mass fraction towards low values.

Figs. 5 and 6 show the results of the analysis of the simulated SZ cluster data. The 
vertical lines show the true values of the parameters. Table 8 also summarises the 
mean, the dispersion and the maximum likelihood of each parameter.
\begin{table}
\caption{Simulated cluster parameters (mean, standard deviation and Maximum likelihood) 
estimated using isothermal $\beta$-model--parameterisation I assuming $h=0.7$.}
\begin{tabular}{@{}ccc@{} }
\hline
Parameter & $\mu \pm \sigma$ & $\hat \mu$ \\\hline
$x_{\rm c}$ & $-2.1 \pm 15.3\arcsec$ & $-0.6\arcsec$  \\
$y_{\rm c}$ & $6.3 \pm 13.9\arcsec$ & $5.5\arcsec$  \\
$r_{\rm c}$ & $391.26 \pm 214.98 \,\rm{kpc}$  & $138.77\,\rm{kpc}$  \\
$\beta$     & $1.7 \pm 0.5$ &$0.98$             \\
$M_{\rm g}(r_{\rm 200})$&$(3.41 \pm 3.16)\times 10^{13}\,\rm{M_\odot}$ &$2.86\times 10^
{13}\,\rm{M_\odot}$\\
$T_{\rm g}$ & $10.61 \pm 5.28 \,\rm{keV}$ &$10.15\,\rm{keV}$             \\
$M_{\rm T}(r_{\rm 500})$&$(3.91\pm 3.41)\times 10^{15}\,\rm{M_\odot}$ &$1.49\times 10^
{15}\,\rm{M_\odot}$\\
$r_{\rm 500}$  & $1.96 \pm 0.69\,\rm{Mpc}$ & $1.57\,\rm{Mpc}$   \\
$M_{\rm g}(r_{\rm 500})$ &$(2.71 \pm 1.42) \times 10^{13}\,\rm{M_\odot}$ &$2.31\times 
10^{13}\,\rm{M_\odot}$\\
$f_{\rm g}(r_{\rm 500})$ &$0.15 \pm 2.7$ & $0.014$ \\
$M_{\rm T}(r_{\rm 200})$&$(6.43\pm 5.43)\times 10^{15}\,\rm{M_\odot}$ &$2.37\times 10^
{15}\,\rm{M_\odot}$\\
$r_{\rm 200}$  & $3.14 \pm 1.0\,\rm{Mpc}$ & $2.49\,\rm{Mpc}$   \\
$f_{\rm g}(r_{\rm 200})$ &$(0.14\pm 3.4) $ & $0.012$ \\
\hline
\end{tabular}
\end{table}

In Fig. 5, we notice the strong degeneracy between $r_{\rm c}$ and $\beta$ (Grego et~al.\ 
2001). However, it is apparent that neither $\beta$ nor $T_{\rm g}$ is well-constrained 
using this parameterisation. Also, higher values than the true input parameters are 
preferred for both parameters. This effect leads to two results: firstly it yields a 
higher estimate for $r_{\rm 200}$ and so equation (28) overestimates the total mass; 
secondly, since for this parameterisation there is a negative \mbox{degeneracy} between 
gas mass and temperature, the high temperature therefore leads the marginalised posterior 
distribution for gas mass peaking towards the lower end of the  distribution although 
the recovered mean value of $M_{\rm g}(r_{\rm 200})$ is within $1\sigma$ from its 
corresponding input value for the simulated cluster. As a result of these two effects, 
the gas mass fraction is driven even further to the lower end of the allowed range. 
There is also  a degeneracy between the two free parameters of $\beta$ and $T_{\rm g}$; 
this degeneracy again originates from dependency of $r_{\rm 200}$ on both parameters as 
given in equation (29).
\subsection{Analysis using isothermal $\beta$-model-Parameterisation II}
\begin{figure}
  \includegraphics[width=80mm]{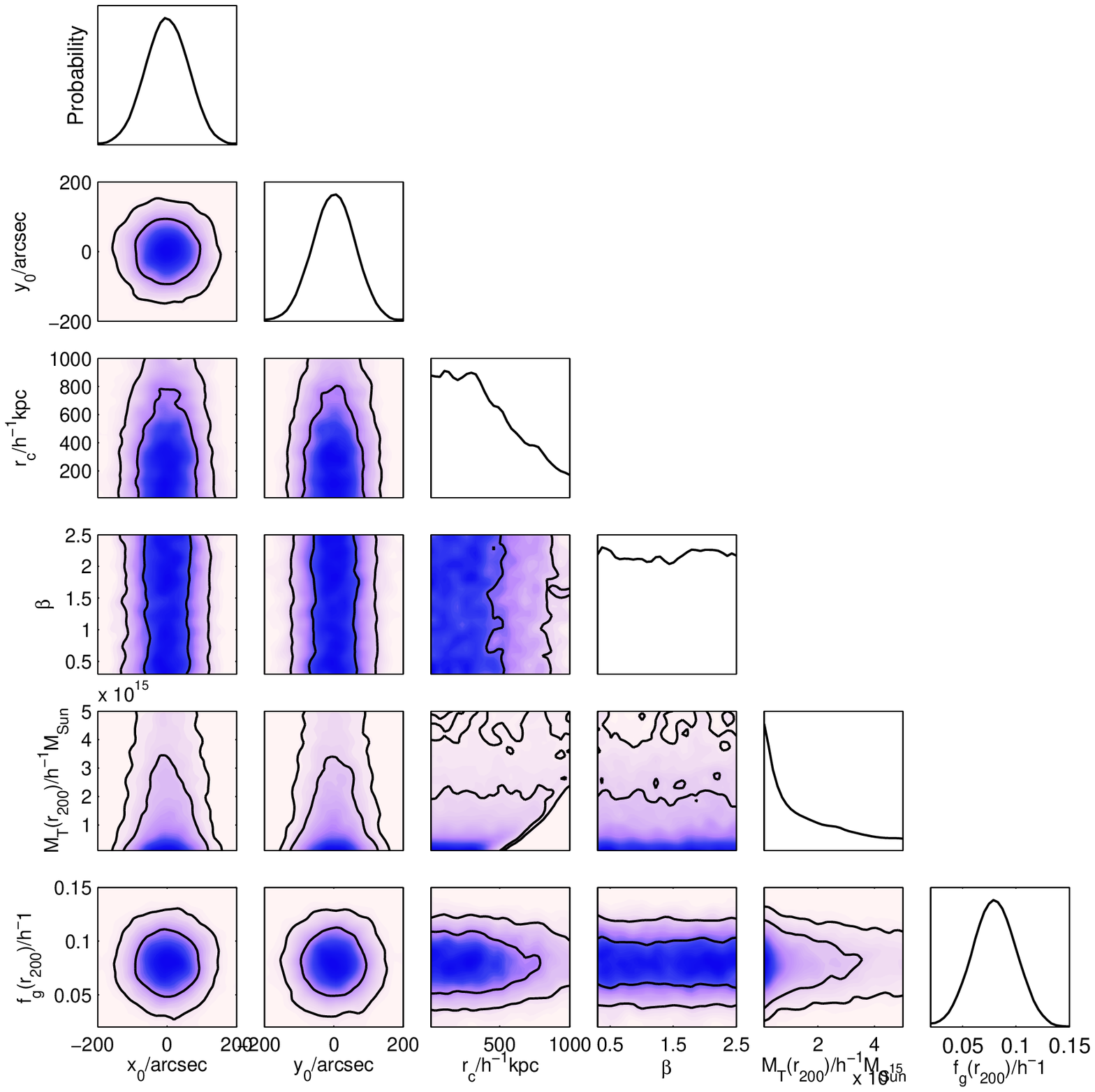}
  \caption{Marginal distributions of the sampling parameters with no data for isothermal 
$\beta$-model- parameterisation II.}
  \medskip
  \includegraphics[width=80mm]{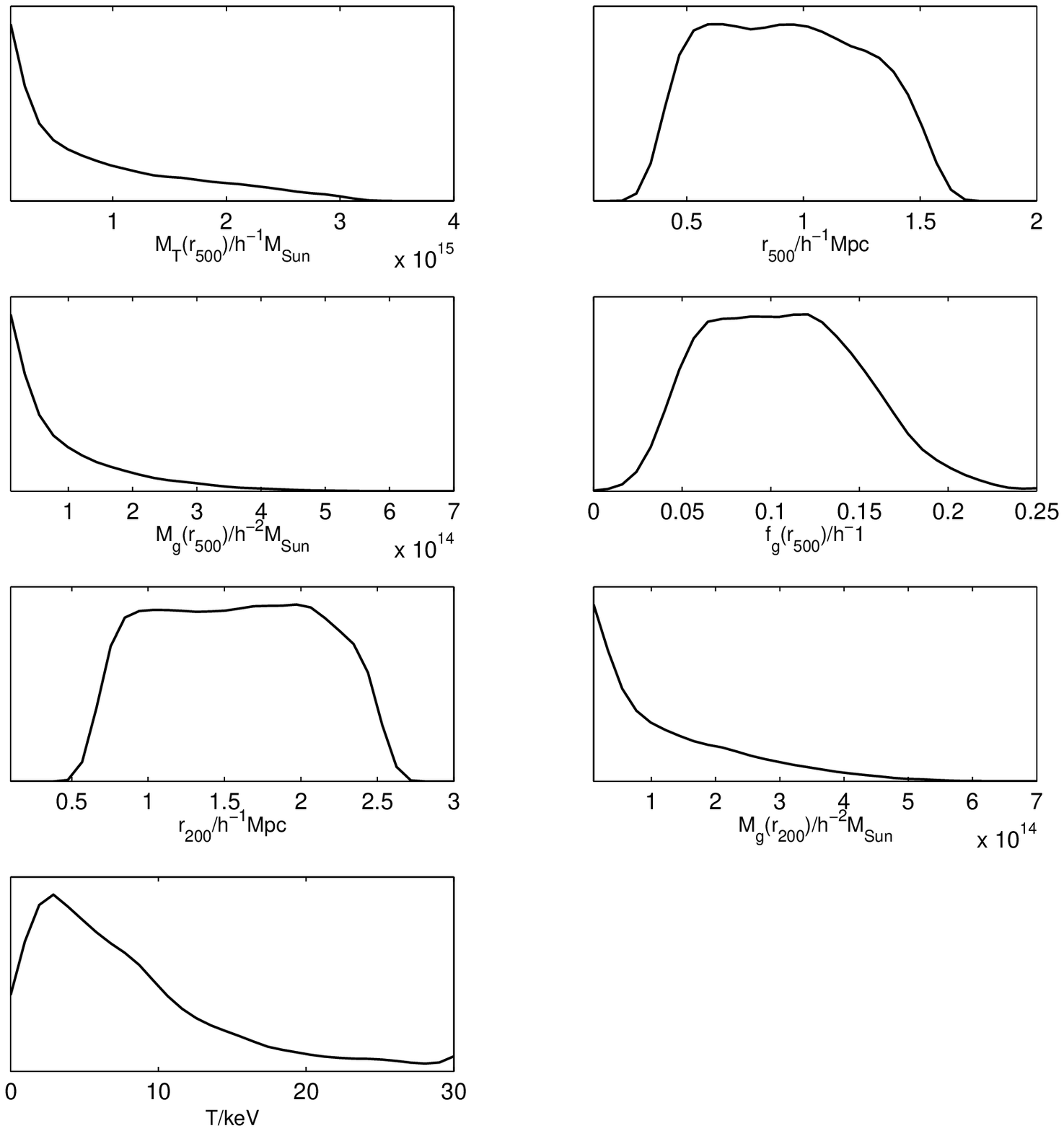}
  \caption{Marginal distributions of the derived cluster physical parameters with no 
data for isothermal $\beta$-model-parameterisation II.}
\end{figure}
\begin{figure}
  \includegraphics[width=80mm]{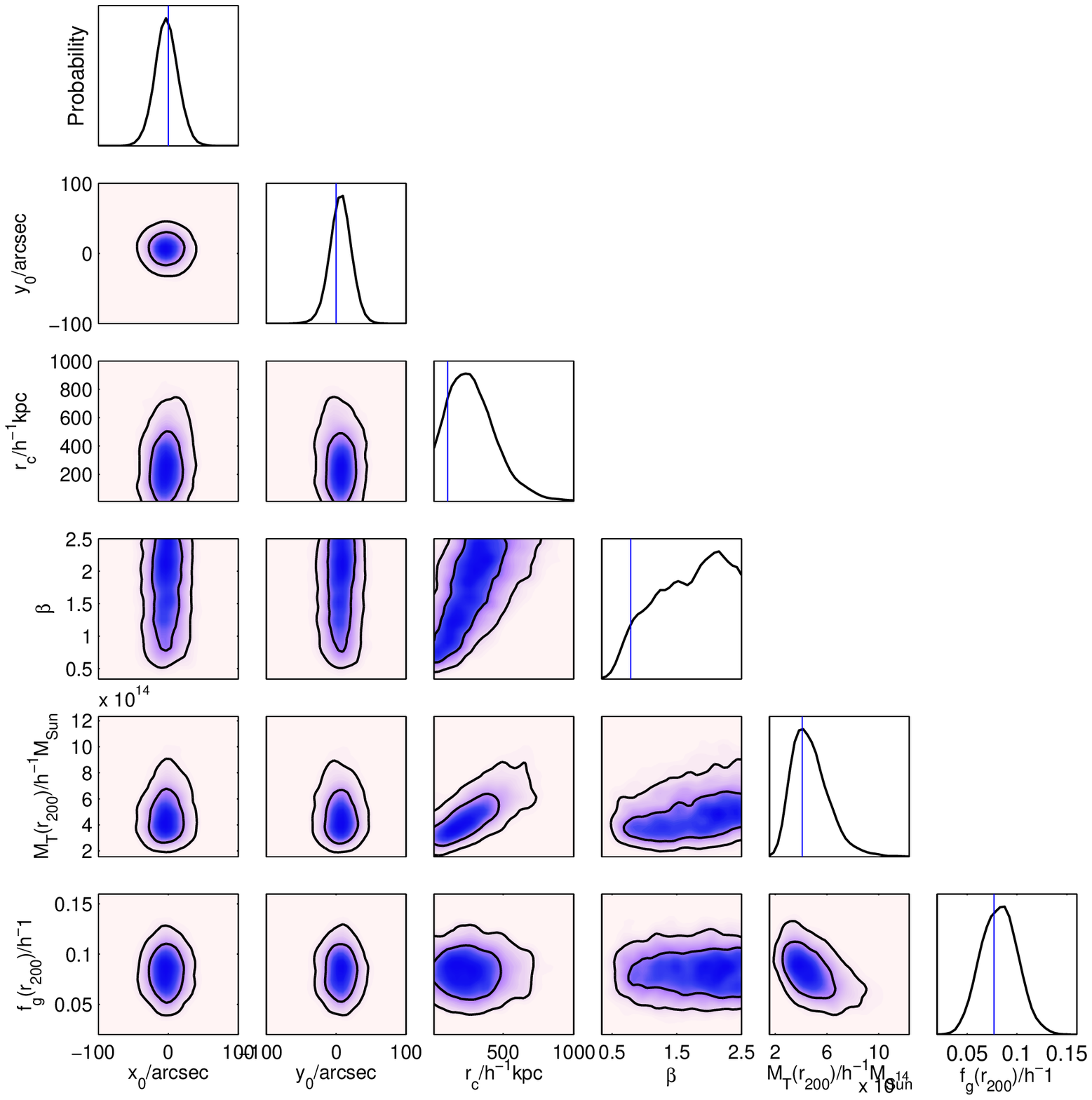}
  \caption{Marginal distributions of the sampling parameters with simulated data for 
isothermal $\beta$-model- parameterisation II.}
  \medskip
  \includegraphics[width=80mm]{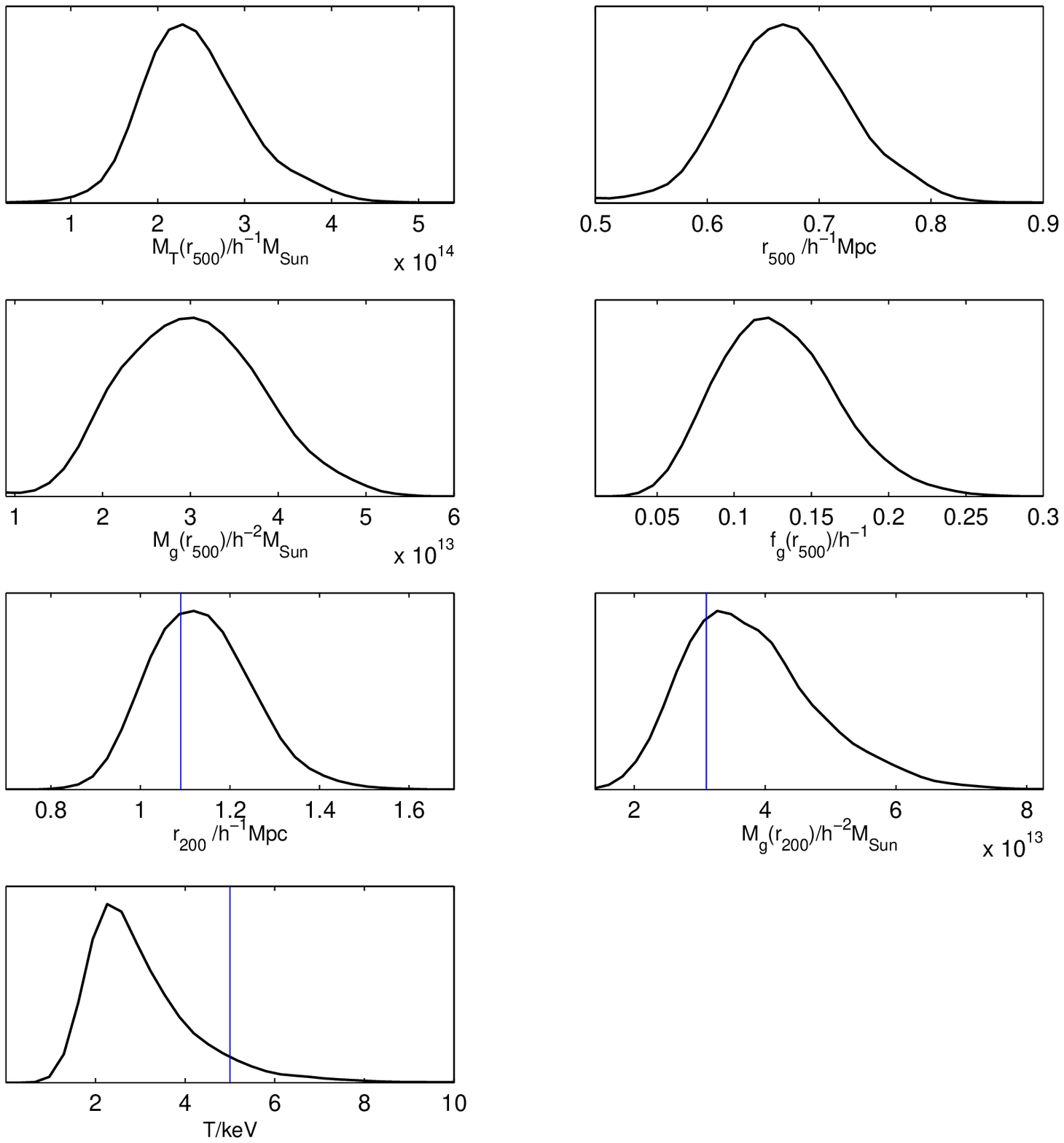}
  \caption{Marginal distributions of the derived cluster physical parameters with 
simulated data for isothermal $\beta$-model-parameterisation II.}
\end{figure}
Figs. 7 and 8 show the results from prior-only analysis for parameterisation
II. We recover the assumed prior probability distributions for cluster
position, $\beta$, total mass and gas mass fraction. There is a similar trend
in the 1-D posterior probability distribution of $r_{\rm c}$ to that mentioned in
the parameterisation I, which leads to a void region in the 2-D marginalised
posterior distribution of $M_{\rm T}(r_{\rm 200}) - r_{\rm c}$ for the same
reason as discussed for the parameterisation I. However, parameterisation II
prefers a lower temperature which arises from the fact that HSE
mass-temperature relation used in this parameterisation (equation 32) is
inversely proportional to $\beta$.

Figs. 9 and 10 show the results of the analysis using simulated SZ cluster
data, with vertical lines representing the true parameter values. Table 9 also 
summarises the mean, the dispersion and the maximum likelihood values of each cluster 
parameter estimated using parameterisation II.
\begin{table}
\caption{Simulated cluster parameters (mean, standard deviation and Maximum likelihood) 
estimated using isothermal $\beta$-model--parameterisation II assuming $h=0.7$.}
\begin{tabular}{@{}ccc@{} }
\hline
Parameter & $\mu \pm \sigma$ & $\hat \mu$ \\\hline
$x_{\rm c}$ & $-2.6 \pm 15.7\arcsec$ & $5.5\arcsec$  \\
$y_{\rm c}$ & $6.4 \pm 14.5\arcsec$ & $5.6\arcsec$  \\
$r_{\rm c}$ & $410.37 \pm 237.24\,\rm{kpc}$  & $135.43\,\rm{kpc}$  \\
$\beta$     & $1.7 \pm 0.5$ &$0.8$             \\
$M_{\rm T}(r_{\rm 200})$&$(6.8 \pm 2.1)\times 10^{14}\,\rm{M_\odot}$ &$5.0\times 10^{14}
\,\rm{M_\odot}$\\
$f_{\rm g}(r_{\rm 200})$ &$0.12\pm 0.03 $ & $0.11$ \\
$M_{\rm T}(r_{\rm 500})$&$(3.5\pm 8.81)\times 10^{13}\,\rm{M_\odot}$ &$3.13\times 10^
{14}\,\rm{M_\odot}$\\
$r_{\rm 500}$  & $0.96 \pm 0.08\,\rm{Mpc}$ & $0.93\,\rm{Mpc}$   \\
$M_{\rm g}(r_{\rm 500})$ &$(6.2 \pm 1.6 )\times 10^{13}\,\rm{M_\odot}$ &$3.9\times 10^
{13}\,\rm{M_\odot}$\\
$f_{\rm g}(r_{\rm 500})$ &$0.18 \pm 0.05$ & $0.12$ \\
$r_{\rm 200}$  & $1.59 \pm 1.57\,\rm{Mpc}$ & $1.47\,\rm{Mpc}$   \\
$M_{\rm g}(r_{\rm 200})$&$(7.76\pm 2.08)\times 10^{13})\,\rm{M_\odot}$ &$5.35\times 10^
{13}\,\rm{M_\odot}$\\
$T_{\rm g}$ & $3.0 \pm 1.2\,\rm{keV}$ &$4.3\,\rm{keV}$             \\
\hline
\end{tabular}
\end{table}
 A tight degeneracy between $ r_{\rm c}$ and $\beta$ is noticeable in the corresponding
2-D marginalised probability distribution. $\beta$ on the other hand is not
well constrained and moves towards higher values which results in the
probability distribution of temperature being driven to lower values again
because of the $1/\beta$ relationship in equation (32). However, this
parameterisation along with the simulated SZ data reliably constrains $M_{\rm T}(r_{\rm 
200})$, $M_{\rm g}(r_{\rm 200})$ and $f_{\rm g}(r_{\rm 200})$. Comparing the 1-D
marginalised posterior distributions of gas mass fractions at two overdensity
radii $r_{\rm 500}$ and $ r_{\rm 200}$ also reveals that we cannot constrain
the radial behaviour of the gas mass fraction using this parameterisation, as
$f_{\rm g}(r_{\rm 500})$ exhibits too wide a probability distribution. For $f_{\rm g}(r_
{\rm 200})$, we seem to have recovered the input prior distribution.
\subsection{Analysis using isothermal $\beta$-model-Parameterisation III}
\begin{figure}
  \includegraphics[width=80mm]{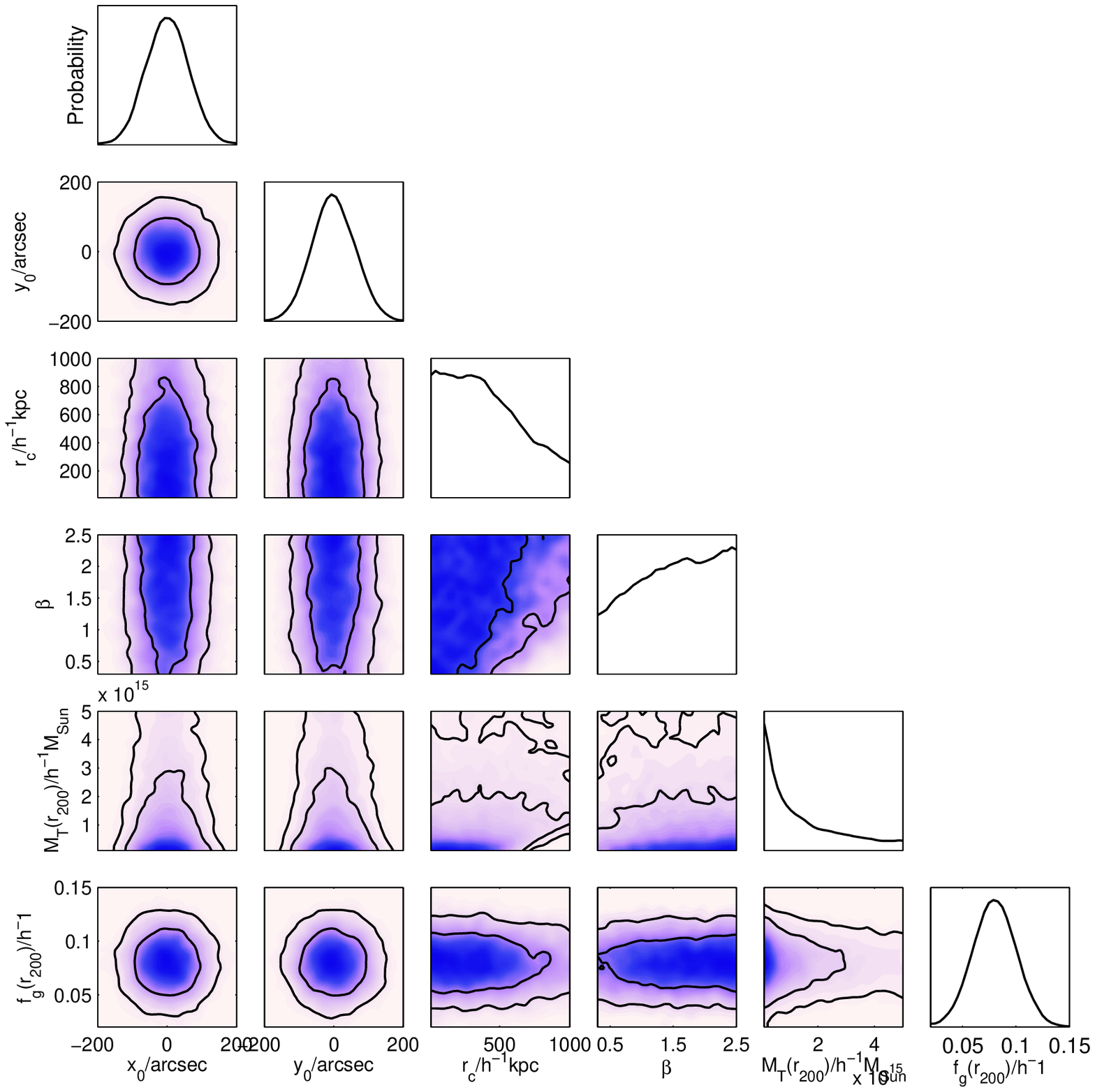}
  \caption{Marginal distributions of the sampling parameters with no data for isothermal 
$\beta$-model- parameterisation III.}
  \medskip
  \includegraphics[width=80mm]{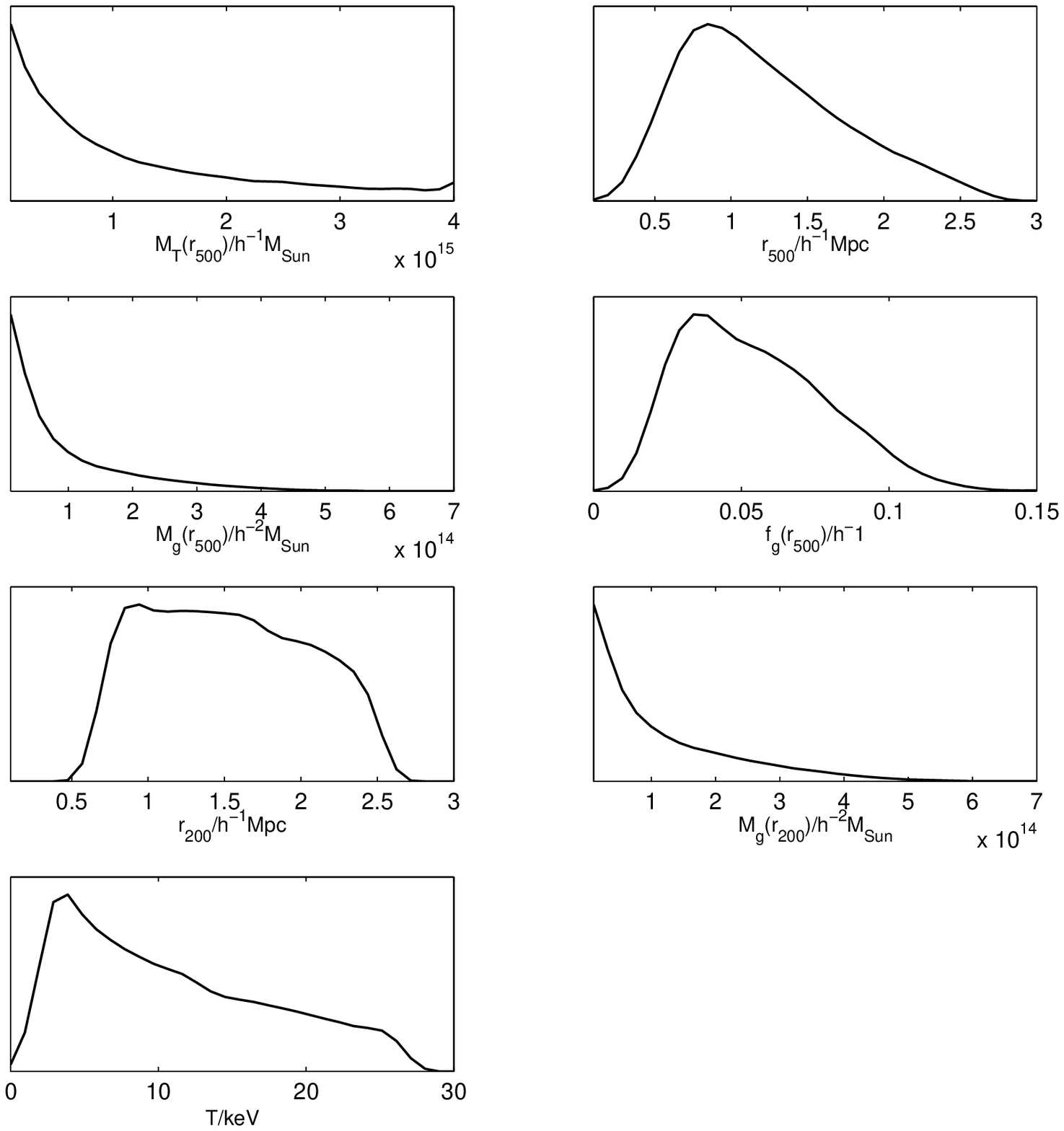}
  \caption{Marginal distributions of the derived cluster physical parameters with no 
data for isothermal $\beta$-model- parameterisation III.}
\end{figure}
\begin{figure}
  \includegraphics[width=80mm]{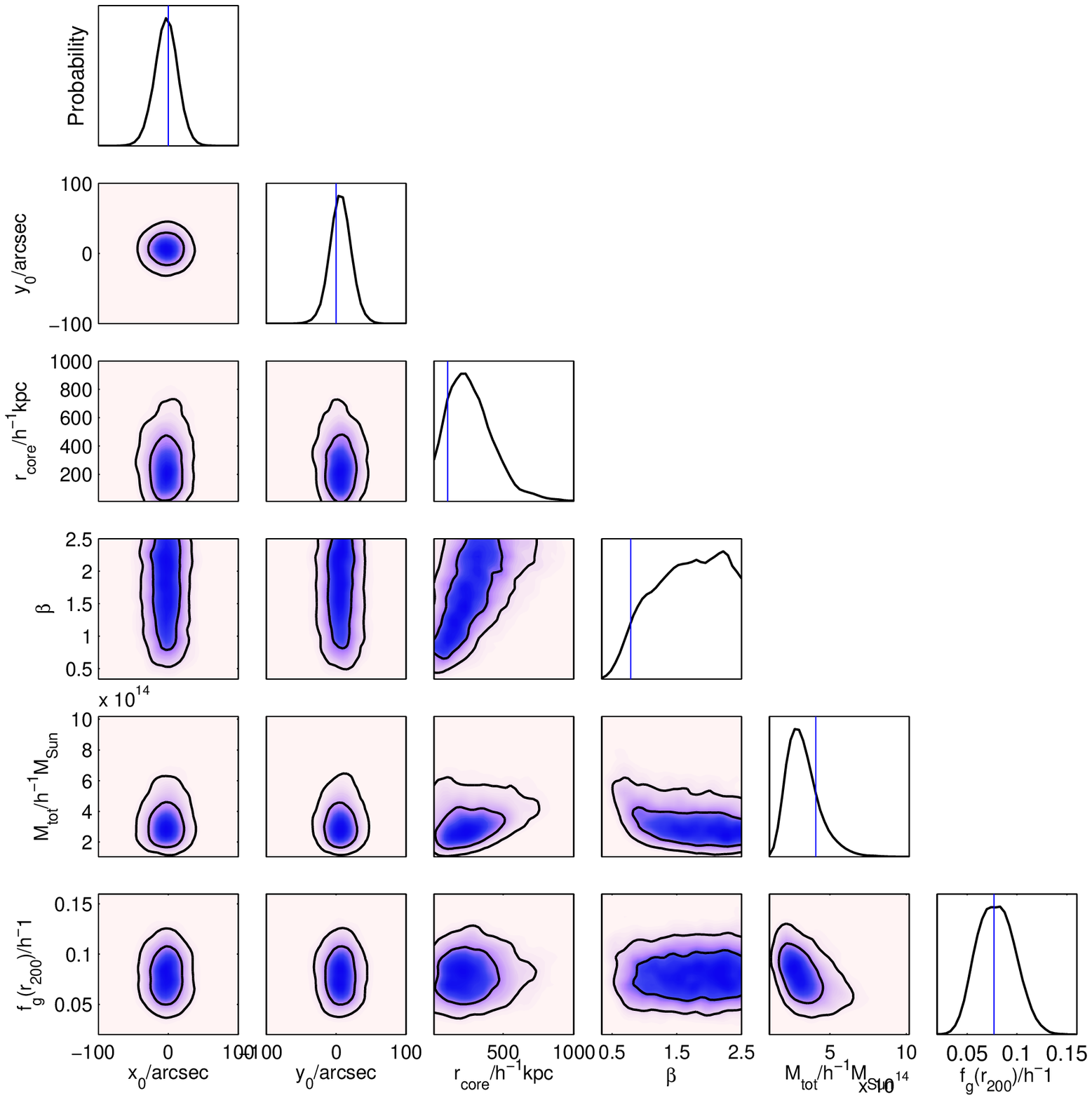}
  \caption{Marginal distributions of the sampling parameters with simulated data for 
isothermal $\beta$-model- parameterisation III.}
  \medskip
  \includegraphics[width=80mm]{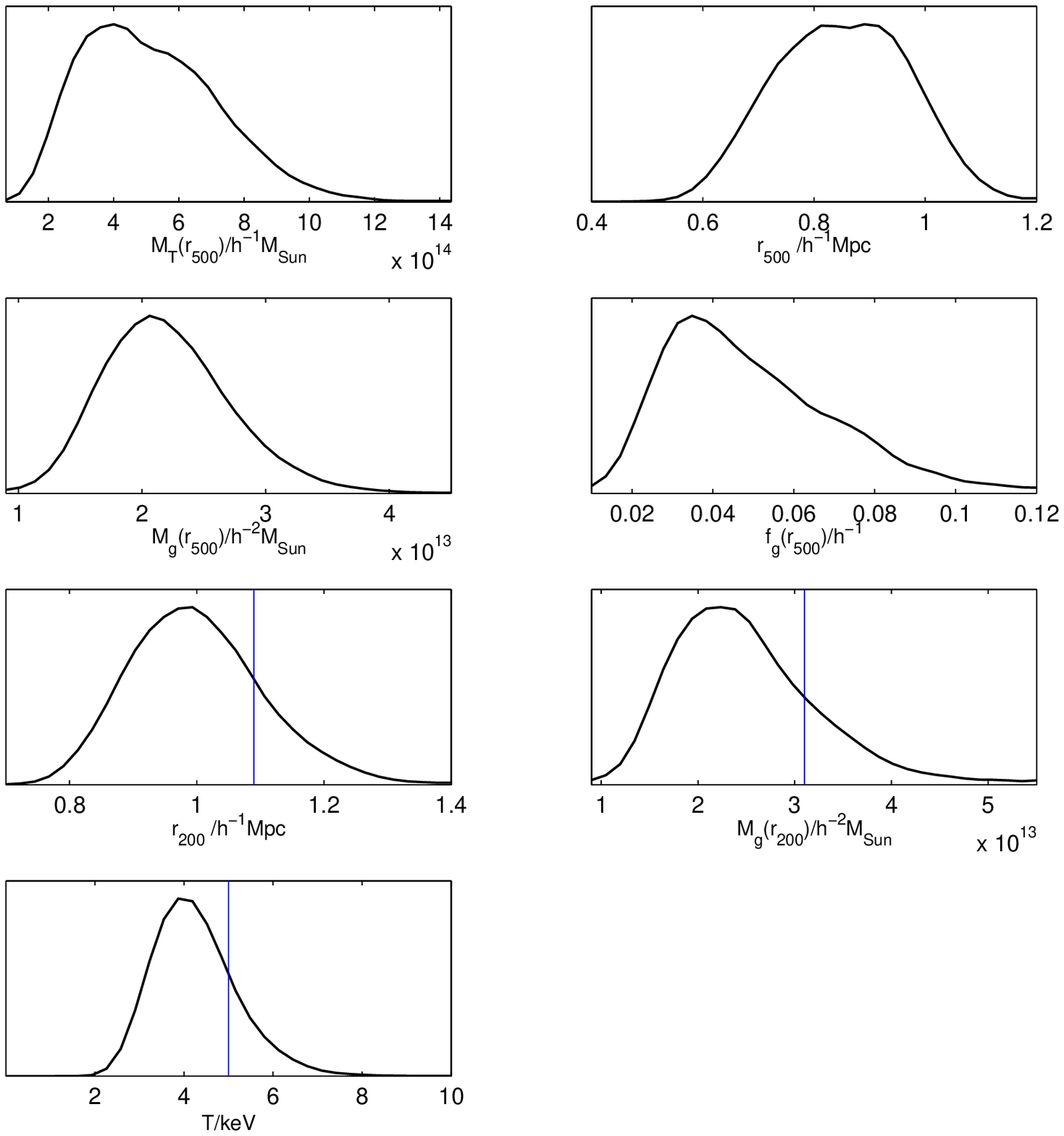}
  \caption{Marginal distributions of the derived cluster physical parameters with 
simulated data for isothermal $\beta$-model- parameterisation III.}
\end{figure}
The results of the analysis with no data are plotted in Figs. 11 and 12.  It is evident 
that, while the assumed prior probability distributions for the cluster
position, total mass and gas mass fraction are recovered,  the two sampling
parameters $r_{\rm c}$ and $\beta$ show the same behaviours as discussed for
the other two parameterisations. We also see a trend towards lower values in
the 1-D posterior probability distribution of temperature. However this
behaviour is due to the direct relationship between the total mass and the temperature
in this parameterisation and the specific prior distribution we have assumed
for the total mass which clearly has a higher probability at the lower masses.

Figs. 13 and 14 represent the marginalised posterior distributions from the analysis of 
simulated SZ cluster data for sampling and derived parameters respectively while in 
Table 10 we present the mean, the dispersion and the maximum likelihood of each 
cluster parameter estimated using parameterisation III.
\begin{table}
\caption{Simulated cluster parameters estimated (mean, standard deviation and Maximum likelihood) using isothermal $\beta$-model--parameterisation III assuming $h=0.7$.}
\begin{tabular}{@{}ccc@{} }
\hline
Parameter & $\mu \pm \sigma$ & $\hat \mu$ \\\hline
$x_{\rm c}$ & $-3.0 \pm 15.6\arcsec$ & $-5.0\arcsec$  \\
$y_{\rm c}$ & $6.4 \pm 14.4\arcsec$ & $5.7\arcsec$  \\
$r_{\rm c}$ & $395.11 \pm 226.21\,\rm{kpc}$  & $142.14\,\rm{kpc}$  \\
$\beta$     & $1.7 \pm 0.5$ &$1.1$             \\
$M_{\rm T}(r_{\rm 200})$&$(4.68 \pm 1.56)\times 10^{14}\,\rm{M_\odot}$ &$4.46\times 10^
{14}\,\rm{M_\odot}$\\
$f_{\rm g}(r_{\rm 200})$ &$0.11\pm 0.03 $ & $0.1$ \\
$M_{\rm T}(r_{\rm 500})$&$(7.35\pm 3.0)\times 10^{14}\,\rm{M_\odot}$ &$4.58\times 10^
{14}\,\rm{M_\odot}$\\
$r_{\rm 500}$  & $1.21 \pm 0.17\,\rm{Mpc}$ & $1.06\,\rm{Mpc}$   \\
$M_{\rm g}(r_{\rm 500})$ &$(4.50 \pm 1.04) \times 10^{13}\,\rm{M_\odot}$ &$4.00\times 
10^{13}\,\rm{M_\odot}$\\
$f_{\rm g}(r_{\rm 500})$ &$0.07 \pm 0.03$ & $0.09$ \\
$r_{\rm 200}$  & $1.43 \pm 1.50\,\rm{Mpc}$ & $1.42\,\rm{Mpc}$   \\
$M_{\rm g}(r_{\rm 200})$&$(5.14\pm 1.6)\times 10^{13}\,\rm{M_\odot}$ &$4.49\times 10^
{13}\,\rm{M_\odot}$\\
$T_{\rm g}$ & $4.3 \pm 0.9\,\rm{keV}$ &$4.2\,\rm{keV}$             \\
\hline
\end{tabular}
\end{table}
The strong degeneracy between $ r_{\rm c}$ and $\beta$ is quite apparent in this 
parameterisation, while $\beta$ is poorly constrained and biased towards higher values. 
We note that, since the SZ analysis constrains cluster total mass internal to the radius
$r_{\rm 200}$ and we use the virial M-T relation (equation (33)) to
derive cluster average temperature within this radius, the result of
temperature estimation is less biased and more reliable than the
parameterisations I and II in recovering the temperature true value. We have used this 
parameterisation in our follow-up analysis of the real data where we studied a joint 
weak gravitational lensing and SZ analysis of six clusters (AMI Consortium: Hurley- 
Walker et~al. 2011) and high and moderate X-ray luminosity sample of LoCuSS clusters 
(AMI Consortium: Rodr\'{i}guez-Gonz\'{a}lvez et~al.\ 2011; AMI Consortium: Shimwell 
et~al. 2011).
\subsection{Analysis using  entropy-GNFW pressure model}
Similar to the isothermal $\beta$-model we first studied our methodology for the 
"entropy"-GNFW pressure model with no data. The results are represented in Figs. 15 
and 16. This analysis again helps us understand which parameters are constrained by SZ 
measurement as well as to check the algorithm in retrieving the prior probability 
distributions. From both 1-D and 2-D marginalised probability distributions it is clear 
that we are able to recover the input priors probability distributions and the 
probability distributions of the derived parameters are according to their
corresponding functional dependencies on the sampling parameters.

Figs. 17, 18 and Table 11 show the results of our analysis for "entropy" -GNFW 
pressure profile using parameterisation II while Figs. 19, 20 and Table 12 show the 
results of the same analysis using parameterisation III. We note that in both analyses 
$r_{\rm c}$ and $\alpha$  the parameters that define the shape of the entropy profile 
are not constrained while the scaling radius, $r_{\rm p}$, which 
defines the GNFW pressure profile is completely constrained. As a result we notice similar 
constraints in the estimation of $r_{500}$ in both parameterisations since we assume a 
fixed $c_{500}$. We also note the degeneracies between $M_{\rm T}(r_{\rm 200})$-$r_{\rm 
c}$ and $M_{\rm T}(r_{\rm 200})$-$\alpha$ which are because of the dependency of $P_
{\rm {ei}}$ on these two free parameters. On the other hand the $M_{\rm T}(r_{\rm 200})$-
$r_{\rm p}$ degeneracy seen in Figs. 17 and 19 is due to the intrinsic degeneracy 
that exists between the cluster size and the volume integrated Comptonisation parameter 
($Y_{\rm SZ}$-$r_{\rm p}$ degeneracy) in the SZ measurements (Planck~ Collaboration 
2011d). Moreover, comparing $T_{\rm g}(r_{\rm 500})$ and $T_{\rm g}(r_{\rm 200})$ 
(Table 11 and 12) confirms a radial decline in the ICM temperature distribution as 
expected.

Overall, both parameterisations could  constrain the cluster physical parameters, 
however, analysis using parameterisation III leads to a tighter constrain on both  $T_
{\rm g}(r_{\rm 500})$ and $T_{\rm g}(r_{\rm 200})$. The results of Parameterisation III 
once more show that this parameterisation can reliably be used in the analysis of 
clusters of galaxies as it is less model dependent and produces unbiased results in 
particular when the assumption of hydrostatic equilibrium breaks, in young or disturbed clusters  
 (parameterisation II).
\begin{figure}
\includegraphics[width=80mm]{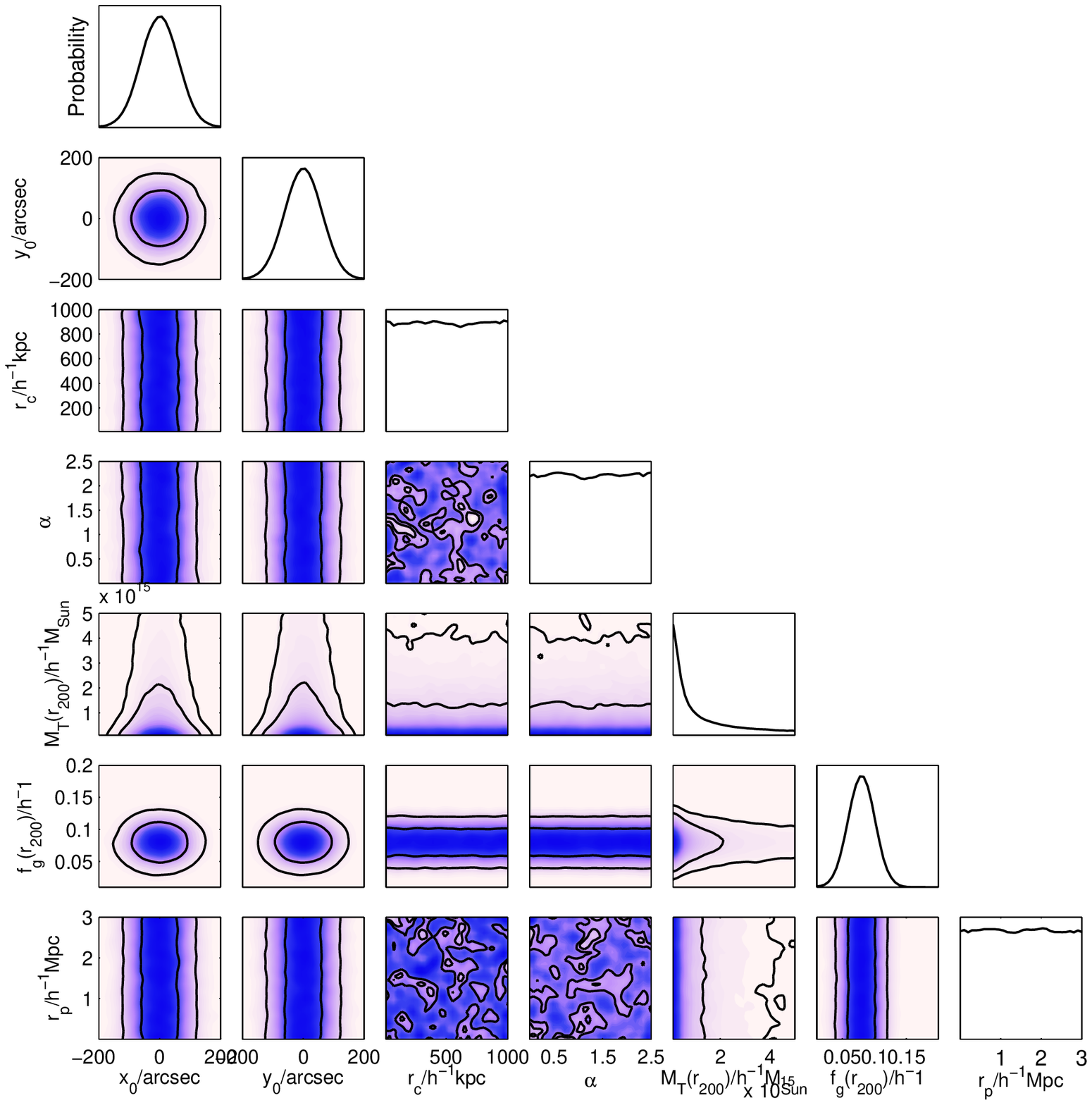}
\caption{Marginal distributions of the sampling parameters with no data for "entropy"-
GNFW pressure model.}
\medskip
\includegraphics[width=80mm]{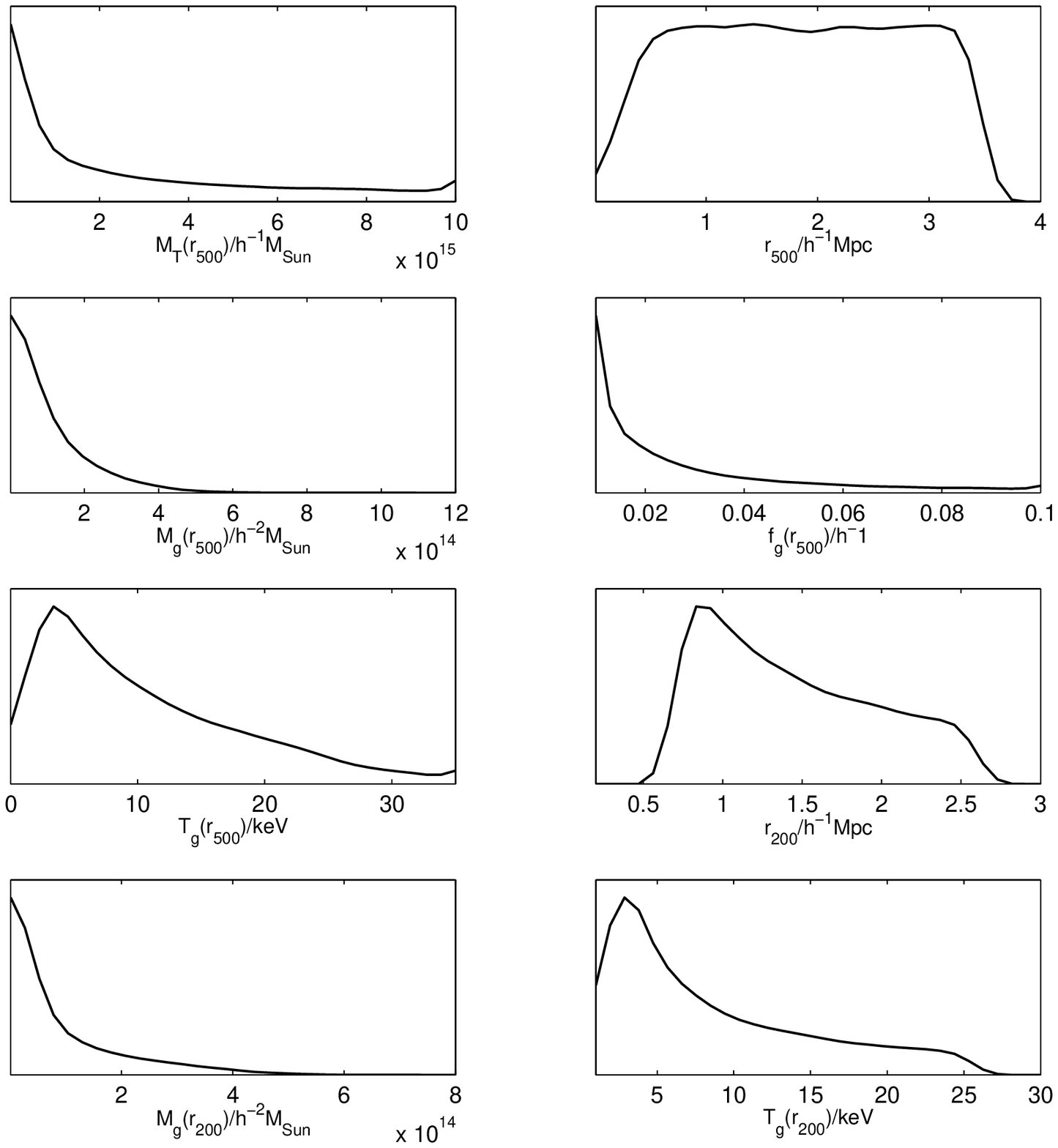}
\caption{Marginal distributions of the derived cluster physical parameters with no 
data for "entropy"-GNFW pressure model.}
\end{figure}
\begin{figure}
  \includegraphics[width=80mm]{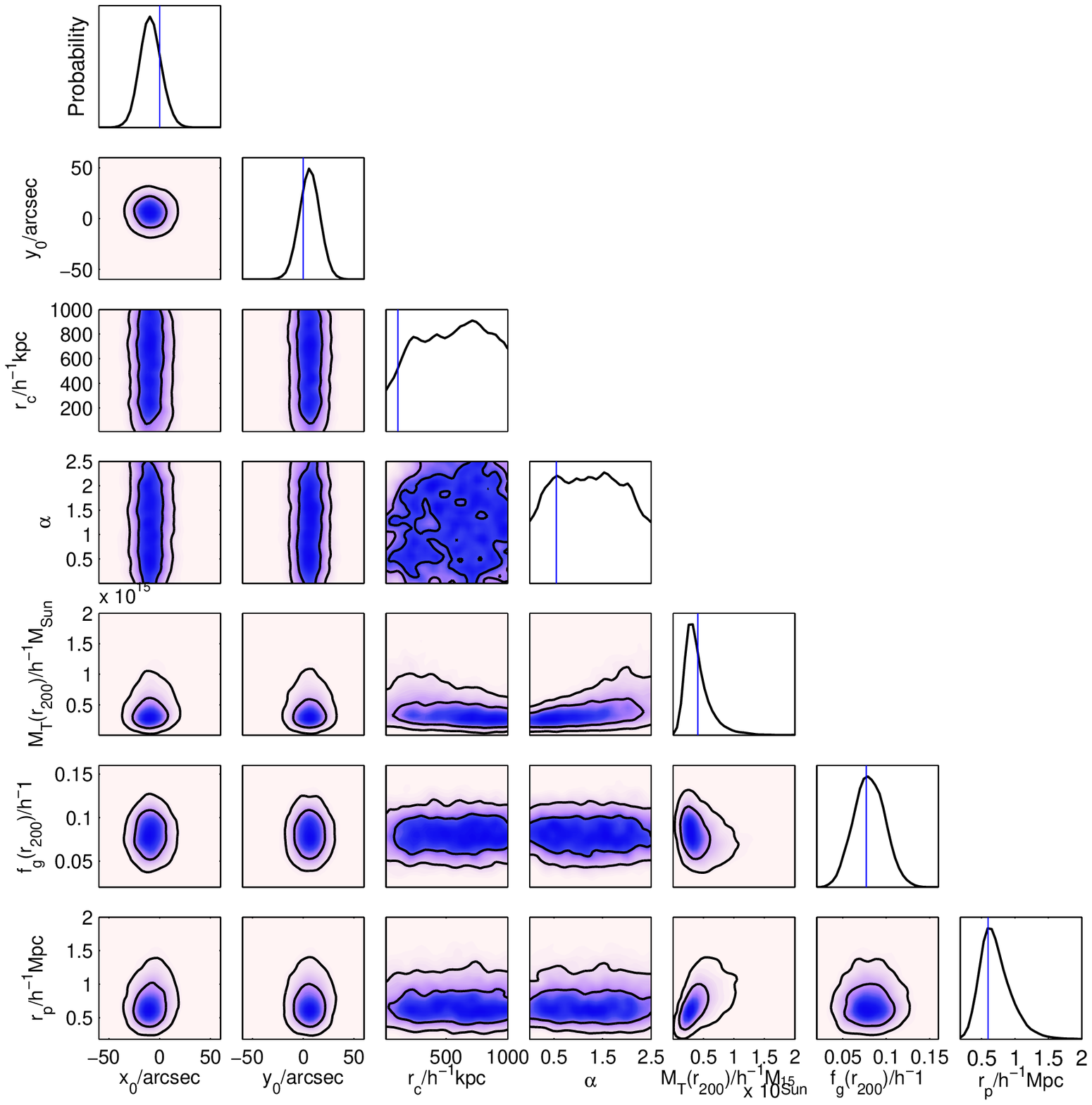}
  \caption{Marginal distributions of the sampling parameters with simulated data for 
"entropy"-GNFW pressure model using parameterisation II.}
  \medskip
  \includegraphics[width=80mm]{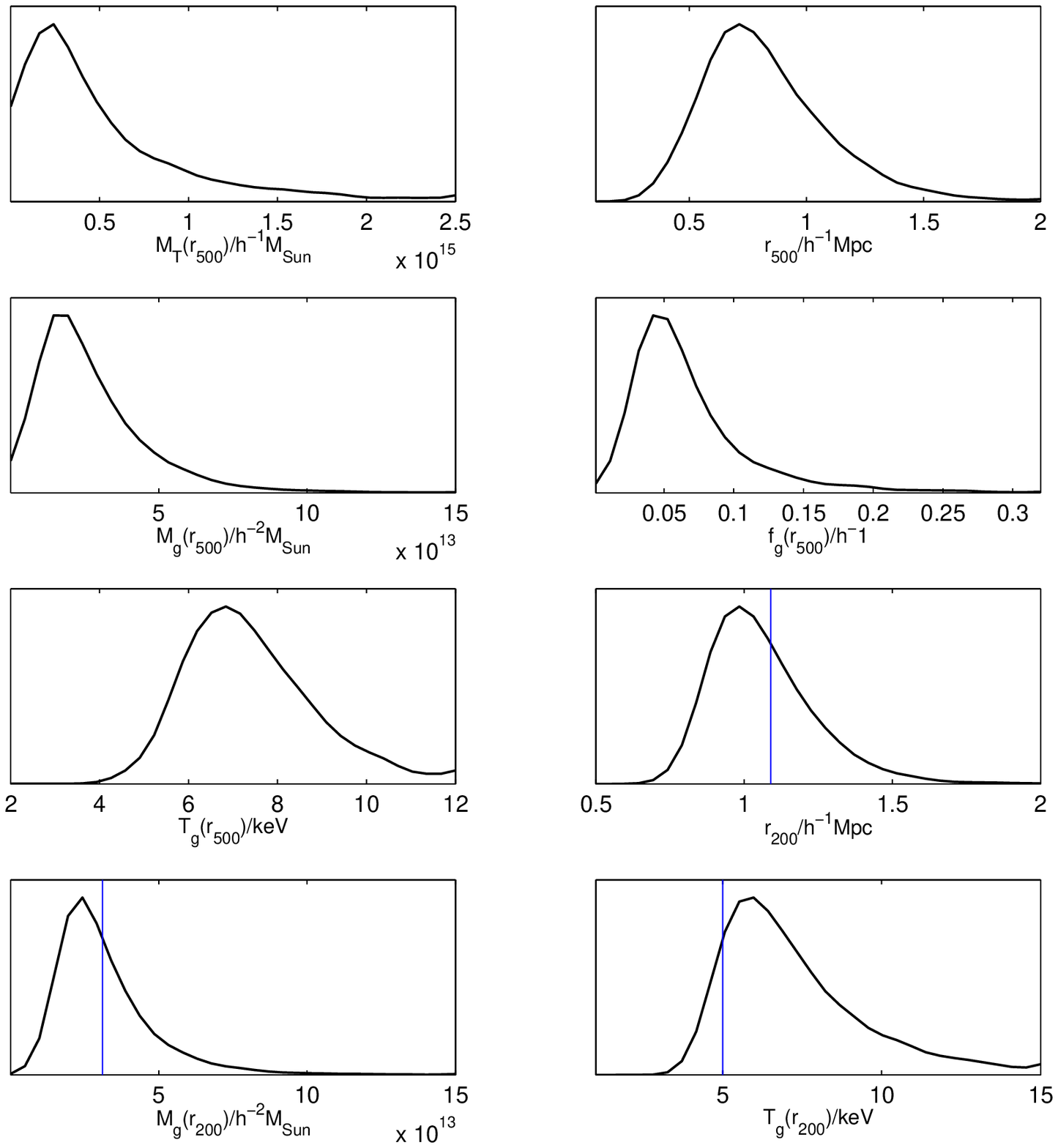}
  \caption{Marginal distributions of the derived cluster physical parameters with 
simulated data for "entropy"-GNFW pressure model using parameterisation II.}
\end{figure}
\begin{table}
\caption{Simulated cluster parameters estimated (mean, standard deviation and Maximum likelihood) using ``entropy''-GNFW 
pressure model-- parameterisation II assuming $h=0.7$.}
\begin{tabular}{@{}ccc@{} }
\hline
Parameter & $\mu \pm \sigma$ & $\hat \mu$ \\\hline
$x_{\rm c}$ & $-8.9 \pm 10.04\arcsec$ & $-9.3\arcsec$  \\
$y_{\rm c}$ & $6.3 \pm 9.5\arcsec$ & $7.8\arcsec$  \\
$r_{\rm c}$ & $ 770.43\pm 382.29\,\rm{kpc}$  & $1358.86\,\rm{kpc}$  \\
$\alpha$    & $1.2 \pm 0.67$ &$1.2$             \\
$M_{\rm T}(r_{\rm 200})$&$(5.86\times 10^{14} \pm 3.43\times 10^{14})\,
\rm{M_\odot}$ &$3.3\times 10^{14}\,\rm{M_\odot}$\\
$f_{\rm g}(r_{\rm 200})$ &$0.11\pm 0.02 $ & $0.13$ \\
$r_{\rm p}$ & $1.03 \pm 0.33\,\rm{Mpc}$  & $0.87\,\rm{Mpc}$  \\
$M_{\rm T}(r_{\rm 500})$&$(8.86\pm 11.57)\times 10^{14}\,\rm{M_\odot}$ 
&$4.0\times 10^{14}\,\rm{M_\odot}$\\
$r_{\rm 500}$  & $1.18 \pm 0.39\,\rm{Mpc}$ & $1.01\,\rm{Mpc}$   \\
$M_{\rm g}(r_{\rm 500})$ &$(5.71 \pm 4.08 )\times 10^{13}\,\rm{M_
\odot}$ &$3.06\times 10^{13}\,\rm{M_\odot}$\\
$f_{\rm g}(r_{\rm 500})$ &$0.1 \pm 0.09$ & $0.07$ \\
$T_{\rm g}(r_{\rm 500})$ & $7.5 \pm 1.8\,\rm{keV}$ &$6.5\,\rm{keV}$\\
$r_{\rm 200}$  & $1.57 \pm 0.28\,\rm{Mpc}$ & $1.3\,\rm{Mpc}$   \\
$M_{\rm g}(r_{\rm 200})$&$(6.53\pm 3.67)\times 10^{13}\,\rm{M_\odot}
$ &$4.14\times 10^{13}\,\rm{M_\odot}$\\
$T_{\rm g}(r_{\rm 200})$ & $7.4 \pm 2.6\,\rm{keV}$ &$5.02\,\rm{keV}$\\
\hline
\end{tabular}
\end{table}
\begin{figure}
  \includegraphics[width=80mm]{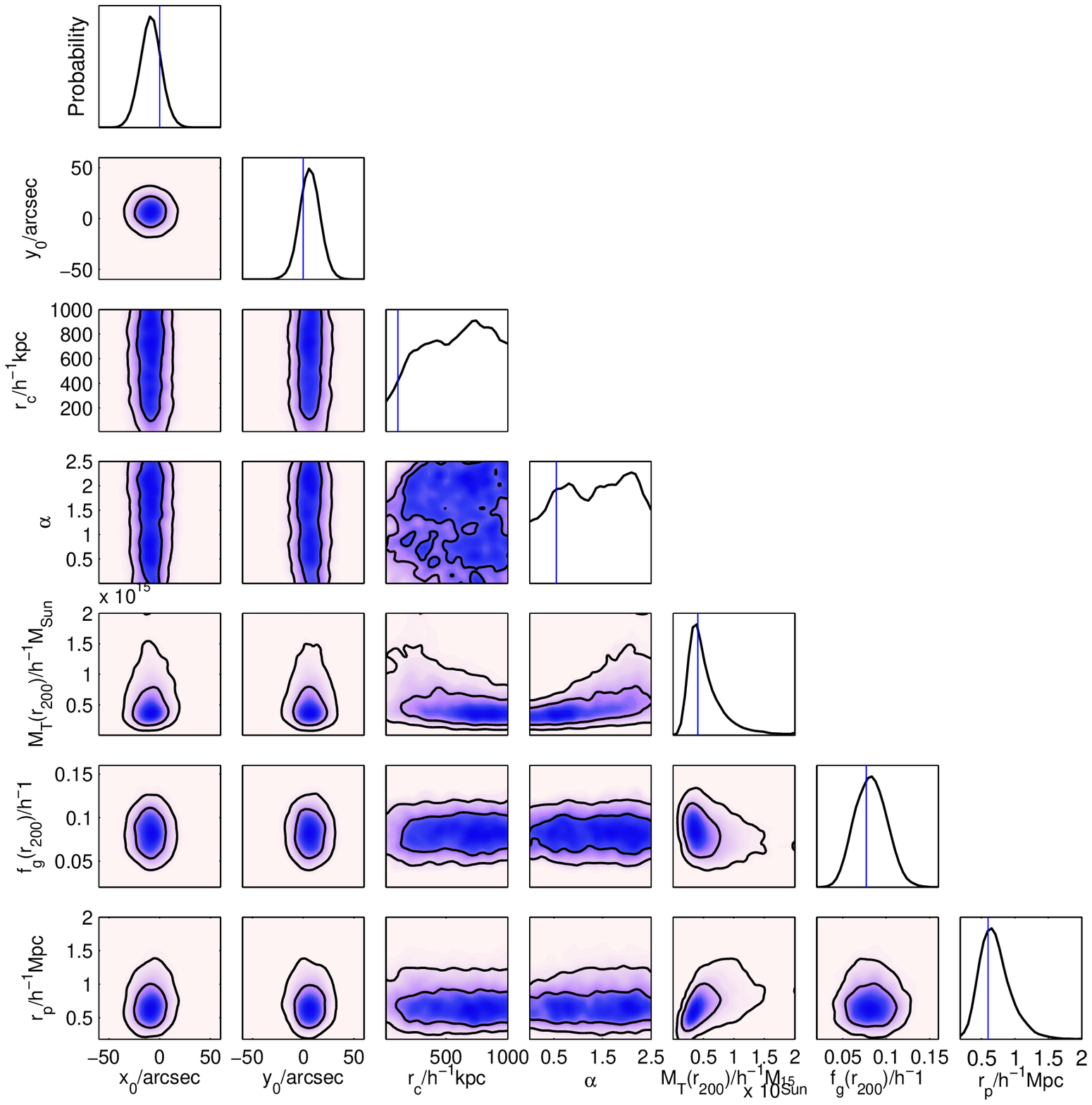}
  \caption{Marginal distributions of the sampling parameters with simulated data for 
"entropy"-GNFW pressure model using parameterisation III.}
  \medskip
  \includegraphics[width=80mm]{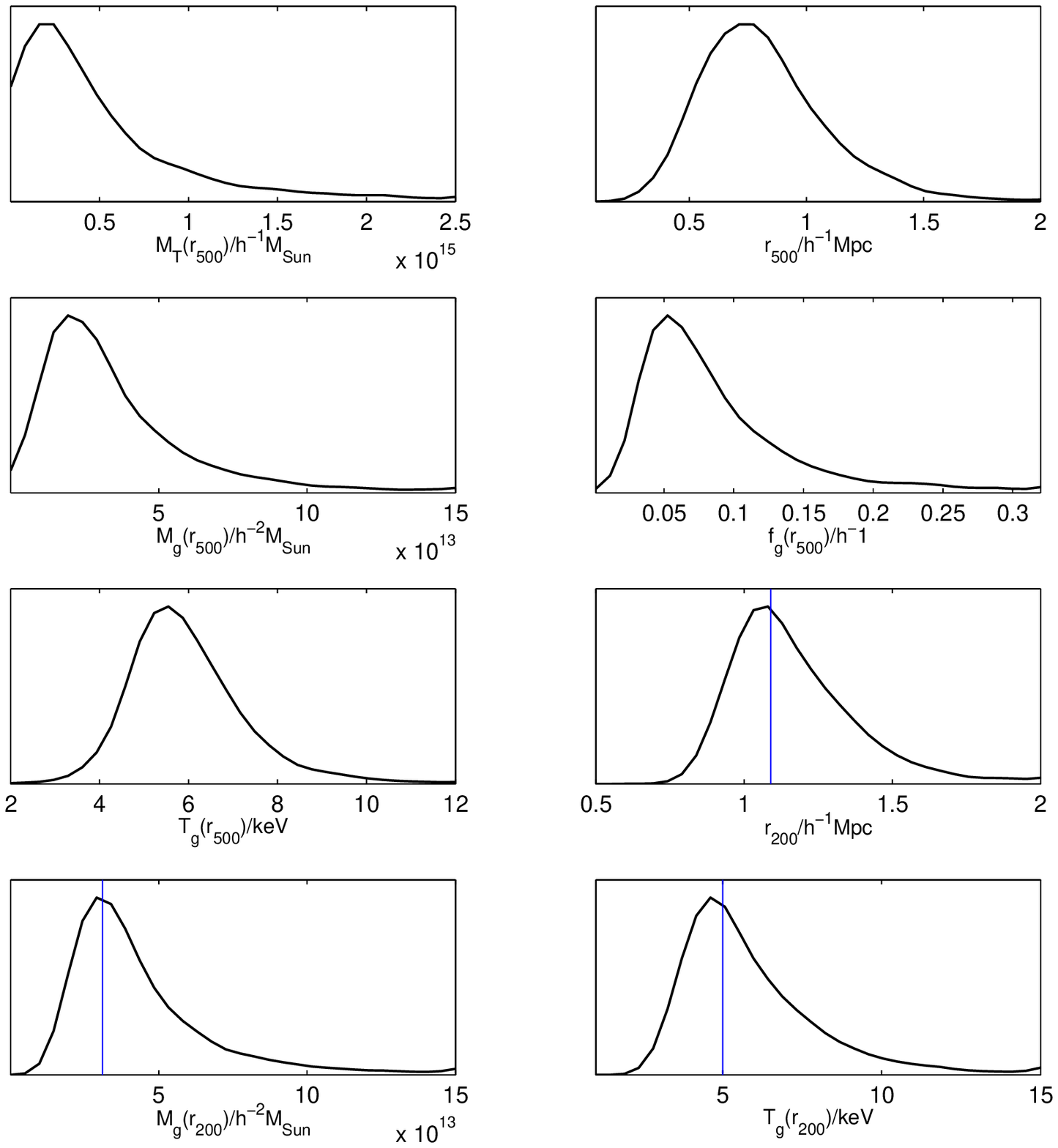}
  \caption{Marginal distributions of the derived cluster physical parameters with 
simulated data for "entropy"-GNFW pressure model using parameterisation III.}
\end{figure}
\begin{table}
\caption{Simulated cluster parameters estimated (mean, standard deviation and Maximum likelihood) using entropy-GNFW pressure model--parameterisation III assuming $h=0.7$.}
\begin{tabular}{@{}ccc@{} }
\hline
Parameter & $\mu \pm \sigma$ & $\hat \mu$ \\\hline
$x_{\rm c}$ & $-8.8 \pm 9.8\arcsec$ & $-10.4\arcsec$  \\
$y_{\rm c}$ & $6.6 \pm 9.5\arcsec$ & $7.2\arcsec$  \\
$r_{\rm c}$ & $798.7 \pm 376.8\,\rm{kpc}$  & $1280.8\,\rm{kpc}$  \\
$\alpha$     & $1.3 \pm 0.69$ &$1.3$             \\
$M_{\rm T}(r_{\rm 200})$&$(8.0\times 10^{14} \pm 5.6\times 10^{14})\,\rm{M_\odot}$ &$4.5
\times 10^{14}\,\rm{M_\odot}$\\
$f_{\rm g}(r_{\rm 200})$ &$0.11\pm 0.03 $ & $0.13$ \\
$r_{\rm p}$ & $1.0 \pm 0.3\,\rm{Mpc}$  & $0.86\,\rm{Mpc}$  \\
$M_{\rm T}(r_{\rm 500})$&$(6.6\pm 1.14)\times 10^{14}\,\rm{M_\odot}$ &$3.4\times 10^{14}
\,\rm{M_\odot}$\\
$r_{\rm 500}$  & $1.14 \pm 0.43\,\rm{Mpc}$ & $1.0\,\rm{Mpc}$   \\
$M_{\rm g}(r_{\rm 500})$ &$(7.35 \pm 5.92 )\times 10^{13}\,\rm{M_\odot}$ &$3.67\times 
10^{13}\,\rm{M_\odot}$\\
$f_{\rm g}(r_{\rm 500})$ &$0.14 \pm 0.14$ & $0.11$ \\
$T_{\rm g}(r_{\rm 500})$ & $6.0 \pm 1.4\,\rm{keV}$ &$5.5\,\rm{keV}$             \\
$r_{\rm 200}$  & $1.7 \pm 0.28\,\rm{Mpc}$ & $1.4\,\rm{Mpc}$   \\
$M_{\rm g}(r_{\rm 200})$&$(9.18\pm 5.71)\times 10^{13}\,\rm{M_\odot}$ &$5.91\times 10^
{13}\,\rm{M_\odot}$\\
$T_{\rm g}(r_{\rm 200})$ & $5.98 \pm 2.43\,\rm{keV}$ &$4.2\,\rm{keV}$             \\
\hline
\end{tabular}
\end{table}
%
\section{Discussion and Conclusions}
We have studied two parameterised models, the traditional isothermal $\beta$-model and 
the ``entropy''-GNFW pressure model, to analyse the SZ effect from galaxy
clusters and extract their physical parameters using AMI SA simulated data. In our 
analysis we have described the current assumptions made on the dynamical
state of the ICM including spherical geometry, hydrostatic equilibrium and the virial 
mass-temperature  relation. In particular we have shown how
different parameterisations which relate the thermodynamical quantities describing the 
ICM to the cluster global properties via these assumptions lead to biases on the cluster 
physical parameters within a particular model. 

In this context, we first generated a simulated cluster using the isothermal
$\beta$-model observed with the AMI SA and used these simulated data to study three 
different parameterisations in deriving the cluster physical parameters.
We showed that in generating AMI simulated data, it is extremely important to select 
the model parameters describing the SZ signal in a way that leads to the consistent 
cluster parameter inferences upon using the three different parameterisation methods.

We found that each parameterisation introduces different constraints and biases
in the posterior probability distribution of the inferred cluster parameters which
arise from the way we implement assumptions about the cluster structure and its
composition. The biases in the posterior probability distributions of the
cluster parameters are more pronounced in parameterisations I and II, as the
results depend strongly on the relatively unconstrained cluster model shape
parameters: $r_{\rm c}$ and $\beta$. However, the biases introduced by the
choice of priors are even worse in parameterisation I, in which the gas temperature is 
assumed to be an independent free parameter. This, along with the assumption of 
isothermality, causes the priors to  dominate in extracting the cluster physical 
parameters regardless the type of prior chosen for the gas temperature (AMI Consortium: 
Rodr\'{i}guez-Gonz\'{a}lvez et~al.\ 2011 and AMI Consortium: Zwart et~al.\ 2010). The 
cluster physical parameters estimated using parameterisation I depend strongly on the
model parameters. Although it can constrain the cluster position and its $M_{\rm g}(r_
{\rm 200})$, it fails to recover the true input values of most of the simulated cluster 
properties. For example the inferred values for mass and temperature at $r_{\rm200}$ are 
$M_{\rm T}(r_{\rm 200})=(6.43\pm 5.43)\times 10^{15}\,\rm{M_\odot}$ and $T_{\rm g}(r_{\rm 
200})= (10.61 \pm 5.28)\,\rm{keV}$ whereas the corresponding input values of simulated 
cluster are: $M_{\rm T}(r_{\rm 200})=5.83\times 10^{14}\,\rm{M_\odot} $ and $T_{\rm g}(r_
{\rm 200})=5\, \rm{keV} $. In terms of the application to the real data, we have noticed 
similar biases in the results of our analysis of 7 clusters using this parameterisation 
(AMI Consortium: Zwart et~al. 2010). In order to improve our analysis methodology in 
parameterisations II and III, the correlation between the cluster total mass and its 
gas temperature is taken into account. In parameterisation II we relate $M_{\rm T}(r_
{\rm 200})$ and $T_{\rm g}(r_{\rm 200})$  using the hydrostatic equilibrium whereas in
parameterisation III we use virial mass-temperature relationship. It should be
noted that the derived $T_{\rm g}(r_{\rm 200})$ in parameterisation II is the
gas temperature at the overdensity radius $r_{\rm 200}$ which is then assumed
to be constant throughout the cluster. In parameterisation III, however,
$T_{\rm g}(r_{\rm 200})$ is the mean gas temperature internal to radius $r_{\rm
200}$ and is assumed to be constant.
We notice that analysing the same simulated data set using parameterisation II can 
constrain the 1-D posterior distribution of the cluster physical parameters better than 
parameterisation I such that $M_{\rm T}(r_{\rm 200})= (6.8 \pm 2.1)\times 10^{14}\,\rm{M_
\odot}$ and $T_{\rm g}(r_{\rm 200})= (3.0 \pm 1.2)\,\rm{keV}$. Since parameterisation II 
uses the full parametric hydrostatic equilibrium, the temperature estimate depends on 
$r_{\rm c}$ and $\beta$  and is therefore biased low. These results were also confirmed 
in our analysis of the bullet like cluster A2146 (AMI Consortium: Rodr\'{i}guez-Gonz
\'{a}lvezet~al.\ 2011). Relating the cluster total mass and its temperature via virial 
theorem in parameterisation III leads to less bias in cluster physical parameters 
compared to the other two parameterisations as it is less model dependent: $M_{\rm T}(r_
{\rm 200})= (4.68 \pm 1.56)\times 10^{14}\,\rm{M_\odot}$ and $T_{\rm g}(r_{\rm 200})= 
(4.3 \pm 0.9)\,\rm{keV}$.

A detailed comparison between our different parameterisations both using simulated data 
and on the bullet like cluster A2146 (AMI Consortium: Rodr\'{i}guez-Gonz\'{a}lvez et~al.
\ 2011) found that parameterisation III can give more reliable results for cluster 
physical properties as it is less dependent on model parameters. Parameterisation
II also gives convincing estimates for the cluster total mass and its gas
content although its temperature estimate is poorly justified, as it depends
strongly on the model parameters. Moreover, young or disturbed clusters are unlikely to 
be well-described by hydrostatic equilibrium. We therefore used parameterisation III as 
our adopted analysis methodology in our follow-up studies of the real clusters including 
the joint SZ and weak lensing analysis of six clusters (AMI Consortium: Hurley-Walker 
et~al.\ 2011) and the analysis of LoCuss cluster sample (AMI Consortium: Rodr\'{i}guez-Gonz
\'{a}lvez et~al.\ 2011; AMI Consortium: Shimwell et~al.\ 2011).

In order to make sure that our results are not biased by one realisation of
primordial CMB, we have studied $\rm {100}$ CMB realisations for the three 
parameterisations. The 1-D marginalised posterior probability distributions of $M_{\rm 
T}(r_{\rm 200})$ and $T_{\rm g}(r_{\rm 200})$ are shown in Figs. 21, 22 and 23 for
each parameterisation. The solid blue line represents the true value
corresponding to the simulated cluster and the dashed red line shows the mean
value of the distributions. Table 13 also presents the numerical results of this 
analysis.
\begin{table}
\caption{The results of $100$ CMB realisations for the three parameterisations assuming h=0.7.}
\begin{tabular}{@{}ccc@{} }
\hline
parameterisation & $M_{\rm T}(r_{\rm 200})\,\rm{M_\odot}$ & $T_{\rm g}(r_{\rm 200})\,\rm
{keV}$ \\\hline
I & $(6.18 \pm 5.23)\times 10^{15}$ & $11.18 \pm 5.16$  \\
II & $(8.067 \pm 2.61)\times 10^{14}$ & $3.94 \pm 1.67$  \\
III & $(5.94 \pm 2.26)\times 10^{14}$  & $4.97 \pm 1.21$  \\
\hline
\end{tabular}
\end{table}
Comparing the 1D posterior distributions along with the mean values of the $M_{\rm T}(r_
{\rm 200})$ estimates in the three parameterisations for these $100$ realisations show 
that parameterisation I can hardly constrain the simulated cluster properties and 
recover the input true values. Parameterisation II  can constrain
the cluster total mass, however, the gas temperature estimate is  biased low as it 
depends on unconstrained model shape parameters. On the other hand, parameterisation 
III can indeed constrain both cluster mass and its gas temperature and the results are 
unbiased.
\begin{figure}
  \includegraphics[width=80mm]{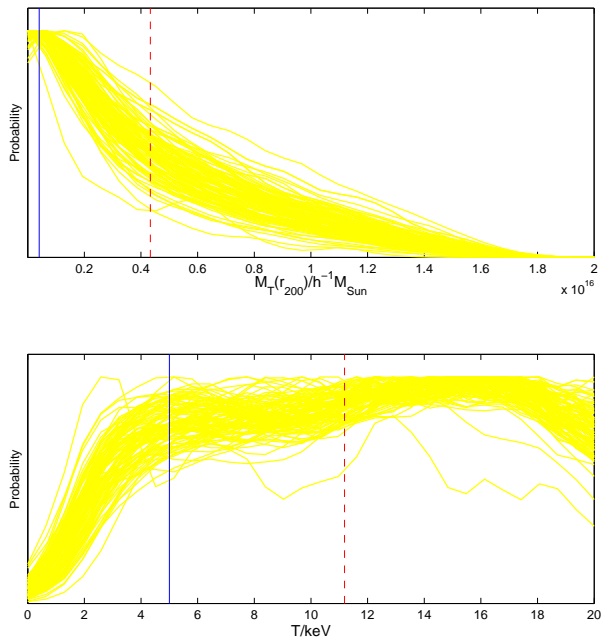}
  \caption{$100$ realisations of 1-D marginalised posterior probability distributions 
of $M_{\rm T}(r_{\rm 200})$ and $T_{\rm g}(r_{\rm 200})$ using isothermal $/beta$-model--parameterisation I. The solid blue line represents the true value corresponding to the simulated cluster and the dashed red line shows the mean value of the distributions.}
 \end{figure}
\begin{figure}
  \includegraphics[width=80mm]{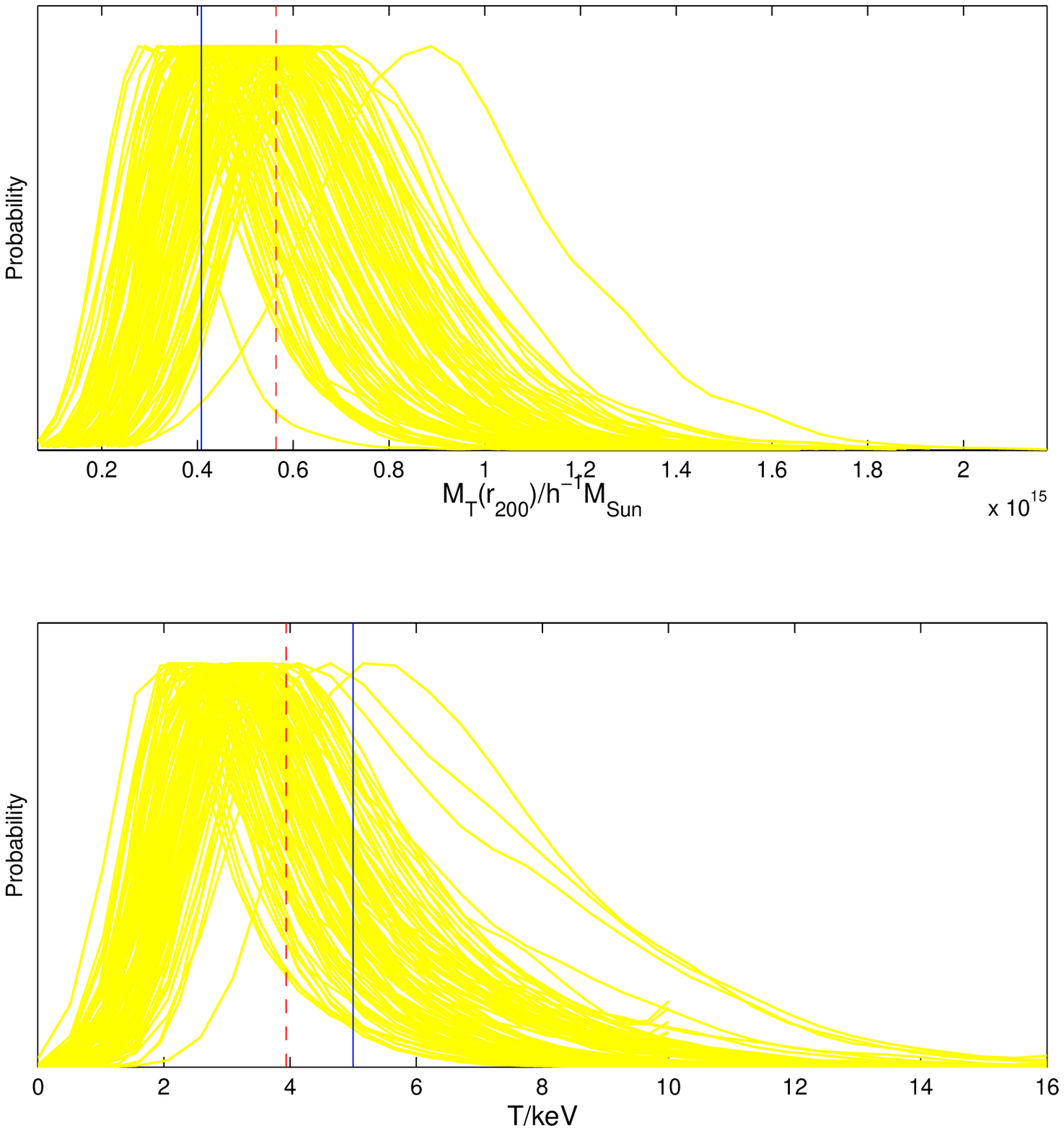}
  \caption{$100$ realisations of 1-D marginalised posterior probability distributions 
of $M_{\rm T}(r_{\rm 200})$ and $T_{\rm g}(r_{\rm 200})$ using isothermal $\beta$-model--
parameterisation II. The solid blue line represents the true value corresponding to the 
simulated cluster and the dashed red line shows the mean value of the distributions.}
\medskip
  \includegraphics[width=80mm]{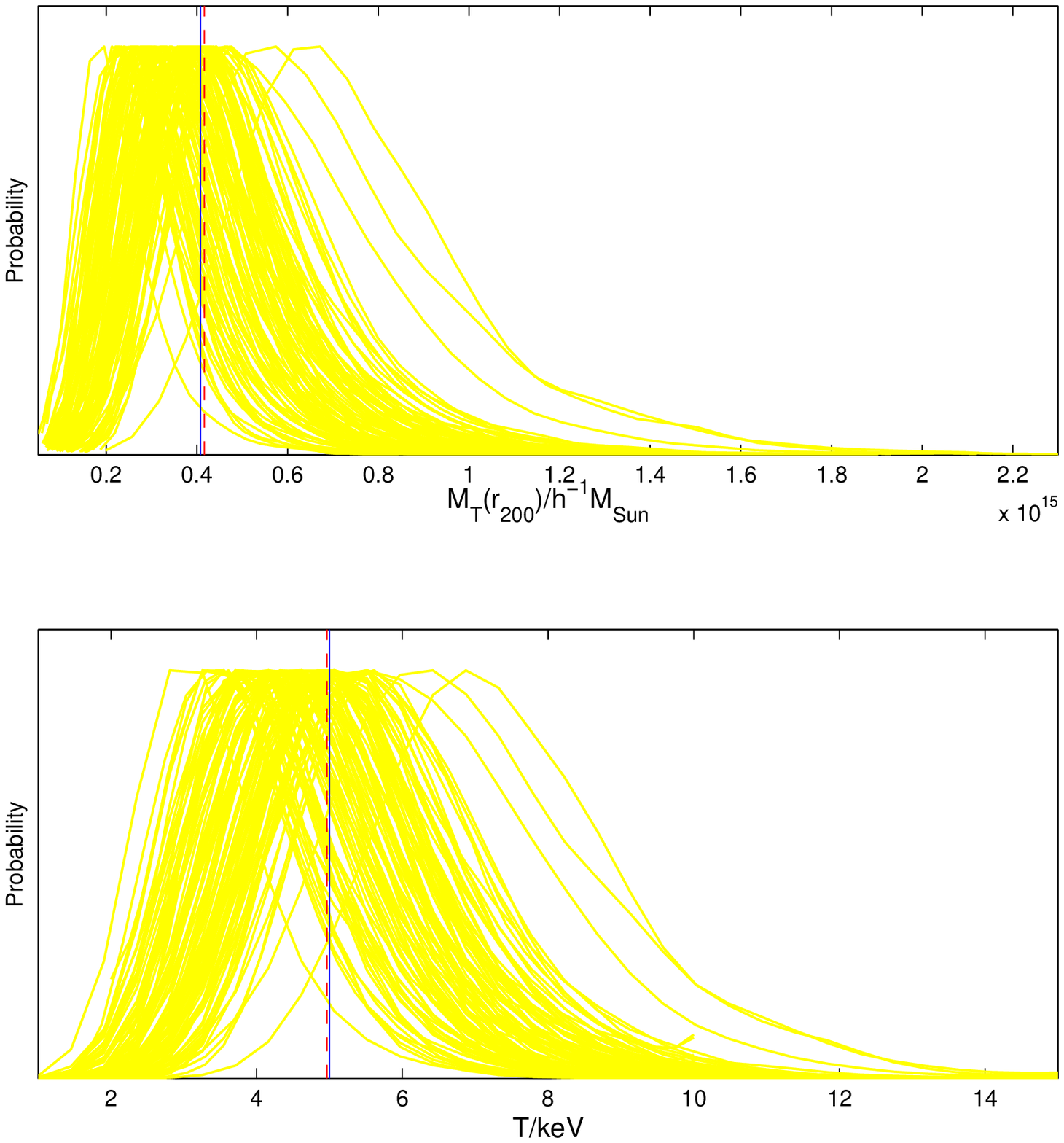}
  \caption{$100$ realisations of 1-D marginalised posterior probability distributions 
of $M_{\rm T}(r_{\rm 200})$ and $T_{\rm g}(r_{\rm 200})$ using isothermal $\beta$-model--
parameterisation III. The solid blue line represents the true value corresponding to the 
simulated cluster and the dashed red line shows the mean value of the distributions.}
\end{figure}

In order to remove the assumption of isothermality which is of course a poor assumption 
both within the cluster inner region and at the large radii and to improve
our analysis model for the cluster ICM which can be fitted accurately throughout the 
cluster, we also studied the SZ effect using ``entropy''-GNFW pressure model. This model 
assumes a 3-D $\beta$-model like radial profile describing the entropy in the ICM as well 
as the GNFW profile for the plasma pressure. This choice is reasonable as the 
entropy is a conserved quantity and describes the structure of the ICM while the pressure 
is related to the dark matter component of the cluster. Moreover, among all the 
thermodynamical quantities describing the ICM, entropy and pressure show more self-
similar distribution in the outskirts of the cluster. The combination of these two 
profiles then allows us to relate the SZ observable properties to the cluster physical 
parameters such as its total mass. This model also allows the electron pressure and its 
number density profiles to have different distributions leading to a 3-D radial 
temperature profile. In this context we simulated a second cluster using an entropy-GNFW 
pressure profile with the same physical parameters and thermal noise as the first cluster 
at $r_{\rm 200}$.

We then analysed the second simulated cluster using "entropy"-GNFW pressure model with 
different parameterisations. In this model temperature is no longer isothermal so that 
we can not use parameterisation I where a single temperature is assumed  as 
an independent input parameter. The results of our analysis using parameterisation II and 
III show that while the characteristic scaling radius describing the GNFW pressure 
profile is constrained, the shape parameters defining the entropy profile remain 
unconstrained. Moreover, all the cluster physical parameters lie within $1 \sigma$ 
errorbars from the corresponding true values of the simulated cluster in the two 
parameterisations. However, parameterisation III provides tighter constrains in 1-D 
marginalised posterior distribution of the temperature and the overall results are less 
model dependent so that it can be reliably used in the analysis of galaxy clusters in 
particular when the assumption of hydrostatic equilibrium breaks (e.g. in disturbed 
clusters and clusters that are going through merging).
 
We conclude that using the ``entropy''-GNFW pressure model overcomes the 
limitations of the isothermal $\beta$-model in fitting cluster parameters over a 
broad radial extent. However, AMI simulated data do not strongly prefer one model over 
the other. We investigated this conclusion further by fitting both GNFW pressure 
profile and isothermal $\beta$-model to a simulated cluster with $\theta_{\rm 500}=2.5
\arcmin$ and $Y_{\rm 500}=2.5 \times 10^{-3}(\rm{arcmin})^{2}$. The result is shown in 
Fig. 24 with blue dashed line representing the fit using the isothermal $\beta$-model 
and the red representing the fit using GNFW pressure profile. However, we aim to 
compare these two models in our future studies using the real data.
\begin{figure}
 \includegraphics[width=80mm]{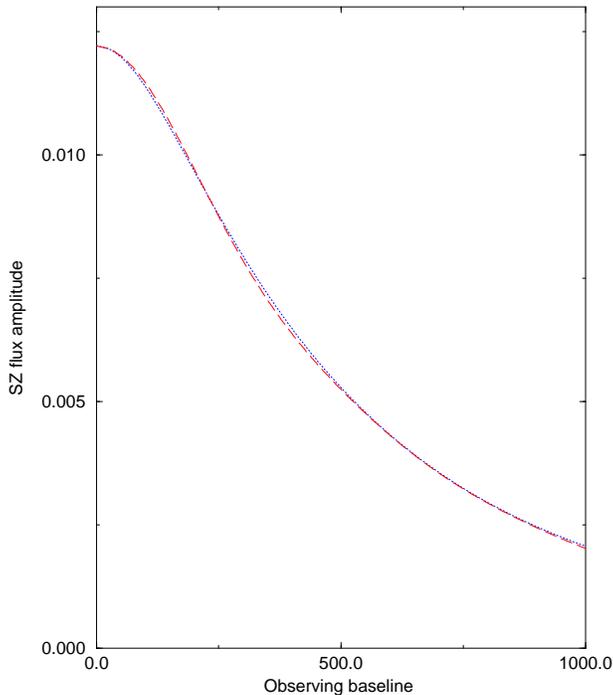}
  \caption{The SZ flux amplitude versus AMI SA observing baseline for a cluster with $\theta_{\rm 500}=2.5\arcmin$ and $Y_{\rm 500}=2.5 \times 10^{-3}(\rm{arcmin})^{2}$. Blue dashed line represents the fit using the isothermal $\beta$-model and the red represents the fit using GNFW pressure profile.}
\end{figure}
\section*{Acknowledgments}
The analysis work was conducted on the Darwin Supercomputer of the University
of Cambridge High Performance Computing Service supported by HEFCE. The authors
thank Stuart Rankin for computing assistance. We also thank Dave Green for his invaluable help with \LaTeX. MO, CRG, MLD, TMOF, MPS and TWS acknowledge PPARC/STFC studentships.
\setlength{\labelwidth}{0pt} 

\label{lastpage}


\begin{thebibliography}{99}
\bibitem[\protect\citeauthoryear{Afshordi
\& Cen}{2002}]{2002ApJ...564..669A} Afshordi N., Cen R., 2002, ApJ, 564, 669

\bibitem[\protect\citeauthoryear{Afshordi et
al.}{2007}]{2007MNRAS.378..293A} Afshordi N., Lin Y.-T., Nagai D.,
Sanderson A.~J.~R., 2007, MNRAS, 378, 293

\bibitem[\protect\citeauthoryear{Allison et
al.}{2011}]{2011MNRAS.410..341A} Allison J.~R., Taylor A.~C., Jones M.~E.,
Rawlings S., Kay S.~T., 2011, MNRAS, 410, 341

\bibitem[\protect\citeauthoryear{AMI Consortium: Hurley-Walker et
al.}{2011}]{2011arXiv1101.5912H} AMI Consortium: Hurley-Walker N. et al., 2011, arXiv,:1101.5912

\bibitem[\protect\citeauthoryear{AMI Consortium: Rodr\'{i}guez-Gonz\'{a}lvez et
al.}{2011}]{2011MNRAS.tmp..661A} AMI Consortium: Rodr\'{i}guez-Gonz\'{a}lvez C. et al., 2011, MNRAS, 661

\bibitem[\protect\citeauthoryear{AMI Consortium: Rodr\'{i}guez-Gonz\'{a}lvez et
al.}{2011}]{2011arXiv1101.5589R} AMI Consortium: Rodr\'{i}guez-Gonz\'{a}lvez C. et al., 2011,
arXiv:1101.5589

\bibitem[\protect\citeauthoryear{AMI Consortium: Shimwell et
al.}{2010}]{2010arXiv1012.4441S}AMI Consortium: Shimwell T. et al., 2010,
arXiv:1012.4441

\bibitem[\protect\citeauthoryear{AMI Consortium: Shimwell et
al.}{2011}]{2011arXiv1101.5590S}AMI Consortium:  Shimwell T. et al., 2011,
arXiv:1101.5590
\bibitem[\protect\citeauthoryear{AMI Consortium: Zwart et al.}{2008}]{2008MNRAS.391.1545Z}
AMI Consortium: Zwart J.~T.~L., et al., 2008, MNRAS, 391, 1545

\bibitem[\protect\citeauthoryear{AMI Consortium: Zwart et al.}{2011}]{2011MNRAS.tmp.1931Z} AMI Consortium: Zwart J.~T.~L., et al., 2011, MNRAS, 1931 

\bibitem[\protect\citeauthoryear{Andersson et 
al.}{2011}]{2011ApJ...738...48A} Andersson K., et al., 2011, ApJ, 738, 48 



\bibitem[\protect\citeauthoryear{Arnaud, Pointecouteau,
\& Pratt}{2005}]{2005A&A...441..893A} Arnaud M., Pointecouteau E., Pratt G.~W., 2005, A\&A, 441, 893

\bibitem[\protect\citeauthoryear{Arnaud et
al.}{2010}]{2010A&A...517A..92A} Arnaud M., Pratt G.~W., Piffaretti R., B{\"o}hringer H., Croston J.~H., Pointecouteau E., 2010, A\&A, 517, A92

\bibitem[\protect\citeauthoryear{Bartlett
\& Silk}{1994}]{1994ApJ...423...12B} Bartlett J.~G., Silk J., 1994, ApJ, 423, 12

\bibitem[\protect\citeauthoryear{Bautz et al.}{2009}]{2009PASJ...61.1117B} 
Bautz M.~W., et al., 2009, PASJ, 61, 1117 

\bibitem[\protect\citeauthoryear{Birkinshaw}{1999}]{1999PhR...310...97B}
Birkinshaw M., 1999, PhR, 310, 97

\bibitem[\protect\citeauthoryear{B{\"o}hringer et 
al.}{2007}]{2007A&A...469..363B} B{\"o}hringer H. et al., 2007, A\&A, 469, 363 

\bibitem[\protect\citeauthoryear{Borgani et 
al.}{2004}]{2004MNRAS.348.1078B} Borgani S., et al., 2004, MNRAS, 348, 1078 

\bibitem[\protect\citeauthoryear{Borgani}{2004}]{2004Ap&SS.294...51B} Borgani S., 2004, Ap\&SS, 294, 51 

\bibitem[\protect\citeauthoryear{Carlstrom, Holder,
\& Reese}{2002}]{2002ARA&A..40..643C} Carlstrom J.~E., Holder G.~P., Reese E.~D., 2002, ARA\&A, 40, 643

\bibitem[\protect\citeauthoryear{Cavaliere
\& Fusco-Femiano}{1976}]{1976A&A....49..137C} Cavaliere A., Fusco-Femiano R., 1976, A\&A, 49, 137
\bibitem[\protect\citeauthoryear{Cavaliere
\& Fusco-Femiano}{1978}]{1978A&A....70..677C} Cavaliere A., Fusco-Femiano R., 1978, A\&A, 70, 677

\bibitem[\protect\citeauthoryear{Challinor
\& Lasenby}{1998}]{1998ApJ...499....1C} Challinor A., Lasenby A., 1998, ApJ, 499, 1


\bibitem[\protect\citeauthoryear{da Silva et
al.}{2004}]{2004MNRAS.348.1401D} da Silva A.~C., Kay S.~T., Liddle A.~R.,
Thomas P.~A., 2004, MNRAS, 348, 1401

\bibitem[\protect\citeauthoryear{Eke et al.}{1998}]{1998MNRAS.298.1145E} 
Eke V.~R., Cole S., Frenk C.~S., Patrick Henry J., 1998, MNRAS, 298, 1145 

\bibitem[\protect\citeauthoryear{Eke, Navarro, 
\& Frenk}{1998}]{1998ApJ...503..569E} Eke V.~R., Navarro J.~F., Frenk C.~S., 1998, ApJ, 503, 569 

\bibitem[\protect\citeauthoryear{Ettori et
al.}{2009}]{2009A&A...501...61E} Ettori S., Morandi A., Tozzi P., Balestra I., Borgani S., Rosati P., Lovisari L., Terenziani F., 2009, A\&A, 501, 61

\bibitem[\protect\citeauthoryear{Evrard, Metzler, 
\& Navarro}{1996}]{1996ApJ...469..494E} Evrard A.~E., Metzler C.~A., Navarro J.~F., 1996, ApJ, 469, 494 

\bibitem[\protect\citeauthoryear{Evrard et al.}{2002}]{2002ApJ...573....7E}
Evrard A.~E., et al., 2002, ApJ, 573, 7

\bibitem[\protect\citeauthoryear{Feroz
\& Hobson}{2008}]{2008MNRAS.384..449F} Feroz F., Hobson M.~P., 2008, MNRAS, 384, 449

\bibitem[\protect\citeauthoryear{Feroz, Hobson,
\& Bridges}{2009}]{2009MNRAS.398.1601F} Feroz F., Hobson M.~P., Bridges M., 2009, MNRAS, 398, 1601

\bibitem[\protect\citeauthoryear{Feroz et al.}{2009}]{2009MNRAS.398.2049F}
Feroz F., Hobson M.~P., Zwart J.~T.~L., Saunders R.~D.~E., Grainge
K.~J.~B., 2009, MNRAS, 398, 2049

\bibitem[\protect\citeauthoryear{Finoguenov, Reiprich, {\ 
B&ouml}hringer}{2001}]{2001A&A...368..749F} Finoguenov A., Reiprich T.~H., B{\"o}hringer H., 2001, A\&A, 368, 749 

\bibitem[\protect\citeauthoryear{Finoguenov}{2002}]{2002ASPC..253...71F} 
Finoguenov A., 2002, ASPC, 253, 71 

\bibitem[\protect\citeauthoryear{Fixsen et al.}{1996}]{1996ApJ...473..576F} 
Fixsen D.~J., Cheng E.~S., Gales J.~M., Mather J.~C., Shafer R.~A., Wright 
E.~L., 1996, ApJ, 473, 576 

\bibitem[\protect\citeauthoryear{George et al.}{2009}]{2009MNRAS.395..657G} 
George M.~R., Fabian A.~C., Sanders J.~S., Young A.~J., Russell H.~R., 
2009, MNRAS, 395, 657 


\bibitem[\protect\citeauthoryear{Grainge et
al.}{2002}]{2002MNRAS.333..318G} Grainge K., Jones M.~E., Pooley G.,
Saunders R., Edge A., Grainger W.~F., Kneissl R., 2002, MNRAS, 333, 318

\bibitem[\protect\citeauthoryear{Grego et al.}{2001}]{2001ApJ...552....2G}
Grego L., Carlstrom J.~E., Reese E.~D., Holder G.~P., Holzapfel W.~L., Joy
M.~K., Mohr J.~J., Patel S., 2001, ApJ, 552, 2

\bibitem[\protect\citeauthoryear{Hallman et 
al.}{2007}]{2007ApJ...665..911H} Hallman E.~J., Burns J.~O., Motl P.~M., 
Norman M.~L., 2007, ApJ, 665, 911 

\bibitem[\protect\citeauthoryear{Hobson
\& Maisinger}{2002}]{2002MNRAS.334..569H} Hobson M.~P., Maisinger K., 2002, MNRAS, 334, 569

\bibitem[\protect\citeauthoryear{Hoshino et 
al.}{2010}]{2010PASJ...62..371H} Hoshino A., et al., 2010, PASJ, 62, 371 

\bibitem[\protect\citeauthoryear{Kawaharada et 
al.}{2010}]{2010ApJ...714..423K} Kawaharada M., et al., 2010, ApJ, 714, 423 



\bibitem[\protect\citeauthoryear{Itoh, Kohyama,
\& Nozawa}{1998}]{1998ApJ...502....7I} Itoh N., Kohyama Y., Nozawa S., 1998, ApJ, 502, 7

\bibitem[\protect\citeauthoryear{Jones et al.}{1993}]{1993Natur.365..320J}
Jones M., et al., 1993, Natur, 365, 320

\bibitem[\protect\citeauthoryear{Kaiser}{1986}]{1986MNRAS.222..323K} Kaiser
N., 1986, MNRAS, 222, 323

\bibitem[\protect\citeauthoryear{Komatsu et 
al.}{2011}]{2011ApJS..192...18K} Komatsu E., et al., 2011, ApJS, 192, 18 


\bibitem[\protect\citeauthoryear{Kravtsov 
\& Gnedin}{2005}]{2005ApJ...623..650K} Kravtsov A.~V., Gnedin O.~Y., 2005, ApJ, 623, 650 

\bibitem[\protect\citeauthoryear{Kravtsov, Nagai, 
\& Vikhlinin}{2005}]{2005ApJ...625..588K} Kravtsov A.~V., Nagai D., Vikhlinin A.~A., 2005, ApJ, 625, 588 

\bibitem[\protect\citeauthoryear{Kravtsov, Vikhlinin,
\& Nagai}{2006}]{2006ApJ...650..128K} Kravtsov A.~V., Vikhlinin A., Nagai D., 2006, ApJ, 650, 128

\bibitem[\protect\citeauthoryear{Larson et al.}{2011}]{2011ApJS..192...16L} 
Larson D., et al., 2011, ApJS, 192, 16 

\bibitem[\protect\citeauthoryear{Lewis, Challinor,
\& Lasenby}{2000}]{2000ApJ...538..473L} Lewis A., Challinor A., Lasenby A., 2000, ApJ, 538, 473


\bibitem[\protect\citeauthoryear{Lloyd-Davies, Ponman,
\& Cannon}{2000}]{2000MNRAS.315..689L} Lloyd-Davies E.~J., Ponman T.~J., Cannon D.~B., 2000, MNRAS, 315, 689

\bibitem[\protect\citeauthoryear{Mason
\& Myers}{2000}]{2000ApJ...540..614M} Mason B.~S., Myers S.~T., 2000, ApJ, 540, 614

\bibitem[\protect\citeauthoryear{Maughan}{2007}]{2007ApJ...668..772M}
Maughan B.~J., 2007, ApJ, 668, 772

\bibitem[\protect\citeauthoryear{Maughan et
al.}{2007}]{2007ApJ...659.1125M} Maughan B.~J., Jones C., Jones L.~R., Van
Speybroeck L., 2007, ApJ, 659, 1125

\bibitem[\protect\citeauthoryear{McCarthy, Bower,
\& Balogh}{2007}]{2007MNRAS.377.1457M} McCarthy I.~G., Bower R.~G., Balogh M.~L., 2007, MNRAS, 377, 1457

\bibitem[\protect\citeauthoryear{Mitchell et 
al.}{2009}]{2009MNRAS.395..180M} Mitchell N.~L., McCarthy I.~G., Bower 
R.~G., Theuns T., Crain R.~A., 2009, MNRAS, 395, 180 

\bibitem[\protect\citeauthoryear{Mroczkowski et
al.}{2009}]{2009ApJ...694.1034M} Mroczkowski T., et al., 2009, ApJ, 694,
1034

\bibitem[\protect\citeauthoryear{Nagai}{2006}]{2006ApJ...650..538N} Nagai
D., 2006, ApJ, 650, 538

\bibitem[\protect\citeauthoryear{Nagai, Kravtsov,
\& Vikhlinin}{2007}]{2007ApJ...668....1N} Nagai D., Kravtsov A.~V., Vikhlinin A., 2007, ApJ, 668, 1

\bibitem[\protect\citeauthoryear{Nagai 
\& Lau}{2011}]{2011ApJ...731L..10N} Nagai D., Lau E.~T., 2011, ApJ, 731, L10 


\bibitem[\protect\citeauthoryear{Nagai}{2011}]{2011MmSAI..82..594N} Nagai 
D., 2011, MmSAI, 82, 594 

\bibitem[\protect\citeauthoryear{Nozawa, Itoh,
\& Kohyama}{1998}]{1998ApJ...508...17N} Nozawa S., Itoh N., Kohyama Y., 1998, ApJ, 508, 17

\bibitem[\protect\citeauthoryear{Piffaretti 
\& Valdarnini}{2008}]{2008A&A...491...71P} Piffaretti R., Valdarnini R., 2008, A\&A, 491, 71 

\bibitem[\protect\citeauthoryear{Plagge et al.}{2010}]{2010ApJ...716.1118P}
Plagge T., et al., 2010, ApJ, 716, 1118

\bibitem[\protect\citeauthoryear{Planck Collaboration et 
al.}{2011}]{2011A&A...536A..12A} Planck Collaboration, et al., 2011, A\&A, 536, A12 


\bibitem[\protect\citeauthoryear{Placnk Collaboration et 
al.}{2011}]{2011A&A...536A..11A} Placnk Collaboration, et al., 2011, A\&A, 536, A11 


\bibitem[\protect\citeauthoryear{Planck Collaboration et 
al.}{2011}]{2011A&A...536A..10A} Planck Collaboration, et al., 2011, A\&A, 536, A10 


\bibitem[\protect\citeauthoryear{Planck Collaboration et 
al.}{2011}]{2011A&A...536A...9A} Planck Collaboration, et al., 2011, A\&A, 536, A9 


\bibitem[\protect\citeauthoryear{Planck Collaboration et 
al.}{2011}]{2011A&A...536A...8A} Planck Collaboration, et al., 2011, A\&A, 536, A8 

\bibitem[\protect\citeauthoryear{Pointecouteau, Giard,
\& Barret}{1998}]{1998A&A...336...44P} Pointecouteau E., Giard M., Barret D., 1998, A\&A, 336, 44

\bibitem[\protect\citeauthoryear{Pointecouteau, Arnaud,
\& Pratt}{2005}]{2005A&A...435....1P} Pointecouteau E., Arnaud M., Pratt G.~W., 2005, A\&A, 435, 1

\bibitem[\protect\citeauthoryear{Ponman, Cannon, 
\& Navarro}{1999}]{1999Natur.397..135P} Ponman T.~J., Cannon D.~B., Navarro J.~F., 1999, Nature, 397, 135

\bibitem[\protect\citeauthoryear{Ponman, Sanderson,
\& Finoguenov}{2003}]{2003MNRAS.343..331P} Ponman T.~J., Sanderson A.~J.~R., Finoguenov A., 2003, MNRAS, 343, 331

\bibitem[\protect\citeauthoryear{Pratt
\& Arnaud}{2002}]{2002A&A...394..375P} Pratt G.~W., Arnaud M., 2002, A\&A, 394, 375

\bibitem[\protect\citeauthoryear{Pratt, Arnaud, 
\& Pointecouteau}{2006}]{2006A&A...446..429P} Pratt G.~W., Arnaud M., Pointecouteau E., 2006, A\&A, 446, 429 

\bibitem[\protect\citeauthoryear{Pratt, Arnaud, 
\& Pointecouteau}{2006}]{2006ESASP.604..695P} Pratt G.~W., Arnaud M., Pointecouteau E., 2006, ESASP, 604, 695 

\bibitem[\protect\citeauthoryear{Pratt et 
al.}{2010}]{2010A&A...511A..85P} Pratt G.~W., et al., 2010, A\&A, 511, A85

\bibitem[\protect\citeauthoryear{Press
\& Schechter}{1974}]{1974ApJ...187..425P} Press W.~H., Schechter P., 1974, ApJ, 187, 425

\bibitem[\protect\citeauthoryear{Reiprich et 
al.}{2009}]{2009A&A...501..899R} Reiprich T.~H., et al., 2009, A\&A, 501, 899 


\bibitem[\protect\citeauthoryear{Rephaeli}{1995}]{1995ARA&A..33..541R} Rephaeli Y., 1995, ARA\&A, 33, 541

\bibitem[\protect\citeauthoryear{Sanderson
\& Ponman}{2003}]{2003MNRAS.345.1241S} Sanderson A.~J.~R., Ponman T.~J., 2003, MNRAS, 345, 1241

\bibitem[\protect\citeauthoryear{Sanderson et
al.}{2003}]{2003MNRAS.340..989S} Sanderson A.~J.~R., Ponman T.~J.,
Finoguenov A., Lloyd-Davies E.~J., Markevitch M., 2003, MNRAS, 340, 989

\bibitem[\protect\citeauthoryear{Sarazin}{1988}]{1988xrec.book.....S}
  Sarazin C.~L., 1988, X--ray Emission from Clusters of Galaxies, Cambridge University Press

\bibitem[\protect\citeauthoryear{Sarazin}{2008}]{2008LNP...740....1S}
Sarazin C.~L., 2008, LNP, 740, 1

\bibitem[\protect\citeauthoryear{Simionescu et 
al.}{2011}]{2011Sci...331.1576S} Simionescu A., et al., 2011, Sci, 331, 
1576 

\bibitem[\protect\citeauthoryear{Skilling}{2004}]{2004 AIP Conf.Proc.....S}
Skilling J., 2004,AIP Conf.Proc.,735, 395

\bibitem[\protect\citeauthoryear{Sunyaev
\& Zeldovich}{1970}]{1970CoASP...2...66S} Sunyaev R.~A., Zeldovich Y.~B., 1970, CoASP, 2, 66

\bibitem[\protect\citeauthoryear{Tozzi 
\& Norman}{2001}]{2001ApJ...546...63T} Tozzi P., Norman C., 2001, ApJ, 546, 63 

\bibitem[\protect\citeauthoryear{Urban et al.}{2011}]{2011MNRAS.414.2101U} 
Urban O., Werner N., Simionescu A., Allen S.~W., B{\"o}hringer H., 2011, 
MNRAS, 414, 2101 

\bibitem[\protect\citeauthoryear{Vikhlinin et 
al.}{2005}]{2005ApJ...628..655V} Vikhlinin A., Markevitch M., Murray S.~S., 
Jones C., Forman W., Van Speybroeck L., 2005, ApJ, 628, 655 

\bibitem[\protect\citeauthoryear{Vikhlinin et 
al.}{2006}]{2006ApJ...640..691V} Vikhlinin A., Kravtsov A., Forman W., 
Jones C., Markevitch M., Murray S.~S., Van Speybroeck L., 2006, ApJ, 640, 
691 
\bibitem[\protect\citeauthoryear{Voit}{2000}]{2000ApJ...543..113V} Voit
G.~M., 2000, ApJ, 543, 113

\bibitem[\protect\citeauthoryear{Voit
\& Ponman}{2003}]{2003ApJ...594L..75V} Voit G.~M., Ponman T.~J., 2003, ApJ, 594, L75

\bibitem[\protect\citeauthoryear{Voit}{2004}]{2004ogci.conf..253V} Voit 
G.~M., 2004, IAU Colloq. 195: Outskirts of Galaxy Clusters: Intense Life in the Suburbs, ogci.conf, 253 

\bibitem[\protect\citeauthoryear{Voit}{2005}]{2005RvMP...77..207V} Voit
G.~M., 2005, RvMP, 77, 207


\bibitem[\protect\citeauthoryear{Wadsley, Veeravalli, 
\& Couchman}{2008}]{2008MNRAS.387..427W} Wadsley J.~W., Veeravalli G., Couchman H.~M.~P., 2008, MNRAS, 387, 427 

\bibitem[\protect\citeauthoryear{Yoshikawa, Jing, 
\& Suto}{2000}]{2000ApJ...535..593Y} Yoshikawa K., Jing Y.~P., Suto Y., 2000, ApJ, 535, 593 

\end{thebibliography}
\end{document}